\pdfoutput=1
\documentclass[10pt,twocolumn]{article}

\usepackage[letterpaper,top=0.8in,bottom=0.8in,left=0.7in,right=0.7in,columnsep=0.25in,headheight=13pt]{geometry}
\usepackage[T1]{fontenc}
\usepackage[utf8]{inputenc}
\usepackage{amsmath}
\usepackage{newtxtext,newtxmath}
\usepackage[scaled=0.92]{helvet}
\usepackage{microtype}
\DeclareUnicodeCharacter{2011}{\mbox{-}}
\DeclareUnicodeCharacter{01CE}{\v{a}}

\usepackage{graphicx}
\graphicspath{{figures/}}
\usepackage[dvipsnames]{xcolor}
\usepackage{textcomp}
\usepackage{booktabs}
\usepackage{float}
\usepackage{placeins}
\usepackage{siunitx}
\usepackage{xurl}
\usepackage[numbers,square]{natbib}
\usepackage[labelfont=bf,labelsep=period,font=footnotesize]{caption}
\usepackage{titlesec}
\usepackage{fancyhdr}

\usepackage[most]{tcolorbox}
\tcbset{colback=gray!10,
colframe=gray!50,
boxrule=0.4pt,
arc=1mm,
boxsep=1pt,
  left=2pt,
  right=2pt,
  top=2pt,
  bottom=2pt
  }

\newtcolorbox{promptbox}{
    width=0.8\textwidth,
    fontupper=\small\tt\color{black},
    boxrule=1.5pt,
    colframe=black,
    halign=left,
    arc=5pt
}

\usepackage{hyperref}
\hypersetup{
    colorlinks = true,
    citecolor  = blue,
    linkcolor  = blue,
    urlcolor   = blue,
    allcolors  = blue,
    pdftitle   = {Asymmetric rhetorical strategies drive partisanship in television news},
    pdfauthor  = {Upasana Dutta, Homa Hosseinmardi, Amir Ghasemian, Aaron Clauset, Duncan J. Watts}
}

\titleformat{\section}{\sffamily\large\bfseries}{\thesection.}{0.5em}{}
\titleformat{\subsection}{\sffamily\normalsize\bfseries}{\thesubsection.}{0.5em}{}
\titleformat{\subsubsection}{\sffamily\small\bfseries}{\thesubsubsection.}{0.5em}{}
\titlespacing*{\section}{0pt}{2.2ex plus 1ex minus .2ex}{1ex plus .2ex}
\titlespacing*{\subsection}{0pt}{1.8ex plus 1ex minus .2ex}{0.7ex plus .2ex}
\titlespacing*{\subsubsection}{0pt}{1.4ex plus 1ex minus .2ex}{0.5ex plus .2ex}

\begin{document}

\twocolumn[{%
\begin{flushleft}
{\sffamily\bfseries\LARGE Asymmetric rhetorical strategies drive partisanship in\\television news\par}
\vspace{10pt}
{\sffamily\normalsize Upasana Dutta\textsuperscript{1,$\ast$}, Homa Hosseinmardi\textsuperscript{2}, Amir Ghasemian\textsuperscript{1}, Aaron Clauset\textsuperscript{3,4,5}, and Duncan J. Watts\textsuperscript{1,6,7}\par}
\vspace{6pt}
{\sffamily\footnotesize
\textsuperscript{1}Department of Computer and Information Science, University of Pennsylvania, PA, USA\\
\textsuperscript{2}Department of Communication, University of California, Los Angeles, CA, USA\\
\textsuperscript{3}Department of Computer Science, University of Colorado, CO, USA\\
\textsuperscript{4}BioFrontiers Institute, University of Colorado, CO, USA\\
\textsuperscript{5}Santa Fe Institute, NM, USA\\
\textsuperscript{6}Annenberg School of Communication, University of Pennsylvania, PA, USA\\
\textsuperscript{7}Operations, Information, and Decisions Department, University of Pennsylvania, PA, USA\\
\textsuperscript{$\ast$}To whom correspondence should be addressed: \href{mailto:upasanad@engineering.upenn.edu}{upasanad@engineering.upenn.edu}
\par}
\end{flushleft}
\vspace{2pt}
\noindent\rule{\textwidth}{0.5pt}\par
\vspace{4pt}
{\small\noindent{\sffamily\bfseries Abstract.} Television remains a dominant source of political information in the United States, yet systematic evidence on how partisanship is expressed in televised news remains limited. Prior research has largely treated partisanship as a one-dimensional construct, offering limited insight into the rhetorical strategies that underlie ideological bias. Here, by analyzing over 400,000 news episodes using large language models (LLMs), we provide a multi-dimensional measure of partisanship bias in television news across six major U.S. news stations (2013--2025). We identify eight interpretable linguistic features that explain 73\% of partisanship variance in television news.\ Our findings reveal systematic asymmetries across stations. While MSNBC and CNN shifted leftward during Republican presidencies and moderated under Democratic presidencies, Fox News increased its conservative alignment following each election regardless of the electoral outcome. Beyond aggregate trends, MSNBC and CNN emphasized attacks on Republican actors more than reinforcement of Democratic positions, whereas Fox News reinforced Republican viewpoints substantially more than it attacked Democratic viewpoints. The association between in-group praise and out-group attacks was also threefold stronger on Fox News than on MSNBC. These results are validated against trained human raters and replicated with four additional LLMs with differing political leanings. Together, our study reveals systematic station-level differences in the rhetorical strategies with which ideological slant is communicated in television news, opening new opportunities for future research on media effects, audience attitudes, and political communication.\par
\vspace{5pt}
\noindent{\sffamily\bfseries Keywords:} television news, media bias, partisanship, political communication\par}
\vspace{4pt}
\noindent\rule{\textwidth}{0.5pt}\par
\vspace{14pt}
}]
\thispagestyle{plain}


\section{Introduction}

Partisan media have played a central role in shaping political attitudes and discourse in the United States, shaping how individuals interpret current events and form political identities and opinions~\cite{stroud2011niche, prior2007post, arceneaux2022changing}. \textcolor{black} {Although the contemporary news environment is increasingly fragmented across digital and social media platforms, television remains a major component of Americans' news diets. In 2025, 64\% of U.S. adults reported getting news from television (either sometimes or often), and 34\% identified television as their preferred platform for news---more than news websites or apps (21\%), social media (14\%), or podcasts (6\%)~\cite{pew2025platform}. %
Importantly, television's prominence is not simply an artifact of self-reported media use. Studies based on behavioral tracking have consistently found television to account for a substantial share of Americans' news consumption~\cite{allen2020evaluating, robinson2018stubborn}. More recently, a multi-platform analysis tracking approximately 55,000 Americans found that television news reached roughly 80\% of panelists---substantially more than news on the web or YouTube---and accounted for more time spent consuming news than either online platform~\cite{yang2026platforms}. Thus, even within an increasingly multiplatform information environment, television remains a large and consequential arena of political communication today.}

\textcolor{black} {Its importance is especially pronounced in the context of partisan news exposure. Audiences of television news are substantially more ideologically segregated than audiences of news on the web~\cite{muise2022quantifying}.} Moreover, exposure to partisan television has measurable downstream effects: it can shift political attitudes and beliefs~\cite{martin2017bias, ash2024viewers, broockman2025consuming}, increase affective polarization~\cite{gill2022testing}, alter policy preferences~\cite{gainous2023surfing}, and reduce trust in opposing viewpoints~\cite{levendusky2013partisan}. \textcolor{black} {Yet despite television's continued reach, ideological segregation, and demonstrated political effects, comparatively little is known empirically about the rhetorical forms through which partisan perspectives are expressed in television news.}

\textcolor{black} {Prior research has developed a range of approaches for quantifying partisan slant and polarization in media content. A foundational approach by Gentzkow and Shapiro~\cite{gentzkow2010drives} measures media slant by comparing the language used by news outlets with that of Democratic and Republican members of Congress~\cite{martin2017bias, martin2019local, hassell2022does, widmer2022media, djourelova2023persuasion}. Other work has characterized partisanship in television through visibility-based measures, such as the campaign donation histories of on-air guests~\cite{kim2022measuring}, while recent longitudinal analyses have documented increasing polarization among cable stations in both the topics they select and the language they use when covering the same topics~\cite{hosseinmardi2025unpacking}. Subsequent studies use sentiment analysis, topic modeling, and manual frame coding~\cite{baum2008new, lin2011more, ding2023same, budak2016fair}, or incorporate lexical and embedding-based features~\cite{preoctiuc2017beyond} for measuring partisanship. While these approaches are valuable for establishing the direction, magnitude, and evolution of partisan differences, they do not reveal the rhetorical composition of those differences, thus obscuring meaningful differences in how partisanship is expressed. Similar levels of partisan slant may arise from very different forms of expression---for example, reinforcing aligned viewpoints, criticizing opposing viewpoints, or attributing blame to political actors. Such distinctions are substantively important. Prior work shows that media effects depend not only on which issues receive attention, but on how political actors and conflicts are framed, who is blamed, and whether coverage reinforces in-group or attacks out-group positions~\cite{entman1993framing, iyengar1994anyone, levendusky2013partisanpoliticalcomm}. Related research further shows that rhetorical style itself can matter: Televised incivility and emotionally charged political discourse can affect political trust and audience responses~\cite{mutz2005new, brader2005striking}. Motivated by these distinctions, we decompose aggregate partisan slant into recurring and interpretable forms of rhetorical strategies, and ask whether politically opposed stations rely on those strategies differently.}

\textcolor{black} {Recent advances in large language models (LLMs) make it possible to capture context-dependent rhetorical and ideological features of political language that are otherwise difficult to represent using predefined lexicons or word-level measures, while doing so at scale.} LLMs have been shown to reliably classify ideology in tweets and legislative documents~\cite{le2024positioning, heseltine2024large}, estimate the ideological positions of political actors~\cite{wu2023large}, perform natural language inference in political contexts~\cite{burnham2026political}, detect frames and stances in text~\cite{gilardi2023chatgpt, burnham2025stance}, and support fine-grained measurement of political lean, tone, and topic selection in news content~\cite{haider2026media}, demonstrating their potential to infer political meaning without relying on manual coding or predefined lexicons~\cite{ziems2024can}. 

Building on these advances in large language models, in this paper we make two high-level contributions to the study of partisan bias in media.

First, using an LLM-based framework for measuring ideological slant in U.S. television news, we assess the partisanship in 408,464 news episodes (3,505,401 segments) from six major U.S. stations---Fox News, MSNBC, CNN, ABC, CBS, and NBC---spanning January 2013 to March 2025. Our analysis reveals some striking trends as well as systematic asymmetries in how partisanship is expressed. Notably, during our study period, while partisanship increased over time on both the left and the right, we find that Fox News was substantially more partisan overall than either MSNBC or CNN, and that its partisanship tended to increase regardless of which party was in power, whereas MSNBC and CNN tended to fluctuate with presidential incumbency.

We then show that this asymmetry between right and left applies consistently across 12 major topics such as immigration, social justice, climate change, healthcare, and U.S. Elections. Across all topics, Fox was more partisan than MSNBC, while MSNBC was, in turn, more partisan than CNN.

Finally, in contrast with the cable stations, we find that the three broadcast stations---ABC, CBS, and NBC---have remained essentially neutral over this period. At the topic level, we find that the broadcast stations' aggregate neutrality stemmed in part from less partisan reporting across all topics we analyze, and in part because they allocated a higher share of their coverage to topics such as crime that are less partisan across all stations.

Second, we propose a model of partisanship that explains our measured partisan scores using eight inductively identified rhetorical strategies, consistent with prior work on the linguistic signatures of partisanship\textcolor{black} {~\cite{entman1993framing, iyengar1994anyone, mutz2005new, brader2005striking}}. Prior research on framing has largely relied on lexicon-based approaches that operationalize frames through predefined keywords associated with particular narratives~\cite{boydstun2014tracking, card2022computational}. While effective, such methods often overlook implicit and context-dependent language. Our LLM-based approach overcomes these limitations by providing more nuanced, semantically grounded measurement. We find that a simple linear model with these eight features---expression of negative sentiment toward Republicans and Democrats, reinforcement of Republican and Democratic viewpoints, blame attribution toward each party, prevalence of inflammatory language, and emphasis on fact-based reporting---explained 73\% of the variance in partisanship \hbox{$(  R^2=0.73,\space \text{out of sample})$}, demonstrating that ideological bias can largely be explained by a compact and interpretable set of rhetorical strategies. \textcolor{black} {Crucially, this decomposition reveals differences across stations that are obscured by a single liberal--conservative score.} Partisanship, therefore, is better understood not as a one-dimensional construct, as commonly operationalized in prior work~\cite{gentzkow2010drives, martin2017bias, gentzkow2019measuring, kim2022measuring}, but as a configuration of recurring rhetorical strategies that capture the linguistic mechanisms through which ideological slant is expressed.

We then leverage the rhetorical strategies to derive additional insights into the partisan asymmetries that we identified at the aggregate level.
For example, we find that while both left and right expressed more negative sentiment to the ``other side'' than their own, Fox News was substantially more negative about Democrats and more positive about Republicans than either MSNBC or CNN were about Republicans and Democrats, respectively. Fox was also distinct in that it devoted more coverage to reinforcing Republican viewpoints than to criticizing Democratic views, whereas MSNBC and CNN both attacked Republican viewpoints more than they affirmed those of Democrats. Fox News was also three times more likely to combine in-group praise with out-group attacks than MSNBC, indicating a stronger coupling of these strategies. In addition, we find that Fox News consistently exhibited the highest levels of inflammatory and derogatory language and the lowest reliance on fact-based reporting among all outlets examined, whereas broadcast stations displayed the opposite pattern, with MSNBC and CNN occupying an intermediate position. Finally, we show that while the overall pattern of in-group support versus out-group criticism remained broadly consistent across topics for Fox News, MSNBC and CNN showed stronger topic dependence in the relative emphasis they place on these strategies.

To ensure that our findings are not artifacts of any single language model's underlying political tendencies~\cite{westwood2025measuring, motoki2024more, santurkar2023whose, feng2023pretraining}, we replicated our main findings on a uniformly random subset of our data using four additional closed-source and open-weight LLMs with varying political leanings. The consistency of our findings across all models provides evidence that our conclusions are not artifacts of our choice of model.

Together, our findings reveal stark asymmetries between television news stations --- not only in their ideological alignment and responsiveness to political events, but also in the rhetorical strategies used to communicate their partisanship.

\section{Results}

\subsection{Trends in Partisanship Bias}
\begin{figure*} [t!]
\centering
\includegraphics[width=\textwidth]{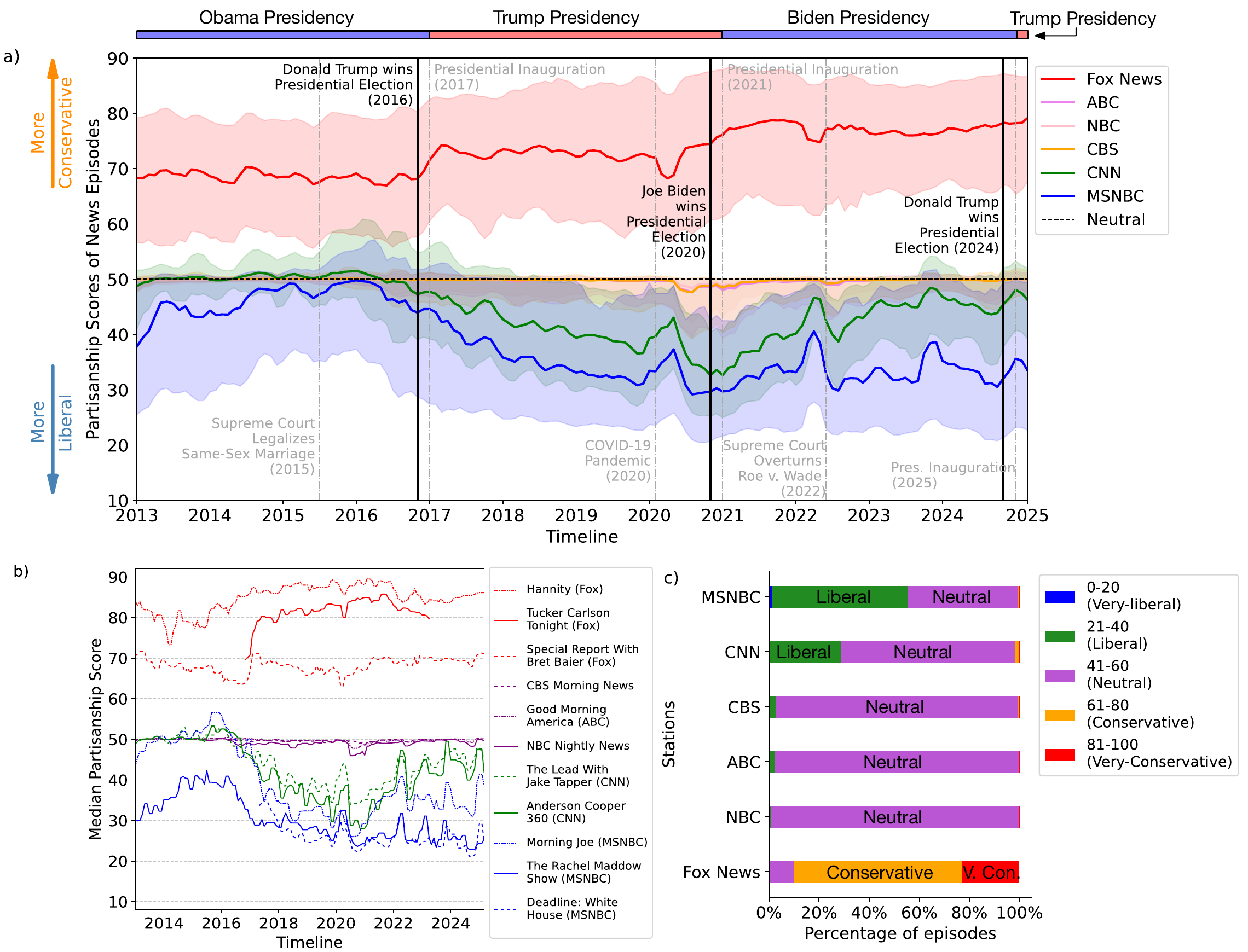}
\caption{\textbf{Partisanship patterns in U.S. television news across time, stations, and programs.}
(a) Monthly partisanship scores for 408,464 television news episodes from 2013--2025, estimated using a neural network trained on segment-level GPT-annotated partisanship labels. Episode-level scores are computed by averaging segment-level scores within each episode. A score of 50 indicates neutral slant; values closer to 0 (100) reflect increasingly liberal (conservative) ideological leanings. Solid lines denote the 3-month moving average of the median partisanship score for each station, while shaded regions represent the 10th to 90th percentile range. Major political events are marked with vertical lines for reference. Fox News consistently exhibited strong conservative alignment. MSNBC and CNN leaned towards liberal, although with a lesser magnitude than Fox leaned conservative, while the broadcast stations remained consistently closer to neutrality. (b) Monthly median partisanship scores for a selection of high-profile programs, plotted as 3-month moving averages, revealing substantial within-station heterogeneity, especially among cable stations: for instance, Fox News includes both highly partisan shows (e.g., Hannity) and relatively moderate ones (e.g., Special Report with Bret Baier). These program-level trajectories are consistent with qualitative expectations and further validate the construct validity of our partisanship scoring approach. c) Distribution of episode-level partisanship scores categorized into five ideological bins. While over 90\% of Fox News content fell within partisan ranges (scores below 40 or above 60), the corresponding share for MSNBC was approximately 55\%, and only 30\% for CNN. These differences challenge the notion of MSNBC as a direct liberal counterpart of Fox News, indicating that it instead had a more moderate and ideologically mixed profile. See SI Appendix Section~\ref{sec:llm_comparison_appendix} for replication of this figure with four alternative LLMs.}
\label{fig:partisanship_trend}
\end{figure*}

To measure the partisanship bias expressed in news language, we prompt GPT-4o to assign a continuous score ranging from 0 (strongly liberal) to 100 (strongly conservative), along with an explanation for the score. The measure demonstrates high agreement with independent human coders (see \hyperref[sec:methods]{Materials and Methods}).

Temporal trends in partisanship across different stations show marked differences (Figure~\ref{fig:partisanship_trend}a). Fox News consistently exhibited the strongest conservative alignment (median partisanship between 68.4--81.2) and increased over time. In contrast, MSNBC consistently leaned liberal (median partisanship between 50.0--26.7), particularly during key political events. CNN showed a similar liberal alignment but with a less pronounced partisan stance than MSNBC (median partisanship between 51.0--31.1). Notably, the wide shaded bands---representing the 10th--90th percentile---reveal substantial variation in partisanship across programs within each station. For instance, Fox News featured both highly partisan programs like `Hannity' (average median score = 84.6) and comparatively moderate ones like `Special Report With Bret Baier' (68.1) (Figure~\ref{fig:partisanship_trend}b). Similarly, MSNBC featured programs like `The Rachel Maddow Show' (30.5) and `The Morning Joe' (40.0). This intra-station heterogeneity not only underscores the multifaceted nature of media bias but also validates the sensitivity of our method in detecting fine-grained within-station ideological variation. Meanwhile, broadcast stations---ABC, NBC, and CBS---remained consistently closer to neutrality  (median partisanship between 49.5--50.0) and displayed significantly narrower variance, suggesting a sustained emphasis on politically balanced reporting compared to their cable counterparts.

The temporal variations observed closely followed significant historical events, underscoring how television news responded dynamically to political events. The period following the 2016 presidential election saw a marked divergence in partisanship trends: MSNBC's median partisanship shifted leftward by -11.8 points, from 46.1 (2013--2016 average) to 34.3 (2017--2020 average); CNN's shifted leftward by \hbox{-8.2} points, from 50.2 to 42.0, while Fox News intensified its conservative stance with a modest rise of +6.3 points from 68.6 to 74.9. Following Joe Biden's election in late 2020, MSNBC and CNN exhibited a sharp shift towards less partisan positions, unlike Fox News, indicating a possible editorial recalibration in response to Trump's departure.

These temporal dynamics highlight a critical difference among the cable stations -- MSNBC and CNN exhibited clear patterns of shifting partisanship aligned closely with presidential incumbency, both becoming markedly more liberal in their coverage during periods of Republican presidency (e.g., Trump's first term, 2016--2020), and moderating their stances during periods of Democratic leadership (e.g., Biden's presidency starting in early 2021). In contrast, Fox News' alignment with conservative positions intensified after every election cycle, regardless of which party won, suggesting a structural editorial bias that persisted irrespective of the party in power.

\begin{figure*}[t!]
\centering
\includegraphics[width=0.8\textwidth]{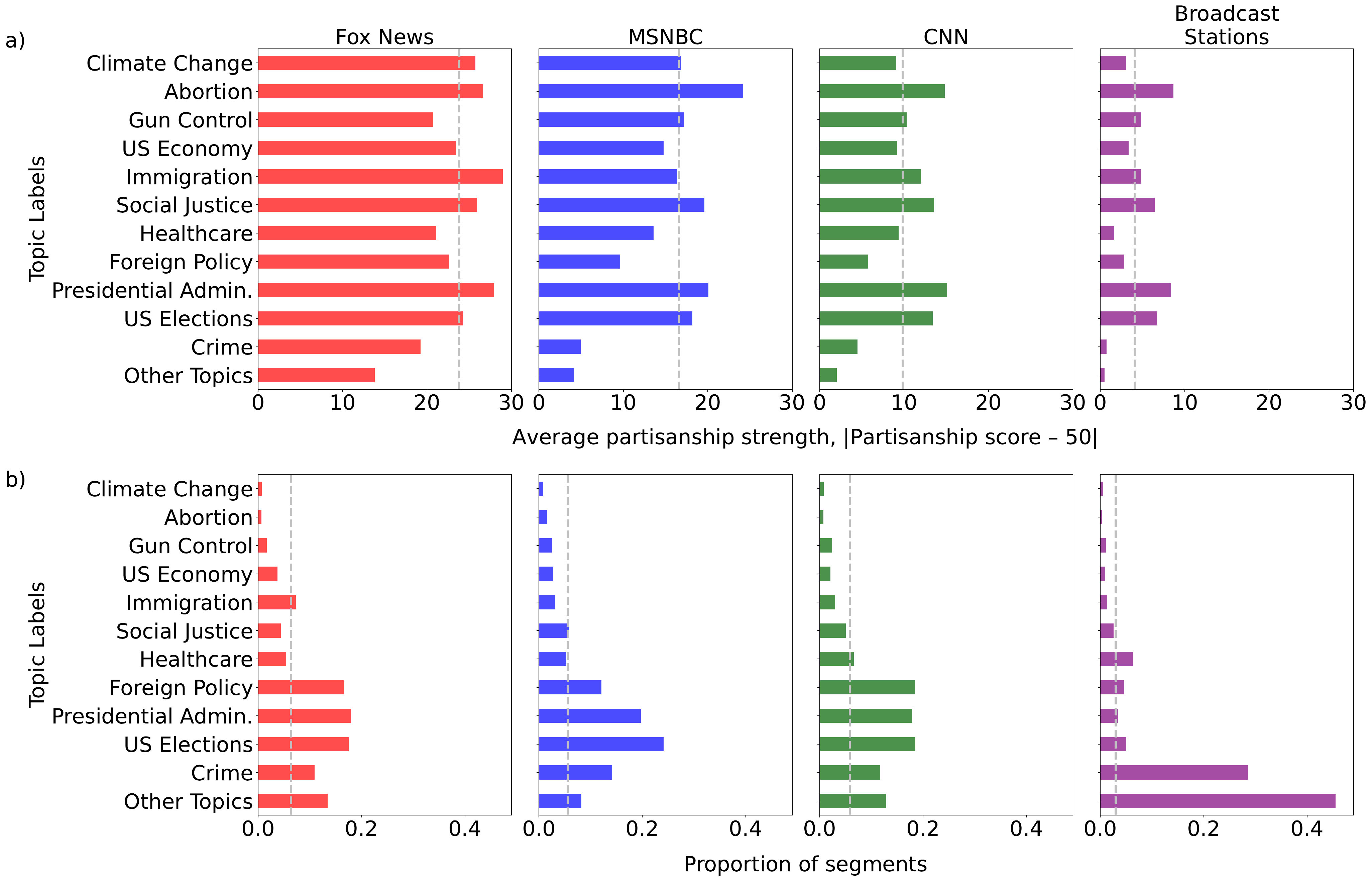}
\caption{\textbf{Topic distribution and topic-specific partisanship across U.S. television news stations (2013--2025).}
(a) Average partisanship strength---the average absolute difference between a segment's partisanship score and ideological neutrality, i.e. ($|$partisanship score -- 50$|$), for 12 topical categories, shown separately for Fox News, MSNBC, CNN, and the combined broadcast stations of ABC, CBS, and NBC. Vertical dashed line denotes the station's median. Fox News displayed uniformly high partisan language across topics, with the sharpest deviations on immigration, presidential administration, abortion, social justice and racial inequality, and climate change. MSNBC was likewise polarized on these ``culture-war" issues but at lower magnitude, CNN was comparatively moderate, and broadcast stations remained near-neutral across all topics. (b) Proportion of total airtime each station devoted to analyzed topics. Vertical dashed line denotes the station's median. Cable stations distributed coverage broadly across politically salient topics, whereas broadcast stations devoted about 75\% of their airtime solely to the two least partisan categories across all stations-- crime and `other topics'. See SI Appendix Section~\ref{sec:llm_comparison_appendix} for replication of this figure with four alternative LLMs.
}
\label{fig:Coverage_and_PartisanshipStrength}
\end{figure*}

A notable pattern also emerged in early 2020 during the initial outbreak of COVID-19, when all cable stations temporarily shifted toward more neutral reporting, suggesting a brief prioritization of factual coverage over partisan reporting. However, this shift was short-lived, with partisan patterns reemerging by mid-2020, ahead of the presidential election later that year. Another major shift occurred in mid-2022, coinciding with the Supreme Court decision to overturn Roe versus Wade, which ended federal protections for abortion rights. Following that ruling, both MSNBC and CNN polarized, with MSNBC shifting more sharply (--5.9 points) than CNN (--4.1 points). CNN gradually returned to a more centrist stance by early 2024, whereas MSNBC maintained its elevated partisanship through 2025. This divergence resulted in a substantial ideological gap between the two liberal-leaning cable stations, averaging 12.2 points over 2023--2025.

Our analyses also diverge from findings from prior work measuring partisanship from campaign contributions of on-air guests~\cite{kim2022measuring}, which found that all three cable stations shifted right after the 2016 election and left after the 2018 midterm elections. Instead, we find that only Fox News moved rightward after 2016 and stayed there through 2018, while MSNBC and CNN sharply shifted left. This contrast highlights that guest-based metrics may capture visibility bias in television news but fail to capture how partisanship is conveyed through language and tone.

Beyond aggregate temporal trends, our analysis reveals a pronounced asymmetry in the ideological composition of news content across stations (Figure~\ref{fig:partisanship_trend}c). While over 90\% of Fox News episodes exhibited clear conservative or very conservative leanings, only approximately 55\% of MSNBC episodes appeared liberal or very liberal, and just 30\% of CNN episodes fell on the liberal side (Figure \ref{fig:partisanship_trend}c). This disparity challenges the conventional portrayal of Fox News and MSNBC as equally polarized opposite ends of the ideological spectrum~\cite{ucsb_2021, hosseinmardi2025unpacking}, suggesting that MSNBC and CNN instead produce less consistently partisan---and more ideologically mixed---content than Fox News.

\subsubsection{Variation of Partisanship with Topics} Having established differences in overall partisanship across stations and over time, we next ask the question of how the stations differed in their partisanship across topics. Existing polling data suggests that public opinion is not equally polarized across all topics; rather, it is particularly acute on certain ``culture war" issues, such as abortion, climate change, and immigration, and less so on others \cite{gallup_polling_matter_215210, gallup_509129, gallup_652970, gallup_395408}. Motivated by such topical disparity, we now investigate whether television news stations exhibited differences across 11 political topics that span hallmark culture-war issues (e.g., Climate change, Abortion and reproductive rights, Gun control, Immigration), core policy domains (e.g., the U.S. economy, Healthcare, Foreign policy), institutional themes (e.g., Presidential administration, U.S. elections, Crime), and broader sociopolitical concerns (e.g., Social justice and racial inequality). A twelfth category (`Other topics') captured segments that did not match any of the predefined topics, which could include non-political news (e.g., weather, sports, natural disasters) and relatively less politicized topics (e.g., education debates, transportation) (see SI Appendix Section~\ref{sec:topic_analysis} for topic classification).

Our analysis reveals that polarization was not evenly distributed across all political topics. A comparison of the absolute differences between each station's average partisanship score for each topic (across 2013-2025) and the neutral partisanship score of 50 (Figure~\ref{fig:Coverage_and_PartisanshipStrength}a) shows that among cable news stations, topics like abortion, presidential administration, climate change, social justice, and immigration were associated with the highest levels of partisanship. In contrast, other topics---such as healthcare and crime---exhibited relatively lower levels of partisanship strength within each station. However, across every topic we analyze, Fox News consistently displayed more pronounced partisanship than MSNBC and CNN. While MSNBC and CNN also showed elevated partisanship on some topics, such as abortion and reproductive rights, and social justice and racial inequality, they did so with relatively less intensity.

Next, we investigate the dynamics behind the neutrality of broadcast stations (ABC, CBS, NBC). Across every topic we analyze, the broadcast stations showed much lower levels of partisanship strength than all three cable stations (Figure~\ref{fig:Coverage_and_PartisanshipStrength}a). Furthermore, a comparison of topic-wise coverage across the cable and broadcast stations reveals that for the broadcast stations, topic selection itself was heavily skewed toward non-partisan content---roughly three-quarters of their programming is concentrated on coverage of crime and segments falling into uncategorized or miscellaneous topics (`other topics' category) (Figure~\ref{fig:Coverage_and_PartisanshipStrength}b), both of which are the least partisan topics consistently across all cable and broadcast stations. This emphasis on de-prioritizing politically divisive topics sets apart the broadcast stations from
the cable stations. That is, not only did they cover all topics with significantly less partisan slant than the cable stations, they also allocated disproportionately more coverage to non-contentious topics such as crime and non-political events.

\subsection{An Interpretable Model of Partisanship Scores}

Although our method produces partisanship scores that are demonstrably accurate and reflect the overall ideological alignment of television news segments, these scores reduce rich rhetorical and stylistic variation to a single dimension. To move beyond this aggregate measure and capture the rhetorical strategies through which partisan slant is expressed, we analyzed the natural language explanations generated by the LLM when providing the partisanship scores (see SI Appendix Figure~\ref{fig:prompt_partisanship} for prompt details). A close reading of these explanations revealed recurring patterns in how partisan alignment was justified. These included negative sentiment toward opposing viewpoints---for instance, ``the transcript criticizes liberal views, ridicules gender inclusion $[\dots]$ and dismisses perceived threats to democracy," or ``the transcript critiques Republican strategies regarding abortion-focused legislation in Ohio." Another prominent theme was reinforcement of preferred narratives, such as: ``the transcript emphasizes struggles faced by undocumented immigrants, highlights fear of deportation," or ``praises market choices $[\dots]$ and praises traditional gasoline infrastructure." Attribution of blame also emerged as a common rhetoric, as illustrated by explanations like: ``the transcript criticizes Republicans and Speaker Boehner, framing them as responsible for the shutdown," or ``the transcript highlights crime spikes under Democratic leadership."

In addition to attitudinal reasoning, the LLM also referenced stylistic features of the reporting. These included emotionally charged language---e.g., ``labeling him a fascist and white supremacist"---and conversely, neutral, fact-focused reporting, for instance, ``the transcript covers factual news on various topics, including Twitter layoffs," or ``the transcript focuses on weather updates and safety measures related to the heat wave."

These recurring elements closely align with foundational theories in political communication, particularly those related to identity construction, blame attribution, and stylistic framing. Past research in political communication has demonstrated the centrality of us-versus-them rhetoric in partisan communication, characterized by in-group reinforcement and out-group criticism~\cite{wirth2016diverse, iyengar2012affect, iyengar2019origins, druckman2021affective, nettasinghe2025group}. In the U.S., these us-versus-them frames have expanded to politicize even ostensibly non-political spheres, such as entertainment, sports, and consumer culture~\cite{dagnes2019us}. Beyond us-vs-them sentiment, partisan rhetoric often relies on explicit blame attribution to political opponents. Prior research shows that partisan identity and media environments shape whom the public blames for negative outcomes, such as natural disasters~\cite{malhotra2008partisan}, COVID-19 response failures~\cite{uttermark2024blame}, and economic crises~\cite{brutger2025framing}.

Partisanship is also shaped by stylistic choices in language.\ Inflammatory or derogatory expressions can intensify polarization by portraying opposing views in morally charged terms and framing political opponents as fundamentally adversarial---an increasingly common rhetorical strategy in contemporary politics~\cite{winberg2017insult, zeitzoff2023nasty}.\ In contrast, fact-based reporting reflects a commitment to journalistic neutrality, where political issues are framed through evidentiary claims rather than moral outrage~\cite{graves2019fact, musolff2016factual}.

Based on our analysis of the LLM-generated explanations, we identified eight interpretable features of partisan language that capture both attitudinal and stylistic dimensions of rhetorical strategy, in line with established frameworks in political communication. The first four features reflect attitudinal strategies: \hbox{(1, 2)}\ negative sentiment toward Democratic/Republican viewpoints, \hbox{(3, 4)} reinforcement of Democratic/Republican viewpoints. Two additional features capture blame attribution, characterizing how political responsibility is assigned: \hbox{(5, 6)} blame attribution towards Democratic/Republican politicians/policies. Finally, we include two stylistic features that describe stylistic strategies: (7) the use of inflammatory or derogatory language, and (8) a focus on fact-based reporting. Together, these eight features provide a framework for analyzing partisanship through a multidimensional lens by examining the rhetorical strategies that shape partisan content in television news.

\begin{table}[t!]
\centering
\caption{Linear regression of partisanship scores using the eight rhetorical strategies as features. These binary features explain 73\% of the variance in partisanship scores, highlighting that political partisanship in television news systematically consists of identifiable linguistic cues---especially those related to ideological reinforcement, negative sentiment, and blame attribution.\label{tab:regression_results}}
\small
\setlength{\tabcolsep}{3pt}
\renewcommand{\arraystretch}{1.1}
\begin{tabular}{@{}lcc@{}}
\toprule
\textbf{Rhetorical Strategy} & \textbf{$\beta$} & SE \\
\midrule
\textit{Attitudinal Strategy} \\
\quad Neg. sentiment toward D. viewpoints & \hspace{1.22em}8.622* & (0.183) \\
\quad Neg. sentiment toward R. viewpoints & \hspace{0.43em}$-$8.221* & (0.156) \\
\quad Reinforcement of D. viewpoints & $-$10.354* & (0.143) \\
\quad Reinforcement of R. viewpoints & \hspace{0.8em}17.131* & (0.158) \\
\quad Blame attribution toward D. & \hspace{1.22em}6.765* & (0.175) \\
\quad Blame attribution toward R. & \hspace{0.43em}$-$9.098* & (0.159) \\
\midrule
\textit{Stylistic Strategy} \\
\quad Use of inflammatory/derogatory language & \hspace{1.1em}1.183* & (0.156) \\
\quad Focus on fact-based reporting & \hspace{0.43em}$-$0.013\phantom{*} & (0.117) \\
\midrule
Intercept & \hspace{0.5em}50.434* & (0.109) \\
\midrule
Observations & \multicolumn{2}{c}{42,679} \\
Test $R^2$ & \multicolumn{2}{c}{0.7261}\\
\bottomrule
\multicolumn{3}{@{}l@{}}{\footnotesize $^{*}p<0.05$}
\end{tabular}
\end{table}

\subsubsection{Partisanship as a Multi-dimensional Construct} 

To assess the degree to which partisanship is explained by these eight rhetorical strategies, we fit a segment-level linear regression model with the eight LLM-labeled binary features as independent variables and the LLM-labeled partisanship score (0-100) as the dependent variable. Despite the model's simplicity, the eight features alone explain 73\% of the variance in partisanship scores on held-out data ($R^2 = 0.73$; see Table~\ref{tab:regression_results})---only modestly lower than a neural network model trained to predict the same partisanship scores using the entire transcript (3,072-dimensional embedding; $R^2$ = 0.80; Figure~\ref{fig:partisanship_humanlabelcomparisons}a; see \hyperref[sec:methods]{Materials and Methods}). In other words, the explained variance $R^2$ in partisanship scores reduces only by 8.8\% (from 0.80 to 0.73), even though the input dimensionality reduces by 99.7\% (from 3072 to 8), suggesting that these eight features capture the dominant dimensions along which partisan slant varies in television news.

The regression coefficients further highlight which features drive ideological slant.
The strongest predictor of conservative partisanship is reinforcement of Republican viewpoints ($\beta = 17.13$), and that of liberal partisanship is reinforcement of Democratic viewpoints ($\beta = -10.35$). Similarly, segments blaming Democrats skew rightward ($\beta = 6.77$), while those blaming Republican actors skew leftward ($\beta = -9.10$). In contrast, stylistic elements exert much weaker effects: the use of inflammatory or derogatory language ($\beta = 1.18$) and fact-based reporting ($\beta = -0.01$) show a negligible association, with both effects far smaller in magnitude than the attitudinal features. The model's intercept term ($\beta = 50.43$) is also conceptually informative, representing the predicted partisanship score for a segment in which none of the eight rhetorical strategies are present---close to the midpoint of the 0--100 scale.

Since both the rhetorical strategies and the partisanship score in this regression are LLM-labeled, we also performed a robustness check to confirm that the association also holds under human labeling. We re-fit the regression using the design-based supervised learning framework (DSL)~\cite{egami2023using, egami2024using}, which combines the LLM and human labels through a doubly-robust procedure. The results remain largely consistent: all six attitudinal coefficients retain their sign and significance, with reinforcement of Republican \hbox{($\beta=21.389$)} and Democratic \hbox{($\beta=-11.204$)} viewpoints remaining the strongest predictors of conservative and liberal slant, respectively, reinforcing that these associations hold under human labeling (see SI Appendix Section~\ref{si:dsl}).

Taken together, these findings show that partisanship in TV news is not merely a black-box construct, but can be systematically disaggregated into a small set of linguistically interpretable features.

\begin{figure*}[t!]
\centering
\includegraphics[width=0.9\textwidth]{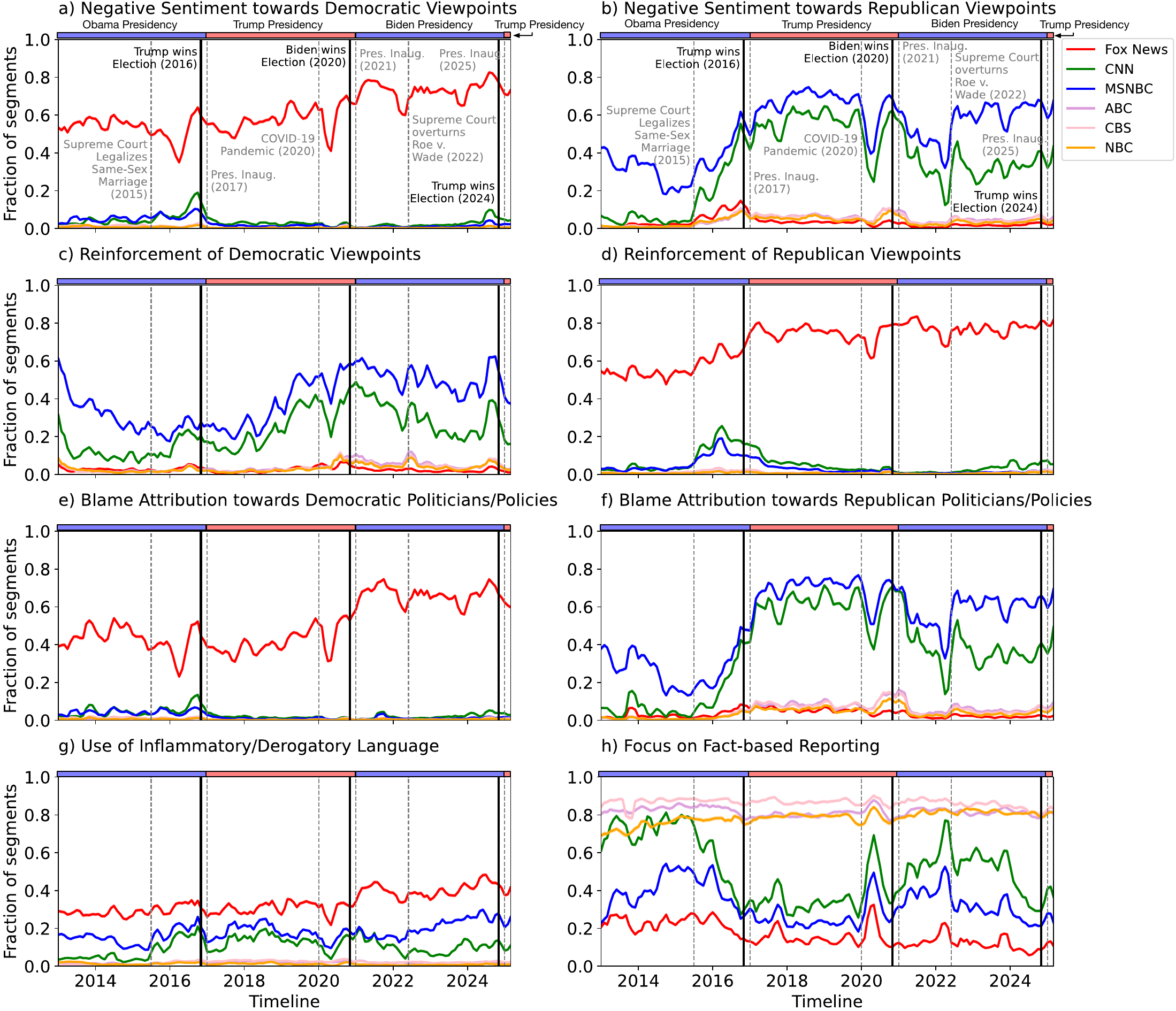}
\caption{\textbf{Temporal trends in rhetorical strategies across broadcast and cable news.} Each panel shows the monthly fraction of segments labeled with one of eight rhetorical features by a neural network model trained on GPT-4o labels, smoothed using a 3-month moving average. These features are: (a--b) negative sentiment towards opposing viewpoints, (c--d) reinforcement of partisan narratives, and (e--f) blame attribution toward partisan actors. Panels (g--h) display broader stylistic strategies---use of inflammatory language and focus on fact-based reporting. Major political events are marked with vertical lines for reference; event annotations appear only in the top row for visual clarity. Notably, Fox News devoted a greater proportion of its coverage to reinforcing Republican viewpoints (mean 69.8\%) than to expressing negative sentiment toward Democratic views (mean 61.6\%), whereas the proportion of MSNBC segments expressing negative sentiment toward Republican views (mean 52.5\%) exceeds the proportion reinforcing Democratic viewpoints~(mean 39.3\%). Broadcast stations consistently exhibited lower levels of partisan strategies and higher levels of fact-based reporting. In contrast to MSNBC and CNN, where sharp shifts follow changes in presidential incumbency, Fox News displayed a more monotonic trajectory in several rhetorical features---either steadily increasing or decreasing over time---largely independent of the party in power. See SI Appendix Section~\ref{sec:llm_comparison_appendix} for replication with four alternative LLMs.}
\label{fig:feature_timetrends}
\end{figure*}

\subsection{Trends in Rhetorical Strategies}

Our analyses reveal several systematic differences between the different news stations, and over time. Fox News exhibited the highest and most sustained levels of negative sentiment toward Democrats (Figure~\ref{fig:feature_timetrends}a), reinforcement of Republican viewpoints (Figure~\ref{fig:feature_timetrends}d), and blame attribution toward Democratic politicians and policies (Figure~\ref{fig:feature_timetrends}e), than any other station. MSNBC and CNN, by contrast, showed the expected ideological reversal---criticizing Republicans and reinforcing Democratic views (Figures~\ref{fig:feature_timetrends}b, c, f). Between MSNBC and CNN, the former exhibited higher levels of partisan features than the latter across all dimensions, though the magnitude of both reinforcing and attacking strategies in MSNBC/CNN remained lower than the corresponding strategies observed in Fox News. However, there is a notable difference in emphasis: Fox News devoted a greater proportion of its coverage to reinforcing Republican viewpoints (69.8\%; 95\% CI: 69.7--69.9\%) than to expressing negative sentiment toward Democratic views (61.6\%; 95\% CI: 61.5--61.7\%; $p < 0.001$, one-sided test of proportions). On the other hand, the proportion of MSNBC/CNN segments expressing negative sentiment toward Republican views (MSNBC: 52.5\%, 95\% CI: 52.4--52.6\%; CNN: 35.0\%, 95\% CI: 34.9--35.1\%) exceeds the proportion reinforcing Democratic viewpoints~(MSNBC: 39.3\%, 95\% CI: 39.2--39.5\%; CNN: 22.7\%, 95\% CI: 22.6--22.8\%; \hbox{$p < 0.001$} for both).

Furthermore, on Fox News, negative sentiment toward Democratic viewpoints and reinforcement of Republican viewpoints co-occurred far more frequently (odds ratio, \hbox{OR = 12.63, $p < 0.001$}; see SI Appendix Section~\ref{sec:co_occurence_analysis}) than negative sentiment toward Republican viewpoints and reinforcement of Democratic viewpoints do on MSNBC (\hbox{OR = 4.20, $p < 0.001$}; difference in log~OR: $Z = 2.98, p < 0.001$). This indicates that Fox News exhibited a significantly stronger tendency of combining attacks on the Democratic party with praise for the Republican party---an "attack-and-praise" pattern that is roughly three times more pronounced than the corresponding pattern on MSNBC.

Fox also stood out for maintaining substantially higher levels of use of inflammatory language on average (33.57\%; 95\% CI: 33.46--33.67\%) than MSNBC (18.13\%; 95\% CI: 18.05--18.21\%) and CNN (10.45\%; 95\% CI: 10.38--10.52\%) throughout our study period (Figure \ref{fig:feature_timetrends} g), and steadily increasing in this dimension over the decade. It was also the least committed to fact-based reporting among the three (Fox News: 16.55\%, 95\% CI: 16.47--16.63\%; MSNBC: 32.90\%, 95\% CI: 32.80--33.0\%; CNN: 49.38\%, 95\% CI: 49.27--49.49\%), and showed a steady decline in focus on fact-based reporting over time (Figure \ref{fig:feature_timetrends}h).

Temporal dynamics further distinguish the editorial strategies of each station. For Fox News, strategies tended to follow unidirectional trends---rising (or falling) steadily across the years, irrespective of the political party in power, whereas MSNBC and CNN exhibited marked responsiveness to electoral outcomes. During the Trump presidency (2017--2020), both stations showed heightened reinforcement of Democratic views \hbox{(Figure~ \ref{fig:feature_timetrends}c)} and increased blame and negativity toward Republicans (Figure~\ref{fig:feature_timetrends}b, f)---patterns that reversed partially after the 2020 election of Biden (panels b, c, f; see SI Appendix Section~\ref{sec:framing_bias_analysis_presidencies} for these strategies aggregated by presidency). Both CNN and MSNBC also showed notable declines in fact-based reporting during the Trump administration years, hinting at changes in editorial priorities during periods of heightened political volatility. Such patterns together suggest that MSNBC and CNN adjusted their rhetorical strategies more responsively to the party in power, whereas Fox maintained a more ideologically anchored style, less contingent on incumbency.

Rhetorical patterns also aligned with salient sociopolitical events. During the onset of the COVID-19 pandemic in 2020, we observe a sharp increase in fact-based reporting by all the cable stations during the early months of the global pandemic (Figure \ref{fig:feature_timetrends}h), and a drop in partisan reporting and style (Figure \ref{fig:feature_timetrends}a-g),  indicating that they temporarily prioritized factual coverage over ideological positioning, before partisan slant returned as the crisis became increasingly politicized. The Supreme Court decision in Dobbs v.\ Jackson Women's Health Organization (597 U.S. 215, 2022), which overturned Roe v.\ Wade, marked another sharp shift, particularly for MSNBC and CNN, where negative sentiment toward Republicans and blame attribution (Figure \ref{fig:feature_timetrends}b, f) sharply increased---highlighting their event-driven spikes in partisan language on socially sensitive issues.

Broadcast stations (ABC, CBS, NBC), on the other hand, provide a clear contrast. Compared to cable stations, they exhibited markedly lower levels of usage across all attitudinal features (Figure \ref{fig:feature_timetrends}a-f) and the use of inflammatory/derogatory language (Figure \ref{fig:feature_timetrends}g), while consistently high focus on fact-based reporting (Figure \ref{fig:feature_timetrends}h). They exhibited limited responsiveness to electoral cycles, underscoring the stark contrast in editorial style between broadcast and cable television news.

\subsubsection{Variation of Rhetorical Strategies across Topics} We next investigate whether these rhetorical strategies showed up uniformly across topics, or whether the topic shaped how news stations frame content. Here, we aggregate data across time and vary by topic. However, because our dataset spans two Democratic presidencies (Obama and Biden) and only one Republican presidency (Trump) for the most part, aggregating over the entire 2013--2025 period could overweight coverage during Democratic administrations. To ensure a more balanced comparison across political contexts, we therefore limit this analysis to the time period spanning the Trump and Biden presidencies (Jan.\ 2017--Jan.\ 2025).

\begin{figure*}[t!]
\centering
\includegraphics[width=\textwidth]{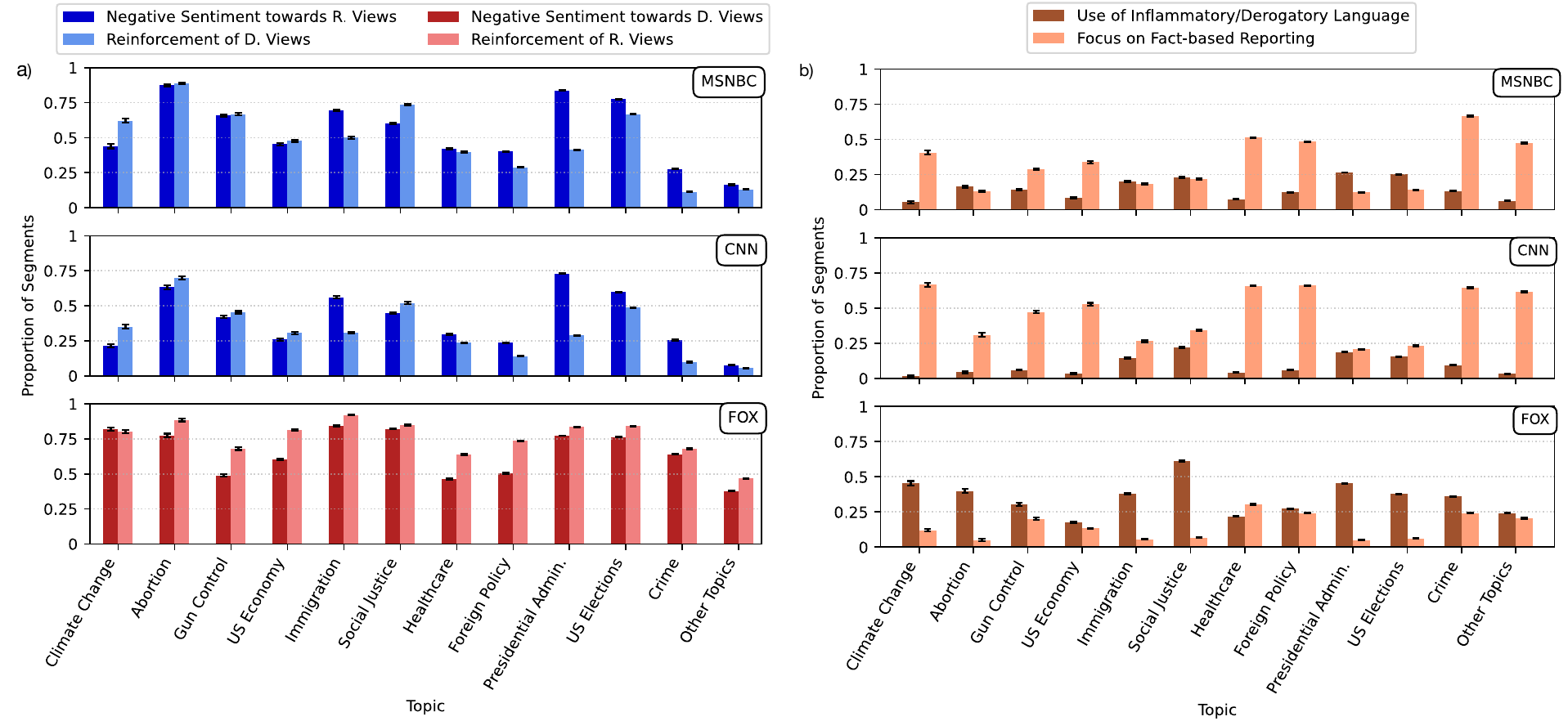}
\caption{\textbf{Topic-specific variation in attitudinal and stylistic strategies across cable news stations.} (a)~Prevalence of in-group reinforcement and out-group criticism across topics, by station (aggregated over two presidencies, Jan.\ 2017--Jan.\ 2025). Light bars indicate reinforcement of aligned viewpoints (Democratic in blue, Republican in red); dark bars indicate negative sentiment toward opposing parties. Fox News consistently emphasized Republican reinforcement over Democratic criticism across topics, with the exception of climate change. In contrast, MSNBC and CNN showed greater topic sensitivity: on salient political issues, such as elections, presidential administration, foreign policy, crime, and immigration, out-group negativity outweighs in-group reinforcement, whereas on other topics, the pattern is more balanced or reversed. Error bars represent 95\% confidence intervals. (b) Prevalence of inflammatory rhetoric versus fact-based reporting across topics, by station. On CNN, fact-based reporting consistently exceeded inflammatory language across all topics. MSNBC exhibited a similar pattern except for presidential administration and U.S. elections. In sharp contrast, Fox News reversed this trend on most topics, except Healthcare and Crime, displaying a higher or much higher share of segments with inflammatory language than with fact-based reporting, particularly on social justice, presidential administration, immigration, and abortion. Error bars represent 95\% confidence intervals. See SI Appendix Section~\ref{sec:llm_comparison_appendix} for replication of this figure with four alternative LLMs.}
\label{fig:topicfeature_interactions}
\end{figure*}

Our topic-level analysis reveals a deeper asymmetry in partisan expression between Fox News and MSNBC/CNN. Fox News's tendency to devote more airtime to affirming Republican views than to criticizing Democratic ones holds across all topics, with climate change as the sole exception (Figure~\ref{fig:topicfeature_interactions}a). By contrast, MSNBC and CNN varied their strategy by topic. On highly salient domains such as elections, presidential administration, foreign policy, crime, and immigration---topics that make up a major share of their programming---their coverage is skewed toward criticizing Republicans. For abortion and healthcare, attack and reinforcement were balanced, whereas on issues like climate change and social justice, Democratic reinforcement dominated. Thus, Fox pursued a consistent "reinforcement-first" strategy across topics, while MSNBC and CNN had a higher tendency to vary their rhetorical emphasis depending on the topic. Furthermore, for MSNBC and CNN, these rhetorical dynamics across topics were largely contingent on which party holds power, whereas Fox News maintains its emphasis on in-party reinforcement over out-party across topics and presidencies (see SI Appendix Section~\ref{sec:balance_presidency}).

Stylistic strategies also varied substantially across topics. Fox News exhibited higher use of inflammatory or derogatory language and lower focus on fact-based reporting than MSNBC/CNN across the majority of the topics (Figure~\ref{fig:topicfeature_interactions}b). For CNN, fact-based reporting predominated across all topics. MSNBC similarly emphasized fact-based reporting overall but showed notable topic-dependent variation. On issues such as abortion, immigration, social justice, elections, and presidential administration, MSNBC featured more inflammatory or derogatory language than fact-based content. In contrast, Fox News reversed this pattern, with inflammatory or derogatory language outweighing fact-based reporting in all but one topic---healthcare.
Fox News was particularly extreme on social justice and racial inequality, with more than half of its segments containing inflammatory language, and it also showed elevated levels of inflammatory/derogatory language on climate change, abortion, and gun control. For focus on fact-based reporting across all cable stations, the sharpest declines were on high-salience issues such as elections, presidential administration, abortion, social justice, and immigration, although this decline was much more significant in Fox News than in MSNBC/CNN.

\subsection{Robustness tests with alternative LLMs}

For our main results, we utilized OpenAI's GPT-4o model, which we selected based on its strong reputation and its application in related work~\cite{le2024positioning, wu2023large, gilardi2023chatgpt, tornberg2025large, haroon2025whose}. However, it has been suggested in recent studies that GPT may exhibit measurable liberal bias on some tasks~\cite{westwood2025measuring, motoki2024more, santurkar2023whose, feng2023pretraining}--a bias that our human research assistants might also share. Because our central contributions depend on how we quantify the degree of liberal and conservative bias in TV news content, it is important to check that our results are not merely artifacts of a measurement procedure that is itself biased. Recent work has found evidence that many users perceive LLMs as left-leaning~\cite{westwood2025measuring}. In contrast, LLMs such as Google's Gemini appear to exhibit a more neutral stance across tasks~\cite{choudhary2024political, chandra2026evaluation}, while xAI's Grok appears to exhibit a conservative stance~\cite{rozado2024political, yasseri2025similar, valeriepoliticization2025}. To evaluate whether our findings depend on our particular choice of LLM, we performed two sets of robustness checks (see SI Appendix Section~\ref{sec:robustness_analysis}) using four alternative large language models --- two closed, proprietary systems---Grok 4.1 Fast (xAI) and Gemini 2.5 Flash-Lite (Google)---and two open-weight models---Llama 4 Scout (Meta) and DeepSeek-v3.1 (DeepSeek). By selecting models that span proprietary and open-weight architectures and that differ in their reported political orientations, this exercise represents a stringent test of whether our empirical findings depend on the ideological bias of any single model.

\textcolor{black} {First,} we find that all four alternative large language models exhibited agreement with human-coded partisanship labels comparable to that of GPT-4o, despite their differences in political lean reported in previous work (see SI Appendix Section~\ref{sec:llm_human_agreement}). \textcolor{black} {Second, and more importantly, although the models exhibited differences in scale, dispersion, and political orientation (closely corresponding to what previous work has suggested)}, all of our main findings (Figures~\ref{fig:partisanship_trend}-\ref{fig:topicfeature_interactions}) are replicated using each of the four alternative models. \ In particular: (a) the partisan asymmetry between left- and right- leaning stations that appears in Figure~\ref{fig:partisanship_trend} and Figure~\ref{fig:Coverage_and_PartisanshipStrength} persists across all models; and (b) the same eight rhetorical strategies examined in Figures \ref{fig:feature_timetrends} and \ref{fig:topicfeature_interactions} are qualitatively similar (see SI Appendix Section~\ref{sec:llm_comparison_appendix} for full replication results). \textcolor{black} {Because these models differ in architecture and reported political orientation, convergence across them provides evidence that our conclusions are not artifacts of any single model's ideological tendencies. }

\section{Discussion}

Through a comprehensive analysis of over 400,000 news episodes spanning 2013 to 2025, our study uncovers systematic and asymmetric patterns of partisan expression across major cable and broadcast stations in the United States. The findings advance our understanding of partisan television news in three central ways.

First, ideological asymmetry in television news is not merely a matter of degree. Although all three cable stations exhibited partisan tendencies during our study period, Fox News differed not only in magnitude but also in rhetorical strategies: its programming consistently prioritized reinforcement of aligned viewpoints over criticism of opponents and intensified its conservative alignment across successive elections, regardless of which party held the U.S. presidency.\ In contrast, MSNBC and CNN displayed greater responsiveness to presidential incumbency and relied more heavily on out-group criticism than in-group affirmation.\ These patterns suggest distinct editorial strategies employed by the stations rather than symmetric positioning along a single ideological continuum.

Second, our findings show that partisan bias in televised news can be decomposed into a small number of rhetorical strategies---including reinforcement, negative sentiment, and blame attribution---that together account for most variation in ideological slant. Partisanship, therefore, is better understood not as a one-dimensional construct as commonly operationalized in prior work~\cite{gentzkow2010drives, martin2017bias, gentzkow2019measuring, kim2022measuring} but as a configuration of recurring rhetorical strategies. This multi-dimensional perspective lays out the linguistic mechanisms through which the stations communicate their ideological slant to television news audiences.

Third, the relative neutrality of broadcast stations reflects both stylistic restraint and agenda selection. By devoting disproportionate attention to less contentious topics and employing less partisan language consistently, broadcast stations differ institutionally from cable stations in both what they cover and how they cover it.

While our study provides novel insights about the television media landscape, several limitations warrant consideration. First, while the eight rhetorical strategies we identify explain a substantial share of variation in partisanship scores ($R^2 = 0.73$), about 27\% of the variance remains unexplained. Given the complexity and nuance of political language, no feature set, however sophisticated, is likely to capture 100\% of the variance in partisanship. Nonetheless, additional, subtler rhetorical cues may contribute meaningfully to residual variation. Recent advances suggest that large language models can be used to iteratively extract interpretable features for prediction tasks~\cite{ludan2023interpretable, koh2020concept}. Extending our framework to identify further dimensions of partisan expression thus represents a promising direction for future work.

Second, the analysis is correlational and does not establish causal relationships regarding the forces that shape partisan expression in television news or its downstream consequences. Integrating our measures with audience-level data---such as panel surveys or behavioral traces---would enable investigation of how political events and audience demand influence changes in media content.
Relatedly, our findings documenting Fox News' steady rightward trajectory invite examination of whether sustained exposure to strongly partisan narratives increases receptivity to previously marginal or extreme ideological positions~\cite{bolet2025media}. Prior research has shown how threat amplification and delegitimization of political opponents can erode democratic norms and foster support for illiberal movements~\cite{benkler2018network, dunwoody2022fascist}. Connecting patterns of televised partisan expression to broader ideological shifts remains an important avenue for future work.

Third, our dataset includes only three months of coverage during the second Trump presidency, limiting our ability to draw firm conclusions about partisan dynamics during that period, given the substantial temporal variability observed in our data.

Fourth, our annotation procedure relies on a proprietary large language model, \textcolor{black} {whose internal decision process is not fully interpretable. This limitation is not unique to LLM-based measurement: human coders and survey respondents also produce judgments through latent cognitive processes that are only partially observable. As discussed earlier, however, our LLM-based labels closely match trained human judgments, with model--human agreement statistically indistinguishable from inter-human agreement,} and our substantive findings replicate across multiple closed-source and open-weight models with differing political orientations. Together, these tests suggest that the results are robust to both rater identity and model-specific ideological tendencies.

Finally, extending our framework to other media environments, including social media and print news, as well as to international contexts, would further illuminate how partisan communication varies globally across institutional and cultural settings.

In conclusion, this study provides a large-scale account of how partisan perspectives are expressed in television news through distinct rhetorical strategies, shaping how audiences encounter and interpret political information. By identifying the linguistic dimensions that structure partisanship slant and documenting systematic asymmetries across stations, we offer critical insights into the communicative mechanisms that shape political discourse and contribute to the ongoing media polarization in the United States.

\section{Materials and Methods}
\phantomsection\label{sec:methods}

\begin{figure*}[t!]
\centering
\includegraphics[width=0.8\textwidth]{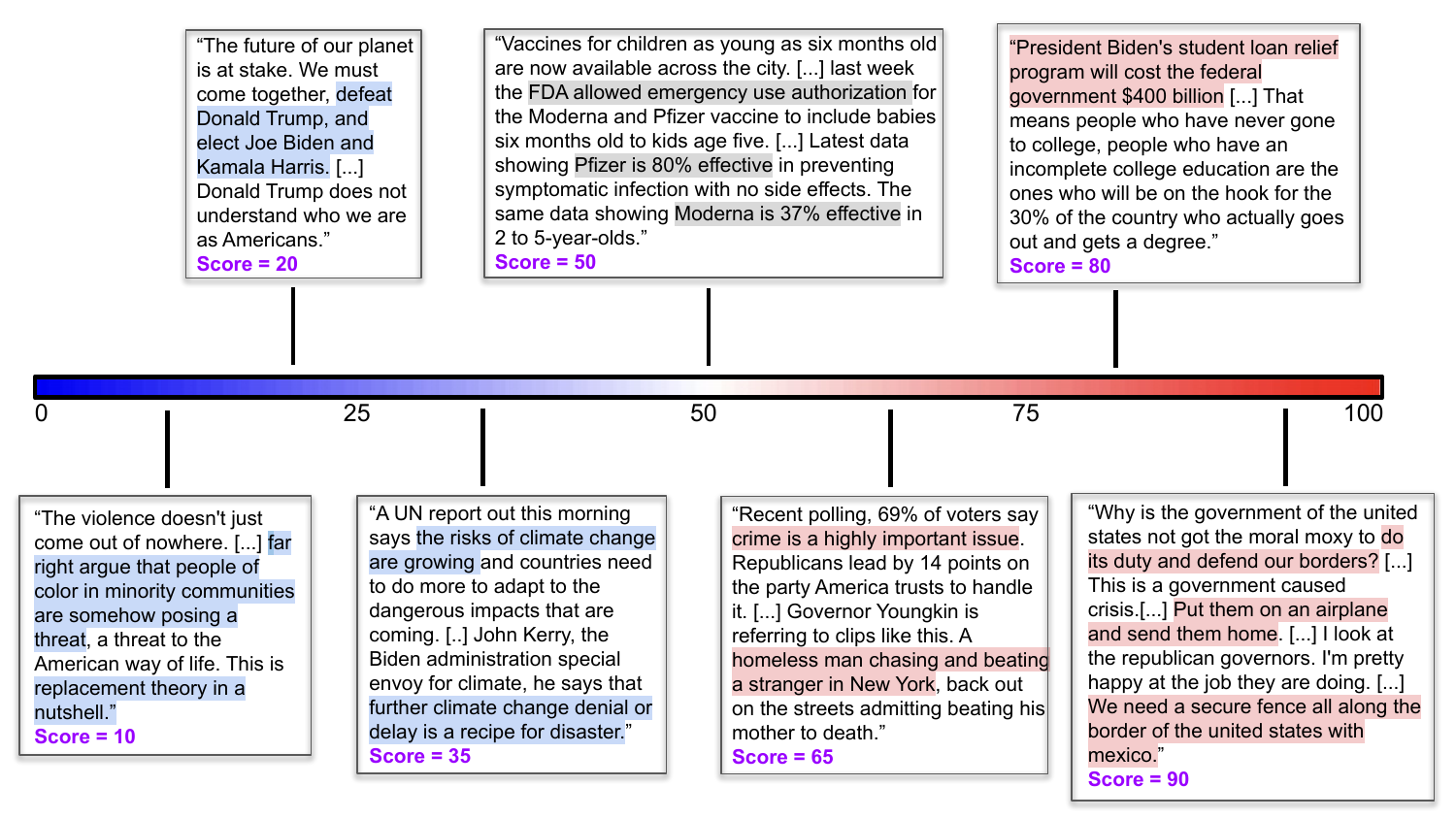}
\caption{\textbf{Example TV news excerpts and their partisanship scores.} The partisanship score is assigned by prompting GPT-4o on a scale of 0-100, where lower (higher) scores indicate alignment with liberal (conservative) viewpoints (see SI Appendix Figure~\ref{fig:prompt_partisanship} for prompt details). For ease of interpretation, key phrases aligning with the assigned scores have been highlighted, illustrating how language choices correspond to partisan slant in television news coverage.}
\label{fig:quotes_figure}
\end{figure*}

\subsection{Data}
We analyze television news transcripts obtained from a commercial media monitoring service that archives broadcast content from major U.S. stations. Our dataset spans 408,464 news episodes aired between January 2013 and March 2025 across six prominent channels: Fox News, CNN, MSNBC, ABC, CBS, and NBC (see \hbox{SI Appendix Section~\ref{Data}} for details). All our analyses are conducted on the population of national TV news data across these six channels, and all advertisements were removed. Given the length and structural heterogeneity of full-length TV episodes, we segmented each transcript into contiguous segments of approximately 800 words, roughly corresponding to five minutes of spoken airtime. This segmentation length was strategically chosen to balance context completeness---ensuring each segment retained sufficient narrative context---and technical feasibility, as longer texts exceed the context window for transformer-based language models used in our analyses. This results in 3,505,401 news segments (median = 279,954 news segments per year).

\subsection{Measuring Partisanship Bias}

To quantify partisan slant in television news we used a hybrid approach that combines the human-level labeling performance of LLMs with the scalability of supervised machine learning. Given a news segment, we prompted OpenAI's GPT-4o model to assign a continuous partisanship score ranging from 0 (strongly liberal) to 100 (strongly conservative) based on its content, along with an explanation for the score, consistent with recent LLM-based approaches to ideological scaling of political text~\cite{le2024positioning, haider2026media} (see SI Appendix Figure~\ref{fig:prompt_partisanship} for prompt details). Figure~\ref{fig:quotes_figure} provides illustrative examples of news segment excerpts and their labels, demonstrating the model's ability to capture nuanced partisan cues.\ For instance, a score of 20 is assigned for explicit advocacy against conservative politicians versus a score of 80 for framing a Democratic policy as an unfair burden on taxpayers.\
Conversely, segments reporting factual information, such as neutral public health announcements, have a score of 50, falling at the center of the partisan scoring scale.

\begin{figure}[t!]
\centering
\includegraphics[width=\columnwidth]{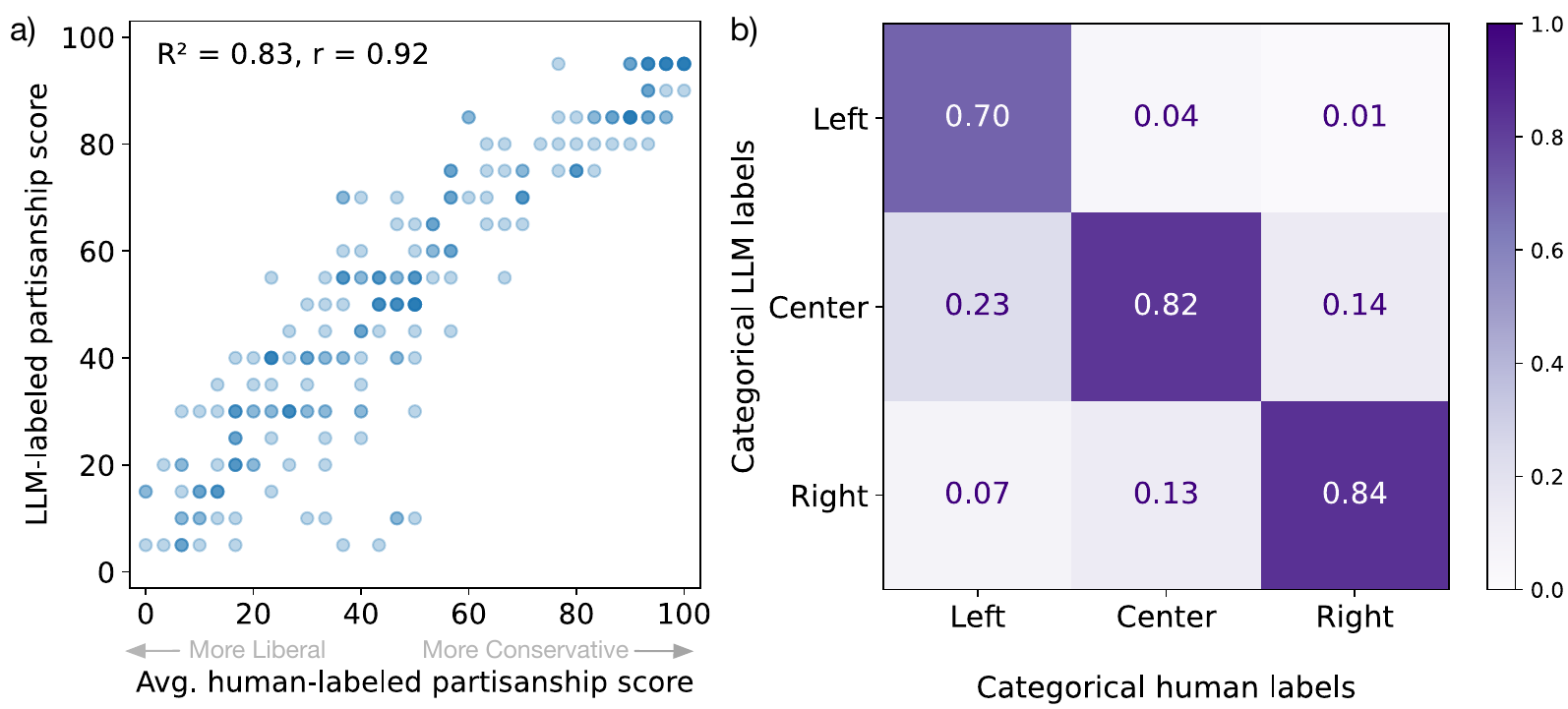}
\caption{\textbf{Comparison of LLM and human-based partisanship labels}
(a)~Scatterplot comparing human-labeled partisanship scores averaged among three trained research assistants with partisanship scores labeled via \hbox{GPT-4o} prompting; \hbox{$R^2=0.83$, $r=0.92$}, indicating high agreement. (b)~Confusion matrix comparing categorical human labels from a previous media bias study~\cite{kim2022measuring} with \hbox{GPT-4o} scores; macro F1=0.783, where continuous 0-100 scores are categorized as: \hbox{Liberal (L) $< 40$}, \hbox{Neutral (N) 40--60}, \hbox{Conservative (C) $> 60$}.}
\label{fig:partisanship_humanlabelcomparisons}
\end{figure}

\begin{figure}[t!]
\centering
\includegraphics[width=\columnwidth]{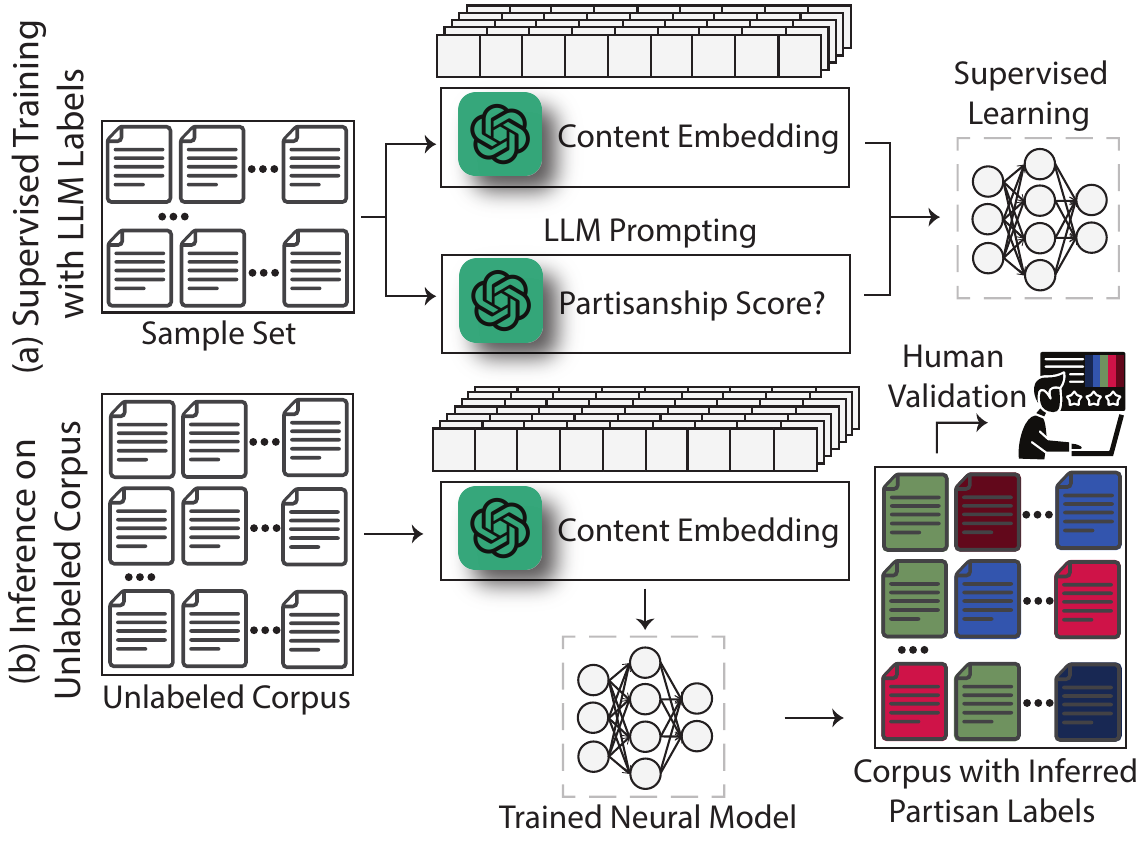}
\caption{\textbf{LLM-guided annotation and neural prediction pipeline.} (a) \textbf{Training:} A stratified sample of 10,000 television news episodes (86,131 segments) is annotated for partisanship scoring (0--100) by prompting GPT-4o. The news segments are embedded as 3072-dimensional vectors and used to train a neural network for partisanship score prediction. (b)~\textbf{Inference:} The trained model is applied to the remaining unlabeled episodes (3,419,270 segments) to generate segment-level partisanship predictions at scale, a subset of which is evaluated against human judgments.}
\label{fig:schematic}
\end{figure}

We validated our LLM-based measures of partisanship by benchmarking against both human annotations and established approaches from prior work. On a stratified sample of 250 news segments labeled by three trained research assistants (see SI Appendix Section~\ref{prompt_gpt_partisanship}), GPT-4o scores showed strong agreement with human judgments (\hbox{Pearson's $r = 0.92$,} \hbox{$R^2 = 0.83$;} Figure~\ref{fig:partisanship_humanlabelcomparisons}a).\ Moreover, \hbox{because partisanship coding} can be inherently subjective---and even trained experts may disagree on some segments---we also assessed whether our LLM-based measures agree with human annotators to the same extent that human annotators agree with each other.\
A nonparametric bootstrap test confirms that the agreement between the LLM-labels and human annotators (average weighted Cohen's $\kappa = 0.87$) is statistically indistinguishable \hbox{($p$-value = 0.53)} from inter-human agreement \hbox{(average weighted Cohen's $\kappa = 0.88$).} Additionally, we assessed whether the LLM's performance simply reflects the model's memorization of preexisting patterns in its training data rather than genuine detection of partisan language.\ To test this, we compared model--human agreement on segments aired before and after GPT-4o's knowledge cutoff of October 1, 2023~\cite{gpt4ocutoffdate}.\ A nonparametric bootstrap test confirms that the model--human agreement on segments aired up to the knowledge cut-off \hbox{(Pearson's $r = 0.92$, $R^2 = 0.82$)} is statistically indistinguishable \hbox{($p$-value = 0.46 and 0.88, respectively)} from that on segments aired \hbox{after (Pearson's $r = 0.94$, $R^2 = 0.88$)}, indicating that GPT-4o's performance is not a result of memorization of its training data.

We next benchmark our LLM-based partisanship measures against a widely used approach in existing literature~\cite{martin2017bias, martin2019local, hassell2022does, widmer2022media, djourelova2023persuasion} -- an elastic net regression model trained on DW-NOMINATE scores~\cite{mccarty1997income} of legislators and their usage of bigrams in congressional floor speeches~\cite{martin2017bias, gentzkow2010drives} to infer partisanship in television news (see SI Appendix Section~\ref{congressional_model}). While this model performs well in-domain on congressional speech (Pearson's $r = 0.85$), it generalizes poorly to television news, yielding a significantly weaker alignment with our human-labeled sample (Pearson's \hbox{$r = 0.225$}). This highlights the domain specificity of partisan language and the limited transferability of models trained on congressional speech to television news---a limitation that our LLM-based approach overcomes. Another influential approach by Gentzkow et al.~\cite{gentzkow2019measuring} measures partisan differences between pairs of speeches as the expected posterior probability that a listener with a neutral prior can correctly infer the source party (e.g., Republican or Democrat) from a brief segment of speech, based on observed word choices (see SI Appendix Section~\ref{gentzkow2019_model_appendix}). While effective at capturing aggregate linguistic polarization, this method shows weaker alignment with human-labeled partisanship scores in our setting (\hbox{$R^2 = 0.636$, $r = 0.819$}) relative to our LLM-based approach (\hbox{$R^2 = 0.83$, $r = 0.92$}). This represents a 30.5\% increase in explained variance ($R^2$) and a 12\% gain in correlation ($r$), demonstrating the value of contextual language modeling for fine-grained, segment-level inference in television news.

Furthermore, we compared our LLM-based approach with the non-textual measure introduced in~\cite{kim2022measuring}, which assigns outlet partisanship using campaign donation scores of individual guests appearing on television shows. Our LLM-based measure achieved a segment-level correlation $r = 0.796$ with human judgments collected in that study (macro F1 = 0.783 on categorical human labels; Figure~\ref{fig:partisanship_humanlabelcomparisons}b), where news segments were labeled as either Liberal, Neutral, or Conservative by human annotators. On the other hand, the study reported a correlation of $r = 0.72$ after aggregating labels on weekly basis~\cite{kim2022measuring}, while the congressional speech based method achieved $r = 0.119$ (macro F1 = 0.373 after categorizing continuous scores using thresholds from a validation set).
Taken together, our analyses show that our LLM-based measure of partisanship not only aligns closely with human judgments but also outperforms both textual and non-textual benchmarks.

However, applying our LLM-based measure directly to all 3,505,401 segments of the 408,464 news episodes would be prohibitively expensive. To address this challenge, we adopt a scalable hybrid annotation strategy inspired by teacher--student frameworks in machine learning~\cite{gu2024minillm, tian2025beyond, bucilua2006model, chen2022knowledge, ganti2023narrative, wan2024tnt}. In our setup, the LLM serves as the ``teacher" providing high-quality partisanship scores on a subset of the data (Figure~\ref{fig:schematic}a), while a neural network ``student" model is trained on these labels to predict the partisanship scores across the full dataset (Figure~\ref{fig:schematic}b), allowing us to transfer the ability of the LLM to a smaller model that can be applied efficiently at scale without much sacrifice of performance (see SI Appendix Section~\ref{neural_network_training}). The trained neural network demonstrated strong performance in predicting LLM labels on a held-out set of 5,000 episodes ($R^2 = 0.803$, Pearson $r =0.896$), as well as on the human annotations ($R^2 = 0.83$, Pearson $r = 0.923$). Episode-level partisanship scores, computed as the average of segment-level scores within each episode, exhibit stronger agreement with the LLM labels ($R^2 = 0.95$, $r = 0.976$), apparently due to noise reduction from imperfect segmentation and non-partisan filler content through averaging. Our evaluation results demonstrate that the neural network (student model) is effective in closely emulating the LLM (teacher model).
Applying the trained neural networks to our corpus (2013-2025) produces a fully labeled dataset, leading to our detailed investigation of how partisan dynamics evolved over the period of 2013--2025.

\subsection{Measuring rhetorical strategies}
To measure the presence of the eight rhetorical features in the news transcripts, we first prompted GPT-4o to assign a binary label for the presence of each of the eight features in the same subset of 250 news segments (\hbox{see SI Appendix Section~\ref{sec:rhetorical_llm_measurement}} for prompt details). The LLM labels show a strong agreement with human labels (F1 scores 0.705--0.853, precision 0.794--0.889, recall 0.611--0.821). Furthermore, the agreement between the LLM labels and human-coded labels is statistically indistinguishable (bootstrap test $p$-values between 0.4872--0.7358) from inter-human agreement for each of the eight features. Then, to scale our measurement across the full dataset, we adopt the teacher--student framework as before, where for each rhetorical strategy, a neural network (student model) is trained to emulate the LLM (teacher model) for labeling news segments. The neural network predictions show high agreement with the LLM labels (F1 scores on held-out segments ranged from 0.571 to 0.819, and from 0.7073 to 0.8609 on our human-coded segments; see SI Appendix Section~\ref{sec:rhetorical_llm_measurement}).

\section*{Acknowledgments}
We appreciate research support from Robin Anand, Lily McLean, and Clarence Guo, as well as helpful discussions with Abhinav Chakraborty and Yph Lelkes. This research was supported in part by Richard Jay Mack and the Knight Foundation (Grant ID G-2023-67859).

\section*{Competing interest}
The authors declare no competing interests.

\section*{Funding}
This research was supported in part by Richard Jay Mack and the Knight Foundation (Grant ID G-2023-67859).

\section*{Author contributions}
U.D., H.H., A.G., A.C., and D.J.W. designed the research. U.D. implemented and carried out the experiments and analyses. U.D., A.C., and D.J.W. wrote the paper.

\section*{Data availability}
The data were provided by a commercial media monitoring company under licensing agreements with the University of Pennsylvania. Researchers interested in access should contact the corresponding author for additional details.

\bibliographystyle{unsrt}
\bibliography{references}

@book{stroud2011niche,
  title={Niche news: The politics of news choice},
  author={Stroud, Natalie Jomini},
  year={2011},
  publisher={Oxford University Press}
}

@book{prior2007post,
  title={Post-broadcast democracy: How media choice increases inequality in political involvement and polarizes elections},
  author={Prior, Markus},
  year={2007},
  publisher={Cambridge University Press}
}

@book{arceneaux2022changing,
  title={Changing minds or changing channels?: Partisan news in an age of choice},
  author={Arceneaux, Kevin and Johnson, Martin},
  year={2022},
  publisher={University of Chicago Press}
}

@misc{pew2025platform,
  author = {{Pew Research Center}},
  title  = {News Platform Fact Sheet},
  year   = {2025},
  note   = {Survey of U.S. adults conducted August 18--24, 2025.
            Accessed: 2026-08-28},
  howpublished = {\url{https://www.pewresearch.org/journalism/fact-sheet/news-platform-fact-sheet/}}}

@article{allen2020evaluating,
  title={Evaluating the fake news problem at the scale of the information ecosystem},
  author={Allen, Jennifer and Howland, Baird and Mobius, Markus and Rothschild, David and Watts, Duncan J},
  journal={Science advances},
  volume={6},
  number={14},
  pages={eaay3539},
  year={2020},
  publisher={American Association for the Advancement of Science}
}

@article{robinson2018stubborn,
  title={The stubborn pervasiveness of television news in the digital age and the field’s attention to the medium, 2010--2014},
  author={Robinson, Nicholas W and Zeng, Chen and Holbert, R Lance},
  journal={Journal of Broadcasting \& Electronic Media},
  volume={62},
  number={2},
  pages={287--301},
  year={2018},
  publisher={Taylor \& Francis}
}

@article{yang2026platforms,
  author  = {Yang, Tian and Gonz{\'a}lez-Bail{\'o}n, Sandra},
  title   = {More platforms, less attention to news? A multi-platform
             analysis of news exposure across {TV}, web, and {YouTube}
             in the {United States}},
  journal = {New Media \& Society},
  volume  = {28}, number = {7}, pages = {3140--3159}, year = {2026},
  doi     = {10.1177/14614448251341496}}

@article{muise2022quantifying,
  title={Quantifying partisan news diets in Web and TV audiences},
  author={Muise, Daniel and Hosseinmardi, Homa and Howland, Baird and Mobius, Markus and Rothschild, David and Watts, Duncan J},
  journal={Science advances},
  volume={8},
  number={28},
  pages={eabn0083},
  year={2022},
  publisher={American Association for the Advancement of Science}
}

@article{martin2017bias,
  title={Bias in cable news: Persuasion and polarization},
  author={Martin, Gregory J and Yurukoglu, Ali},
  journal={American Economic Review},
  volume={107},
  number={9},
  pages={2565--2599},
  year={2017},
  publisher={American Economic Association 2014 Broadway, Suite 305, Nashville, TN 37203}
}

@article{ash2024viewers,
  title={From viewers to voters: Tracing Fox News’ impact on American democracy},
  author={Ash, Elliott and Galletta, Sergio and Pinna, Matteo and Warshaw, Christopher S},
  journal={Journal of Public Economics},
  volume={240},
  pages={105256},
  year={2024},
  publisher={Elsevier}
}

@article{broockman2025consuming,
  title={Consuming cross-cutting media causes learning and moderates attitudes: A field experiment with Fox News viewers},
  author={Broockman, David E and Kalla, Joshua L},
  journal={The Journal of Politics},
  volume={87},
  number={1},
  pages={246--261},
  year={2025},
  publisher={The University of Chicago Press Chicago, IL}
}

@article{gill2022testing,
  title={Testing the effect of cross-cutting exposure to cable TV news on affective polarization: Evidence from the 2020 US presidential election},
  author={Gill, Hyungjin},
  journal={Journal of Broadcasting \& Electronic Media},
  volume={66},
  number={2},
  year={2022},
  publisher={Taylor \& Francis}
}

@article{gainous2023surfing,
  title={Surfing to the political extremes: Digital media, social media, and policy attitude polarization},
  author={Gainous, Jason and Wagner, Kevin M},
  journal={Social Science Quarterly},
  volume={104},
  number={4},
  pages={547--558},
  year={2023},
  publisher={Wiley Online Library}
}

@article{levendusky2013partisan,
  title={Why do partisan media polarize viewers?},
  author={Levendusky, Matthew S},
  journal={American journal of political science},
  volume={57},
  number={3},
  pages={611--623},
  year={2013},
  publisher={Wiley Online Library}
}

@article{gentzkow2010drives,
  title={What drives media slant? Evidence from US daily newspapers},
  author={Gentzkow, Matthew and Shapiro, Jesse M},
  journal={Econometrica},
  volume={78},
  number={1},
  pages={35--71},
  year={2010},
  publisher={Wiley Online Library}
}

@article{martin2019local,
  title={Local news and national politics},
  author={Martin, Gregory J and McCrain, Joshua},
  journal={American Political Science Review},
  volume={113},
  number={2},
  year={2019},
  publisher={Cambridge University Press}
}

@article{hassell2022does,
  title={Does the ideology of the newsroom affect the provision of media slant?},
  author={Hassell, Hans JG and Miles, Matthew R and Reuning, Kevin},
  journal={Political Communication},
  volume={39},
  number={2},
  pages={184--201},
  year={2022},
  publisher={Taylor \& Francis}
}

@article{widmer2022media,
  title={Media slant is contagious},
  author={Widmer, Philine and Galletta, Sergio and Ash, Elliott},
  journal={arXiv:2202.07269},
  year={2022}
}

@article{djourelova2023persuasion,
  title={Persuasion through slanted language: Evidence from the media coverage of immigration},
  author={Djourelova, Milena},
  journal={American economic review},
  volume={113},
  number={3},
  pages={800--835},
  year={2023},
  publisher={American Economic Association 2014 Broadway, Suite 305, Nashville, TN 37203}
}

@article{kim2022measuring,
  title={Measuring dynamic media bias},
  author={Kim, Eunji and Lelkes, Yphtach and McCrain, Joshua},
  journal={Proceedings of the National Academy of Sciences},
  volume={119},
  number={32},
  pages={e2202197119},
  year={2022},
  publisher={National Academy of Sciences}
}

@article{hosseinmardi2025unpacking,
  title={Unpacking media bias in the growing divide between cable and network news},
  author={Hosseinmardi, Homa and Wolken, Samuel and Rothschild, David M and Watts, Duncan J},
  journal={Scientific Reports},
  volume={15},
  number={1},
  pages={17607},
  year={2025},
  publisher={Nature Publishing Group UK London}
}

@article{baum2008new,
  title={New media and the polarization of American political discourse},
  author={Baum, Matthew A and Groeling, Tim},
  journal={Political Communication},
  volume={25},
  number={4},
  pages={345--365},
  year={2008},
  publisher={Taylor \& Francis}
}

@inproceedings{lin2011more,
  title={More voices than ever? quantifying media bias in networks},
  author={Lin, Yu-Ru and Bagrow, James and Lazer, David},
  booktitle={Proceedings of the international AAAI conference on web and social media},
  year={2011}
}

@inproceedings{ding2023same,
  title={Same words, different meanings: Semantic polarization in broadcast media language forecasts polarity in online public discourse},
  author={Ding, Xiaohan and Horning, Michael and Rho, Eugenia H},
  booktitle={Proceedings of the International AAAI Conference on Web and Social Media},
  volume={17},
  pages={161--172},
  year={2023}
}

@article{budak2016fair,
  title={Fair and balanced? Quantifying media bias through crowdsourced content analysis},
  author={Budak, Ceren and Goel, Sharad and Rao, Justin M},
  journal={Public Opinion Quarterly},
  volume={80},
  number={S1},
  pages={250--271},
  year={2016},
  publisher={Oxford University Press US}
}

@inproceedings{preoctiuc2017beyond,
  title={Beyond binary labels: political ideology prediction of twitter users},
  author={Preo{\c{t}}iuc-Pietro, Daniel and Liu, Ye and Hopkins, Daniel and Ungar, Lyle},
  booktitle={Proceedings of the 55th annual meeting of the association for computational linguistics (volume 1: long papers)},
  pages={729--740},
  year={2017}
}

@article{entman1993framing,
  title={Framing: Towards clarification of a fractured paradigm},
  author={Entman, Robert M and others},
  journal={McQuail's reader in mass communication theory},
  volume={390},
  pages={397},
  year={1993}
}

@book{iyengar1994anyone,
  title={Is anyone responsible?: How television frames political issues},
  author={Iyengar, Shanto},
  year={1994},
  publisher={University of Chicago Press}
}

@article{levendusky2013partisanpoliticalcomm,
  title={Partisan media exposure and attitudes toward the opposition},
  author={Levendusky, Matthew},
  journal={Political communication},
  volume={30},
  number={4},
  pages={565--581},
  year={2013},
  publisher={Taylor \& Francis}
}

@article{mutz2005new,
  title={The new videomalaise: Effects of televised incivility on political trust},
  author={Mutz, Diana C and Reeves, Byron},
  journal={American Political Science Review},
  volume={99},
  number={1},
  pages={1--15},
  year={2005},
  publisher={Cambridge University Press}
}

@article{brader2005striking,
  title={Striking a responsive chord: How political ads motivate and persuade voters by appealing to emotions},
  author={Brader, Ted},
  journal={American journal of political science},
  volume={49},
  number={2},
  pages={388--405},
  year={2005},
  publisher={Wiley Online Library}
}

@article{le2024positioning,
  title={Positioning political texts with Large Language Models by asking and averaging},
  author={Le Mens, Ga{\"e}l and Gallego, Aina},
  journal={Political Analysis},
  pages={1--9},
  year={2024},
  publisher={Cambridge University Press}
}

@article{heseltine2024large,
  title={Large language models as a substitute for human experts in annotating political text},
  author={Heseltine, Michael and Clemm von Hohenberg, Bernhard},
  journal={Research \& Politics},
  volume={11},
  number={1},
  pages={20531680241236239},
  year={2024},
  publisher={SAGE Publications Sage UK: London, England}
}

@article{wu2023large,
  title={Large language models can be used to estimate the latent positions of politicians},
  author={Wu, Patrick Y and Nagler, Jonathan and Tucker, Joshua A and Messing, Solomon},
  journal={arXiv preprint arXiv:2303.12057},
  year={2023}
}

@article{burnham2026political,
  title={Political debate: Efficient zero-shot and few-shot classifiers for political text},
  author={Burnham, Michael and Kahn, Kayla and Wang, Ryan Yang and Peng, Rachel X},
  journal={Political Analysis},
  volume={34},
  number={3},
  pages={329--343},
  year={2026},
  publisher={Cambridge University Press}
}

@article{gilardi2023chatgpt,
  title={ChatGPT outperforms crowd workers for text-annotation tasks},
  author={Gilardi, Fabrizio and Alizadeh, Meysam and Kubli, Ma{\"e}l},
  journal={Proceedings of the National Academy of Sciences},
  volume={120},
  number={30},
  pages={e2305016120},
  year={2023},
  publisher={National Academy of Sciences}
}

@article{burnham2025stance,
  title={Stance detection: a practical guide to classifying political beliefs in text},
  author={Burnham, Michael},
  journal={Political Science Research and Methods},
  volume={13},
  number={3},
  pages={611--628},
  year={2025},
  publisher={Cambridge University Press}
}

@article{haider2026media,
  title={The Media Bias Detector: A framework for annotating and analyzing the news},
  author={Haider, Samar and Tohidi, Amir and Wang, Jenny S and D{\"o}rr, Timothy and Rothschild, David M and Callison-Burch, Chris and Watts, Duncan J},
  journal={Science Advances},
  volume={12},
  number={37},
  pages={eaea7456},
  year={2026},
  publisher={American Association for the Advancement of Science}
}

@article{ziems2024can,
  title={Can large language models transform computational social science?},
  author={Ziems, Caleb and Held, William and Shaikh, Omar and Chen, Jiaao and Zhang, Zhehao and Yang, Diyi},
  journal={Computational Linguistics},
  volume={50},
  number={1},
  pages={237--291},
  year={2024},
  publisher={MIT Press One Broadway, 12th Floor, Cambridge, Massachusetts 02142, USA~…}
}

@inproceedings{boydstun2014tracking,
  title={Tracking the development of media frames within and across policy issues},
  author={Boydstun, Amber E and Card, Dallas and Gross, Justin H and Resnik, Philip and Smith, Noah A},
  booktitle={APSA 2014 annual meeting paper},
  pages={1--25},
  year={2014}
}

@article{card2022computational,
  title={Computational analysis of 140 years of US political speeches reveals more positive but increasingly polarized framing of immigration},
  author={Card, Dallas and Chang, Serina and Becker, Chris and Mendelsohn, Julia and Voigt, Rob and Boustan, Leah and Abramitzky, Ran and Jurafsky, Dan},
  journal={Proceedings of the National Academy of Sciences},
  volume={119},
  number={31},
  pages={e2120510119},
  year={2022},
  publisher={National Academy of Sciences}
}

@article{gentzkow2019measuring,
  title={Measuring group differences in high-dimensional choices: method and application to congressional speech},
  author={Gentzkow, Matthew and Shapiro, Jesse M and Taddy, Matt},
  journal={Econometrica},
  volume={87},
  number={4},
  pages={1307--1340},
  year={2019},
  publisher={Wiley Online Library}
}

@article{westwood2025measuring,
  title={Measuring perceived slant in large language models through user evaluations},
  author={Westwood, Sean J and Grimmer, Justin and Hall, Andrew B},
  journal={Working paper},
  year={2025}
}

@article{motoki2024more,
  title={More human than human: measuring ChatGPT political bias},
  author={Motoki, Fabio and Pinho Neto, Valdemar and Rodrigues, Victor},
  journal={Public Choice},
  volume={198},
  number={1},
  pages={3--23},
  year={2024},
  publisher={Springer}
}

@inproceedings{santurkar2023whose,
  title={Whose opinions do language models reflect?},
  author={Santurkar, Shibani and Durmus, Esin and Ladhak, Faisal and Lee, Cinoo and Liang, Percy and Hashimoto, Tatsunori},
  booktitle={International conference on machine learning},
  pages={29971--30004},
  year={2023},
  organization={PMLR}
}

@inproceedings{feng2023pretraining,
  title={From pretraining data to language models to downstream tasks: Tracking the trails of political biases leading to unfair NLP models},
  author={Feng, Shangbin and Park, Chan Young and Liu, Yuhan and Tsvetkov, Yulia},
  booktitle={Proceedings of the 61st Annual Meeting of the Association for Computational Linguistics (Volume 1: Long Papers)},
  pages={11737--11762},
  year={2023}
}

@misc{ucsb_2021,
  author = {Winnie Lam},
  title = {Breaking News: Fox News and MSNBC in a Divided America},
  howpublished = {\url{https://undergradjournal.history.ucsb.edu/our-journal/past-issues/fall-2021/lam/}},
  year = {2021},
  note = {Accessed: 2024-03-10}
}

@misc{gallup_polling_matter_215210, 
author = {Newport, Frank and Dugan, Andrew}, 
title = {Partisan Differences Growing on a Number of Issues},
year = {2017}, 
howpublished = {Gallup, \url{https://news.gallup.com/opinion/polling-matters/215210/partisan-differences-growing-number-issues.aspx}}, 
note = {Accessed: 2025-02-10} 
}

@misc{gallup_509129, 
author = {Newport, Frank}, 
title = {Update: Partisan Gaps Expand Most on Government Power, Climate},
year = {2023}, 
howpublished = {Gallup, \url{https://news.gallup.com/poll/509129/update-partisan-gaps-expand-government-power-climate.aspx}}, 
note = {Accessed: 2025-02-10} 
}

@misc{gallup_652970, 
author = {Brenan, Megan}, 
title = {Economy, Immigration, Abortion, Democracy Driving Voters},
year = {2023}, 
howpublished = {Gallup, \url{https://gallup.com/poll/652970/economy-immigration-abortion-democracy-driving-voters.aspx}}, 
note = {Accessed: 2025-02-10} 
}

@misc{gallup_395408, 
author = {Newport, Frank}, 
title = {Abortion Moves Up on "Most Important Problem" List},
year = {2022}, 
howpublished = {Gallup, \url{https://news.gallup.com/poll/395408/abortion-moves-important-problem-list.aspx}}, 
note = {Accessed: 2025-02-10} 
}

@article{wirth2016diverse,
  author    = {Victoria Wirth‑Kolíba},
  title     = {The Diverse and Dynamic World of ‘Us’ and ‘Them’ in Political Discourse},
  journal   = {CADAAD Journal (Critical Approaches to Discourse Analysis across Disciplines)},
  volume    = {8},
  number    = {1},
  pages     = {23--37},
  year      = {2016}, 
}

@article{iyengar2012affect,
  title={Affect, not ideology: A social identity perspective on polarization},
  author={Iyengar, Shanto and Sood, Gaurav and Lelkes, Yphtach},
  journal={Public opinion quarterly},
  volume={76},
  number={3},
  pages={405--431},
  year={2012},
  publisher={Oxford University Press US}
}

@article{iyengar2019origins,
  title={The origins and consequences of affective polarization in the United States},
  author={Iyengar, Shanto and Lelkes, Yphtach and Levendusky, Matthew and Malhotra, Neil and Westwood, Sean J},
  journal={Annual review of political science},
  volume={22},
  number={1},
  pages={129--146},
  year={2019},
  publisher={Annual Reviews}
}

@article{druckman2021affective,
  title={Affective polarization, local contexts and public opinion in America},
  author={Druckman, James N and Klar, Samara and Krupnikov, Yanna and Levendusky, Matthew and Ryan, John Barry},
  journal={Nature human behaviour},
  volume={5},
  number={1},
  pages={28--38},
  year={2021},
  publisher={Nature Publishing Group UK London}
}

@inproceedings{nettasinghe2025group,
  title={In-Group Love, Out-Group Hate: A Framework to Measure Affective Polarization via Contentious Online Discussions},
  author={Nettasinghe, Buddhika and Rao, Ashwin and Jiang, Bohan and Percus, Allon G and Lerman, Kristina},
  booktitle={Proceedings of the ACM on Web Conference 2025},
  pages={560--575},
  year={2025}
}

@incollection{dagnes2019us,
  title={Us vs. them: Political polarization and the politicization of everything},
  author={Dagnes, Alison},
  booktitle={Super mad at everything all the time: Political media and our national anger},
  year={2019},
  publisher={Springer}
}

@article{malhotra2008partisan,
  title={Partisan polarization and blame attribution in a federal system: The case of Hurricane Katrina},
  author={Malhotra, Neil},
  journal={Publius: The Journal of Federalism},
  volume={38},
  number={4},
  pages={651--670},
  year={2008},
  publisher={Oxford University Press}
}

@article{uttermark2024blame,
  title={Blame Attribution, Partisanship, and Federalism: Evidence from a Panel Survey},
  author={Uttermark, Matthew J and Mewhirter, Jack and Sanders, Rebecca and McLaughlin, Danielle M},
  journal={Public Opinion Quarterly},
  volume={88},
  number={4},
  pages={1296--1308},
  year={2024},
  publisher={Oxford University Press}
}

@article{brutger2025framing,
  title={Framing layoffs: Media coverage, blame attribution, and trade-related policy responses},
  author={Brutger, Ryan and Guisinger, Alexandra},
  journal={Political Behavior},
  volume={47},
  number={2},
  pages={479--501},
  year={2025},
  publisher={Springer}
}

@article{winberg2017insult,
  title={Insult politics: Donald Trump, right-wing populism, and incendiary language},
  author={Winberg, Oscar},
  journal={European journal of American studies},
  volume={12},
  number={12-2},
  year={2017},
  publisher={European Association for American Studies}
}

@book{zeitzoff2023nasty,
  title={Nasty politics: The logic of insults, threats, and incitement},
  author={Zeitzoff, Thomas},
  year={2023},
  publisher={Oxford University Press}
}

@article{graves2019fact,
  title={Fact-checking as idea and practice in journalism},
  author={Graves, Lucas and Amazeen, Michelle},
  year={2019},
  journal={Oxford University Press}
}

@article{musolff2016factual,
  title={Factual narrative and truth in political discourse},
  author={Musolff, Andreas},
  journal={Washington Post},
  year={2016}
}

@article{egami2023using,
  title={Using imperfect surrogates for downstream inference: Design-based supervised learning for social science applications of large language models},
  author={Egami, Naoki and Hinck, Musashi and Stewart, Brandon and Wei, Hanying},
  journal={Advances in Neural Information Processing Systems},
  volume={36},
  pages={68589--68601},
  year={2023}
}

@article{egami2024using,
  title={Using large language model annotations for the social sciences: A general framework of using predicted variables in downstream analyses},
  author={Egami, Naoki and Hinck, Musashi and Stewart, Brandon M and Wei, Hanying},
  journal={Preprint from November},
  volume={17},
  pages={2024},
  year={2024}
}

@article{tornberg2025large,
  title={Large language models outperform expert coders and supervised classifiers at annotating political social media messages},
  author={T{\"o}rnberg, Petter},
  journal={Social Science Computer Review},
  volume={43},
  number={6},
  pages={1181--1195},
  year={2025},
  publisher={Sage Publications Sage CA: Los Angeles, CA}
}

@inproceedings{haroon2025whose,
  title={“Whose Side Are You On?” Estimating Ideology of Political and News Content Using Large Language Models and Few-shot Demonstration Selection},
  author={Haroon, Muhammad and Wojcieszak, Magdalena and Chhabra, Anshuman},
  booktitle={Proceedings of the 14th International Joint Conference on Natural Language Processing and the 4th Conference of the Asia-Pacific Chapter of the Association for Computational Linguistics},
  pages={224--243},
  year={2025}
}

@article{choudhary2024political,
  title={Political bias in large language models: a comparative analysis of ChatGPT-4, Perplexity, Google Gemini, and Claude},
  author={Choudhary, Tavishi},
  journal={IEEE Access},
  volume={13},
  pages={11341--11379},
  year={2024},
  publisher={IEEE}
}

@article{chandra2026evaluation,
  title={An evaluation of LLMs for political bias in Western media: Israel-Hamas and Ukraine-Russia wars},
  author={Chandra, Rohitash and Chen, Haoyan and Zhang, Yaqing and Chen, Jiacheng and Wu, Yuting},
  journal={arXiv preprint arXiv:2601.06132},
  year={2026}
}

@article{rozado2024political,
  title={The political preferences of LLMs},
  author={Rozado, David},
  journal={PloS one},
  volume={19},
  number={7},
  pages={e0306621},
  year={2024}
  }

@article{yasseri2025similar,
  title={How Similar Are Grokipedia and Wikipedia? A Multi-Dimensional Textual and Structural Comparison},
  author={Yasseri, Taha and Mohammadi, Saeedeh},
  journal={arXiv preprint arXiv:2510.26899},
  year={2025}
}

@misc{valeriepoliticization2025,
  author = {Wirtschafter, Valerie and Nadgir, Nitya},
  title = {Is the politicization of generative AI inevitable?},
  year = {2025},
  howpublished = {Brookings, \url{https://www.brookings.edu/articles/is-the-politicization-of-generative-ai-inevitable/}},
  note = {Accessed: 2026-02-22}
}

@article{ludan2023interpretable,
  title={Interpretable-by-design text classification with iteratively generated concept bottleneck},
  author={Ludan, Josh Magnus and Lyu, Qing and Yang, Yue and Dugan, Liam and Yatskar, Mark and Callison-Burch, Chris},
  journal={arXiv preprint arXiv:2310.19660},
  year={2023}
}

@inproceedings{koh2020concept,
  title={Concept bottleneck models},
  author={Koh, Pang Wei and Nguyen, Thao and Tang, Yew Siang and Mussmann, Stephen and Pierson, Emma and Kim, Been and Liang, Percy},
  booktitle={International conference on machine learning},
  pages={5338--5348},
  year={2020},
  organization={PMLR}
}

@article{bolet2025media,
  title={Media platforming and the normalisation of extreme right views},
  author={Bolet, Diane and Foos, Florian},
  journal={British Journal of Political Science},
  volume={55},
  pages={e103},
  year={2025},
  publisher={Cambridge University Press}
}

@book{benkler2018network,
  title={Network propaganda: Manipulation, disinformation, and radicalization in American politics},
  author={Benkler, Yochai and Faris, Robert and Roberts, Hal},
  year={2018},
  publisher={Oxford University Press}
}

@article{dunwoody2022fascist,
  title={The fascist authoritarian model of illiberal democracy},
  author={Dunwoody, Philip T and Gershtenson, Joseph and Plane, Dennis L and Upchurch-Poole, Territa},
  journal={Frontiers in Political Science},
  volume={4},
  pages={907681},
  year={2022},
  publisher={Frontiers Media SA}
}

@misc{gpt4ocutoffdate, 
author = {OpenAI Developers}, 
title = {GPT-4o; Fast, intelligent, flexible GPT model},
year = {2023}, 
howpublished = {\url{https://developers.openai.com/api/docs/models/gpt-4o}}, 
note = {Accessed: 2025-02-10} 
}

@book{mccarty1997income,
  title={Income redistribution and the realignment of American politics},
  author={McCarty, Nolan M and Poole, Keith T and Rosenthal, Howard},
  year={1997},
  publisher={AEI Press, publisher for the American Enterprise Institute}
}

@inproceedings{gu2024minillm,
  title={Minillm: Knowledge distillation of large language models},
  author={Gu, Yuxian and Dong, Li and Wei, Furu and Huang, Minlie},
  booktitle={International Conference on Learning Representations},
  volume={2024},
  pages={32694--32717},
  year={2024}
}

@inproceedings{tian2025beyond,
  title={Beyond answers: Transferring reasoning capabilities to smaller llms using multi-teacher knowledge distillation},
  author={Tian, Yijun and Han, Yikun and Chen, Xiusi and Wang, Wei and Chawla, Nitesh V},
  booktitle={Proceedings of the Eighteenth ACM International Conference on Web Search and Data Mining},
  year={2025}
}

@inproceedings{bucilua2006model,
  title={Model compression},
  author={Buciluǎ, Cristian and Caruana, Rich and Niculescu-Mizil, Alexandru},
  booktitle={Proceedings of the 12th ACM SIGKDD international conference on Knowledge discovery and data mining},
  year={2006}
}

@inproceedings{chen2022knowledge,
  title={Knowledge distillation with the reused teacher classifier},
  author={Chen, Defang and Mei, Jian-Ping and Zhang, Hailin and Wang, Can and Feng, Yan and Chen, Chun},
  booktitle={Proceedings of the IEEE/CVF conference on computer vision and pattern recognition},
  pages={11933--11942},
  year={2022}
}

@inproceedings{ganti2023narrative,
  title={Narrative style and the spread of health misinformation on twitter},
  author={Ganti, Achyutarama and Hussein, Eslam Ali Hassan and Wilson, Steven and Ma, Zexin and Zhao, Xinyan},
  booktitle={Findings of the Association for Computational Linguistics: EMNLP 2023},
  pages={4266--4282},
  year={2023}
}

@inproceedings{wan2024tnt,
  title={Tnt-llm: Text mining at scale with large language models},
  author={Wan, Mengting and Safavi, Tara and Jauhar, Sujay Kumar and Kim, Yujin and Counts, Scott and Neville, Jennifer and Suri, Siddharth and Shah, Chirag and White, Ryen W and Yang, Longqi and others},
  booktitle={Proceedings of the 30th ACM SIGKDD conference on knowledge discovery and data mining},
  pages={5836--5847},
  year={2024}
}

@misc{congressional_record, 
author = {Matthew Gentzkow and Jesse M. Shapiro and Matt Taddy}, 
title = {Congressional Record for the 43rd-114th Congresses: Parsed Speeches and Phrase Counts},
year = {2017}, 
howpublished = {\url{https://data.stanford.edu/congress_text}}, 
note = {Accessed: 2025-02-10} 
}

@misc{voteview, 
author = {Lewis, Jeffrey B. and Poole, Keith and Rosenthal, Howard and Boche, Adam and Rudkin, Aaron and Sonnet, Luke}, 
title = {Voteview: Congressional Roll-Call Votes Database.},
year = {2026}, 
howpublished = {\url{https://voteview.com/data}}, 
note = {Accessed: 2025-02-10} 
}

@inproceedings{haller2025leveraging,
  title={Leveraging in-context learning for political bias testing of llms},
  author={Haller, Patrick and Vamvas, Jannis and Sennrich, Rico and J{\"a}ger, Lena Ann},
  booktitle={Proceedings of the 63rd Annual Meeting of the Association for Computational Linguistics (Volume 1: Long Papers)},
  pages={24718--24738},
  year={2025}
}

@misc{llama4releaseMeta,
  author = {Meta},
  title = {The Llama 4 herd: The beginning of a new era of natively multimodal AI innovation},
  year = {2025},
  howpublished = { \url{https://ai.meta.com/blog/llama-4-multimodal-intelligence/}},
  note = {Accessed: 2026-02-22}
}

@article{gupta2025comparative,
  title={Comparative analysis of DeepSeek R1, ChatGPT, Gemini, Alibaba, and LLaMA: Performance, reasoning capabilities, and political bias},
  author={Gupta, Ragav},
  journal={Authorea Preprints},
  year={2025},
  publisher={Authorea}
}

@misc{deepseekcensorshipguardian,
  author = {Booth, Robert and Milmo, Dan},
  title = {Chinese AI chatbot DeepSeek censors itself in realtime, users report},
  year = {2025},
  howpublished = {The Guardian, \url{https://www.theguardian.com/technology/2025/jan/28/chinese-ai-chatbot-deepseek-censors-itself-in-realtime-users-report}},
  note = {Accessed: 2026-02-22}
}

\clearpage
\newgeometry{margin=1in,headheight=13pt}
\onecolumn
\setcounter{section}{0}
\setcounter{subsection}{0}
\setcounter{figure}{0}
\setcounter{table}{0}
\setcounter{secnumdepth}{3}
\renewcommand{\thesection}{\arabic{section}}
\renewcommand{\thesubsection}{\thesection\Alph{subsection}}
\renewcommand{\thesubsubsection}{\thesubsection.\arabic{subsubsection}}
\renewcommand{\thefigure}{S\arabic{figure}}
\renewcommand{\thetable}{S\arabic{table}}
\renewcommand{\theHsection}{SI.\arabic{section}}
\renewcommand{\theHfigure}{SI.\arabic{figure}}
\renewcommand{\theHtable}{SI.\arabic{table}}

\phantomsection
\pdfbookmark[0]{SI Appendix}{si-appendix}
\begin{center}
{\sffamily\Large\bfseries SI Appendix\par}
\vspace{0.6em}
{\sffamily\large Asymmetric rhetorical strategies drive partisanship in television news\par}
\vspace{0.4em}
{\sffamily\normalsize Upasana Dutta, Homa Hosseinmardi, Amir Ghasemian, Aaron Clauset, Duncan J. Watts\par}
\end{center}
\vspace{1em}

\section{Data}
\label{Data}

We analyze television news transcripts obtained from a commercial media monitoring service that archives broadcast content from major U.S. stations. Our dataset spans 408,464 news episodes aired between January 2013 and March 2025 across six prominent channels: Fox News, CNN, MSNBC, ABC, CBS, and NBC. For the broadcast networks, news programming is drawn from their New York, NY affiliates (WABC, WCBS, WNBC). \textcolor{black}{Each episode corresponds to a single airing of a program (e.g., one broadcast of a given show). The transcripts originate as closed captions produced by the news stations that encode the caption text into the broadcast signal. The media-monitoring provider captures these signals off-air through receivers located in each designated market area (DMA) and extracts the embedded caption text. The transcripts therefore reflect the broadcasters' own caption output rather than speech recognition performed by the provider or by us.}

\begin{figure*}[!htbp]
\centering
\includegraphics[width=0.47\textwidth]{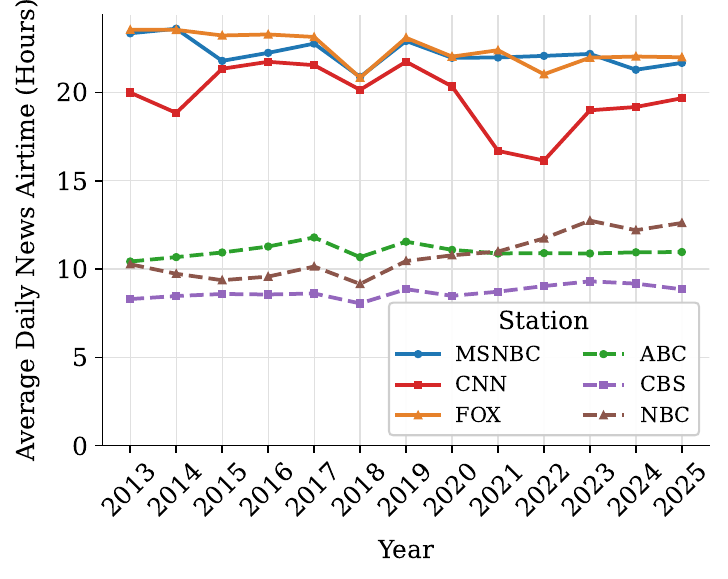}
\caption{\textcolor{black}{\textbf{Average daily news airtime by station, 2013--2025.} Each line shows a station's mean duration of news content per day, aggregated by year. Cable stations (MSNBC, CNN, Fox News) sustain roughly twice the daily volume of the broadcast networks (ABC, CBS, NBC) throughout the period.}}
\label{fig:airtime}
\end{figure*}

\textcolor{black}{To balance context and model input limits, we segmented each episode transcript into contiguous $\sim$800-word segments (roughly five minutes of airtime), yielding 3,505,401 segments. Table~\ref{sample_by_station} shows the total number of episodes and segments by station. Coverage volumes are broadly stable across the study period (Figure~\ref{fig:airtime}), with the cable stations averaging roughly 20--24 hours of daily news content and the broadcast stations 8--13 hours. 46 days (out of 4,473 calendar days; 1.03\%) lack data from the 6 stations. This missingness is concentrated in 2013 (5.9\% of station-days) and 2020 (5.5\%); all other years are below 1.4\%.}

\begin{table}[htbp]
\centering
\caption{Sample size by station (2013-2025)}
\label{tab:coverage_summary}
\begin{tabular}{lccr}
\toprule
\textbf{News Station} & \textbf{Total Episodes} & \textbf{Total Segments} \\
\midrule
FOX News & 84,746 & 801,884 \\
MSNBC & 87,691 & 853,726 \\
CNN & 80,332 & 731,555 \\
ABC & 53,441 & 410,215 \\
CBS & 44,283 & 305,214 \\
NBC & 57,971 & 402,807 \\
\bottomrule
\end{tabular}
\label{sample_by_station}
\end{table}

\FloatBarrier
\section{Partisanship Measurement}
\subsection{LLM-based partisanship measurement}
\label{prompt_gpt_partisanship}

\begin{figure*}[htbp]
\centering
\begin{promptbox}
\footnotesize	
Here is a news transcript from a TV show. \\
Transcript:\{Transcript\}

\vspace*{0.4cm}

Evaluate the political partisanship of the transcript's content based on the language, framing, or viewpoints expressed in the text. \\

Assign a score on a continuous scale from 0 to 100, where:

\textcolor{black}{
- 0 = Strong bias towards liberal viewpoints\\
- 50 = Neutral / non-partisan\\
- 100 = Strong bias towards conservative viewpoints\\
}

\vspace*{0.4cm}

Then, provide an explanation for your score in fewer than 30 words.\\

\vspace*{0.4cm}

Answer only in the following format: \\
Score: <score between 0 and 100> \\
Explanation: <explanation>
\end{promptbox}
\caption{Prompt used with GPT-4o model to assess partisanship in TV news transcripts.}
\label{fig:prompt_partisanship}
\end{figure*}

We use OpenAI's GPT‑4o model to assign a partisanship score to each news segment on a 0–100 scale, where 0 indicates a strong liberal bias and 100 indicates a strong conservative bias (Figure~\ref{fig:prompt_partisanship}). This prompt-based approach builds on recent work demonstrating the effectiveness of large language models for estimating political ideology in textual data~\cite{le2024positioning}. In our teacher–student framework, GPT‑4o serves as the ``teacher": we collect its partisanship scores for a subset of 10,000 news episodes ($\approx 2.5\%$ of the full corpus), stratified by year. These episodes, comprising 86,131 individual news segments, form the training and evaluation data for the neural network ``student" model.

\begin{figure*}[h]
\centering
\includegraphics[width=0.9\textwidth]{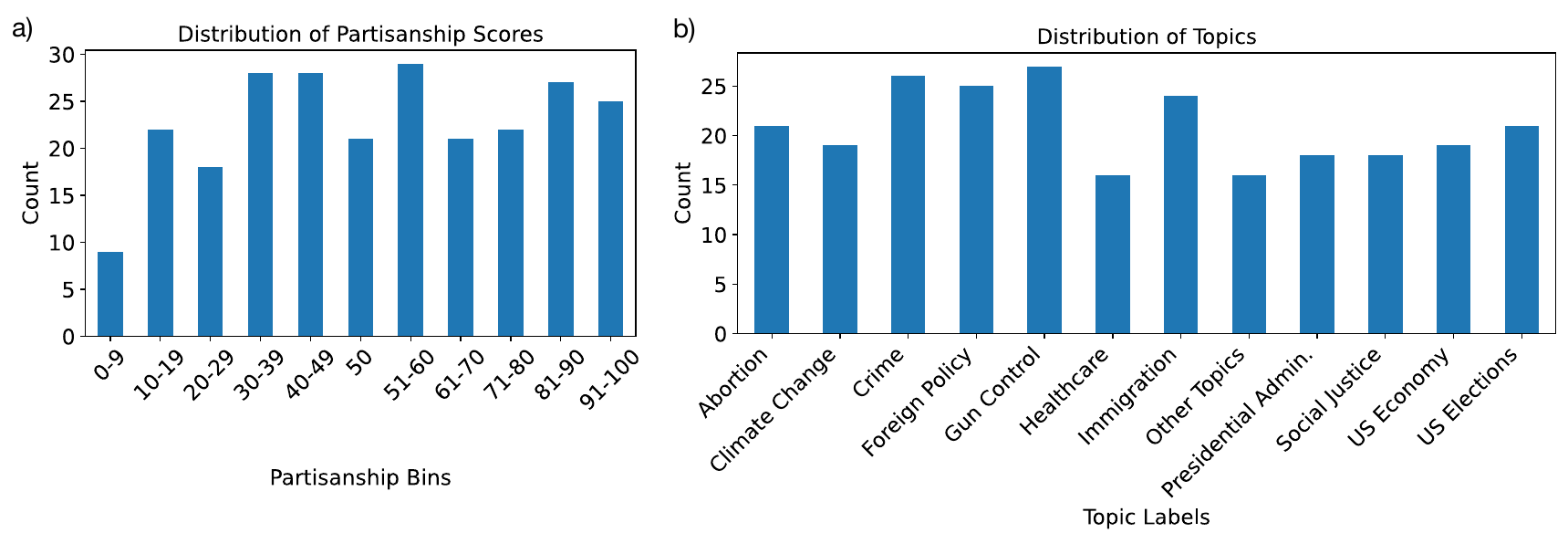}
\caption{\textbf{Distribution of partisanship scores and topics in the 250 segments selected for measurement validation.} The sample spans a wide distribution of partisanship scores and topics. The only exception is the [0–9] partisanship bin, which was due to the dearth of strongly liberal content in the corpus of 10,000 LLM-labeled episodes: only 9 out of the 86,131 segments received a partisanship score in this range.}
\label{fig:distributions_250Segments}
\end{figure*}

Before we train the neural network, we first assess the validity of the LLM-generated partisanship scores. We compare them with human-coded labels on a stratified sample of 250 news segments. The set of 250 news segments was sampled in a way that ensured balanced coverage across the full range of partisanship scores and topical categories. Specifically, for each segment, we first selected a decile range of partisanship scores uniformly at random (e.g., [0–9], [10–19], $\dots$, [91–100]), and then uniformly selected a topic from among the topics represented within that decile (see Section~\ref{sec:topic_analysis} for topic classification). A segment was then randomly selected from the set matching both criteria. The resulting sample spans a wide distribution of partisanship scores and topics (see Figure~\ref{fig:distributions_250Segments}). The only exception is the [0–9] bin, which contains significantly fewer samples due to the rarity of strongly liberal content in the corpus---only 9 out of the 86,131 segments received a partisanship score in this range.

Each of these 250 segments was then independently rated by three trained research assistants for partisanship on an 11-point Likert scale ($-5$ = strongly liberal, $+5$ = strongly conservative). The human labels are linearly scaled to the [0-100] range for direct comparison with LLM labels. The LLM partisanship scores showed strong agreement with human judgments on these 250 segments (Pearson’s $r = 0.92$, $R^2 = 0.83$; Figure~\ref{fig:partisanship_humanlabelcomparisons}a in main text). 

In addition to assigning partisanship scores, the research assistants also classified each news segment into one of 12 topic categories and labeled the presence or absence of each of the eight rhetorical strategies analyzed in our study.

\FloatBarrier

\subsection{Congressional speech based partisanship model}
\label{congressional_model}

To benchmark our LLM-based annotation approach, we compare its performance with a widely used dictionary-based method from~\cite{martin2017bias}, adapted from~\cite{gentzkow2010drives} to estimate the ideological slant of print newspapers. This approach relies on the linguistic patterns of U.S. Congress members.

We source the full text of Congressional speeches via Stanford’s Social Science Data Collection (2015–2016)~\cite{congressional_record}. After standard text preprocessing—including stopword removal, stemming, and lemmatization—we extract all bigrams and compute the Gentzkow-Shapiro partisanship statistic~\cite{gentzkow2010drives} for each. This statistic reflects how strongly a bigram is associated with either Democratic or Republican speakers. To ensure relevance to the television news domain, we retain only those bigrams that appear at least 20 times in our news transcript corpus, which filters out several procedural phrases, and then select the 1,000 most partisan bigrams. Each Congressperson is then represented by a frequency vector over these bigrams, normalized to account for individual verbosity.

We train an elastic net regression model using these normalized bigram vectors to predict legislators’ DW-NOMINATE (Dynamic Weighted NOMINAl Three-step Estimation) first dimension scores~\cite{mccarty1997income}— ideological scores ranging from -1 (liberal) to +1 (conservative)—obtained from Voteview~\cite{voteview}. To ensure sufficient signal, we restrict training on legislators who used the top bigrams at least 10 times. The resulting model achieves an $R^2 = 0.678$, Pearson's $r = 0.849$ on held-out Congressional data, confirming strong in-domain performance.

However, this predictive power does not transfer out-of-domain.\ When applied to our human-labeled sample of television news segments, the model's predictions exhibit a weak correlation $r = 0.225$.\ This reduced performance on out-of-domain data mirrors findings in the original study, where they report a correlation of $r = 0.40$ between the model’s predictions and ideology of newspapers from mondotimes.com, conditioned on the newspapers with at least 25,000 mentions of the top 1000 bigrams~\cite{gentzkow2010drives}.\ This shows that patterns learned from congressional speech do not generalize well to TV news and underscores the drawbacks of using such dictionary-based methods for fine-grained, out-of-domain applications.\

\subsection{Bayesian lexical inference model for measuring polarization}
\label{gentzkow2019_model_appendix}

Gentzkow et al.~\cite{gentzkow2019measuring} introduce a Bayesian framework to quantify linguistic polarization between pairs of corpora based on differences in word usage. The method estimates the expected posterior probability that an observer with a neutral prior can correctly infer the source corpus of a randomly sampled phrase, given observed word usage. Intuitively, it captures how revealing a short text segment (e.g., a two-word phrase) is about its origin---such as whether it came from a Democratic or Republican speaker—based solely on lexical patterns.

We evaluated this method on the 250 human-labeled news segments in our study. Since the model is inherently comparative and does not assign partisanship scores to individual segments in isolation, we follow prior work~\cite{hosseinmardi2025unpacking} and use it to estimate the degree to which each cable news segment (from Fox News, MSNBC, or CNN) is distinguishable from a baseline corpus of broadcast news segments (from ABC, NBC, and CBS). This yields valid estimates for 224 cable news segments of the 250 segments.

On this subset, the Bayesian word-choice model achieves an $R^2 = 0.636$ and $r = 0.819$ with human-coded partisanship labels. In contrast, our LLM-prompting method shows significantly stronger alignment both on the full 250-segment set ($r = 0.92$, $R^2 = 0.83$) and on the same subset of 224 cable segments ($r = 0.928$, $R^2 = 0.85$), demonstrating a notable improvement in predictive performance for fine-grained segment-level partisanship estimation.

\subsection{Neural Network Training}
\label{neural_network_training}

To predict partisanship scores from text, we train a neural network model using the LLM-generated partisanship scores. Such teacher-student frameworks that scale up the labeling behavior of larger models by training lightweight classifiers have been used in past literature~\cite{ganti2023narrative, wan2024tnt}. Each news segment is first embedded as a 3,072-dimensional vector using OpenAI’s \textit{text-embedding-3-large} model. The labeled dataset—comprising 10,000 episodes (86,131 segments)—is split evenly into training and test sets: 5,000 episodes (42,679 segments) for training and 5,000 episodes (43,452 segments) for evaluation.

The prediction task is framed as a regression problem, with the neural network learning to approximate the LLM-assigned partisanship scores. The architecture consists of a shallow feedforward network with two hidden layers of 1,024 neurons each and ReLU activation. The final layer contains a single neuron to output the predicted partisanship score. The partisanship scores (ranging from 0 to 100) are linearly rescaled to \([0, 1]\), and the binary cross-entropy (BCE) loss is used to model the task as probabilistic regression with a sigmoid activation function in the final layer that constrains predictions to the \([0, 1]\) range. Key hyperparameters—including learning rate, batch size, dropout rate, and number of training epochs—are selected via 5-fold cross-validation on the training set.

\begin{figure}[h]
\centering
\includegraphics[width=0.68\textwidth]{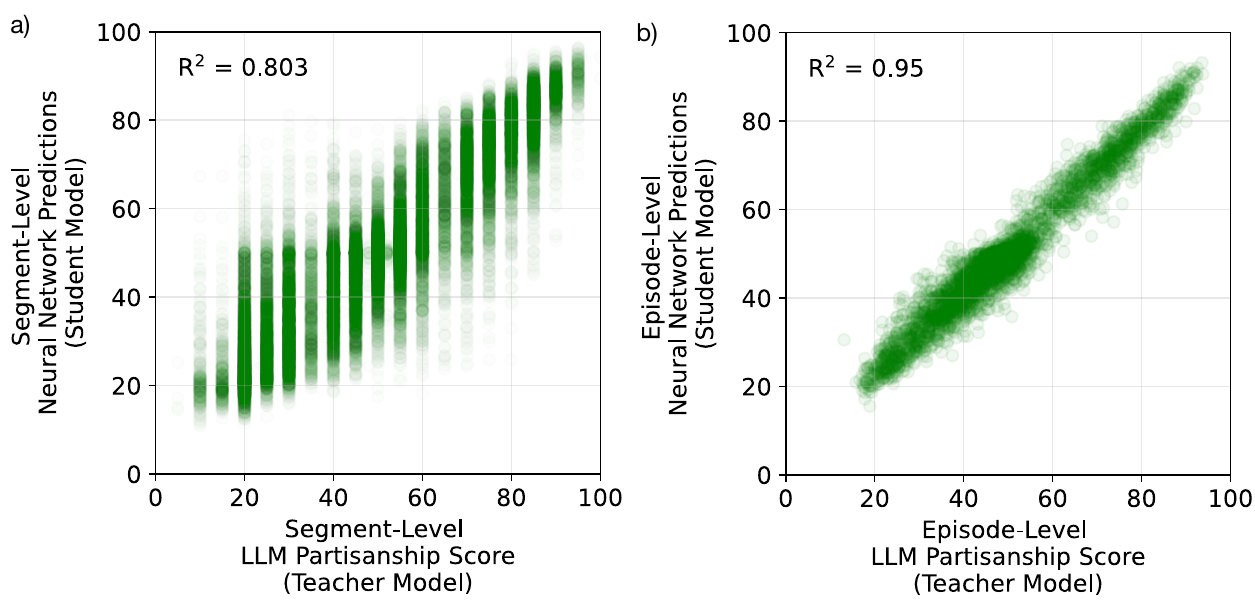}
\caption{\textbf{Agreement between true partisanship scores from LLM (teacher model) and predicted partisanship scores from neural network (student model)} (a) Partisanship labels from LLM and those predicted by the neural network for 43,452 segments in test set show high agreement ($R^2 = 0.803$). (b) Episode-level partisanship scores, obtained by averaging the partisanship scores per episode, also show high agreement ($R^2 = 0.950$).}
\label{fig:nn_agreement_appendix}
\end{figure}

\begin{figure}[h]
\centering
\includegraphics[width=0.35\textwidth]{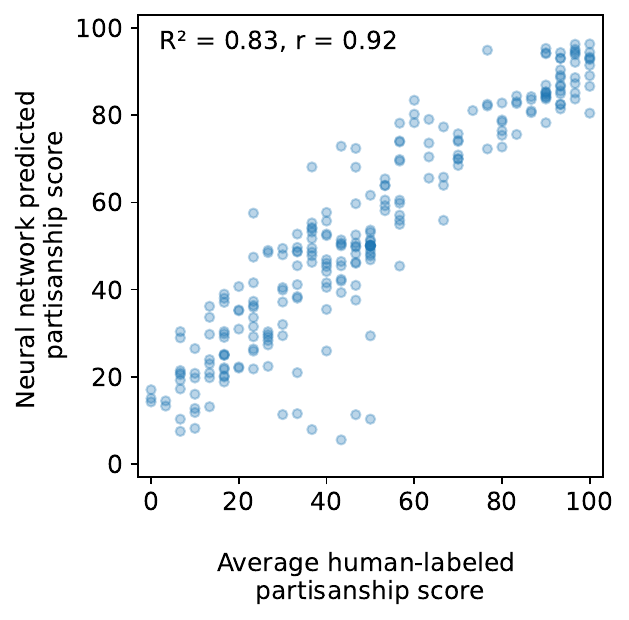}
\caption{\textbf{Agreement between human-labeled partisanship scores and the predicted partisanship scores from the neural network student model.} On the set of 250 segments labeled by trained human coders, the neural network student model shows strong predictive power ($R^2 = 0.83$, Pearson's $r = 0.923$).}
\label{fig:human_NN_agreement_appendix}
\end{figure}

This neural network student model demonstrates strong performance in predicting LLM labels on the held-out test set (43,452 segments; $R^2 = 0.803$; Pearson's $r =0.896$; Figure~\ref{fig:nn_agreement_appendix}a). Episode-level partisanship scores were computed by averaging the segment-level scores within each episode and exhibited stronger agreement with the LLM labels (5,000 episodes; $R^2 = 0.95$, $r = 0.976$; Figure~\ref{fig:nn_agreement_appendix}b). The student model also showed high predictive performance on the set of 250 news segments with human-coded partisanship labels ($R^2 = 0.83$, Pearson's $r = 0.923$; Figure~\ref{fig:human_NN_agreement_appendix}).

\FloatBarrier

\FloatBarrier

\begin{figure*}[h]
\centering
\begin{promptbox}
\footnotesize
    \vspace*{0.1cm}
    Here is a news transcript from a TV show.
    \\~\\ 
    Transcript:
    \\
    \{Transcript\}
    \\

    \vspace*{0.4cm}

    Based on its content, identify which of the following "bias markers" are present in the transcript. 
    \\
    \textcolor{blue}{1. Negative Sentiment towards Democratic Viewpoints} (criticism, mockery, or negative framing of liberal ideas or policies)
    \\
    \textcolor{blue}{2. Negative Sentiment towards Republican Viewpoints} (criticism, mockery, or negative framing of conservative ideas or policies)
    \\
    \textcolor{blue}{3. Reinforcement of Democratic Viewpoints} (support or favorable framing of liberal views or politicians)
    \\
    \textcolor{blue}{4. Reinforcement of Republican Viewpoints} (support or favorable framing of conservative views or politicians)
    \\
    \textcolor{blue}{5. Blame Attribution towards Democratic Politicians/Policies} (assigns fault to Democratic figures or policies for negative outcomes)
    \\
    \textcolor{blue}{6. Blame Attribution towards Republican Politicians/Policies} (assigns fault to Republican figures or policies for negative outcomes)
    \\
    \textcolor{blue}{7. Use of Inflammatory/Derogatory Language} (emotionally charged, mocking, or hostile language)
    \\
    \textcolor{blue}{8. Focus on Fact-based Reporting} (emphasis on factual information, statistics, or neutral descriptions without strong opinion)
    \\

    \vspace*{0.4cm}
    
    You should assess the presence/absence of each feature independently. First respond with the labels of the bias markers that are present in the transcript (do not provide their numbers). If multiple bias markers are present, separate them with commas. A bias marker should be said to be present only if it is substantially present in the transcript, not simply mentioned once. If none of the bias markers are present, your answer should be an empty string. In the next line, provide a short explanation for your answer in fewer than 30 words. Do not make up new bias markers. \\~\\
    
    Answer only in the following format: \\
    Bias markers: <bias marker label(s) present, separated by commas> \\
    Explanation: <explanation>
    \vspace*{0.1cm}
\end{promptbox}
\caption{\textbf{Prompt for rhetorical strategy classification.} Prompt used with GPT-4o for detecting the presence or absence of eight linguistic features that capture rhetorical strategies in TV news transcripts. The ordering of the 8 strategies in the prompts was randomized in every prompt so that the output is not biased by its order.}
\label{fig:prompt_features}
\end{figure*}
\FloatBarrier

\section{Rhetorical Strategies}
\label{sec:interpretable_features}

\subsection{LLM-based measurement}
\label{sec:rhetorical_llm_measurement}

The prompt for detecting each of the eight rhetorical strategies is shown in Figure~\ref{fig:prompt_features}. We randomize the order of the eight strategies each time the prompt is run so that the outputs are not biased by the inherent ordering of the strategies in the prompt. The explanations from the LLM are used for qualitatively validating the LLM labels. Like before, these LLM labels were collected for the subset of 10,000 news episodes.

\begin{figure*}[h]
\centering
\includegraphics[width=0.95\textwidth]{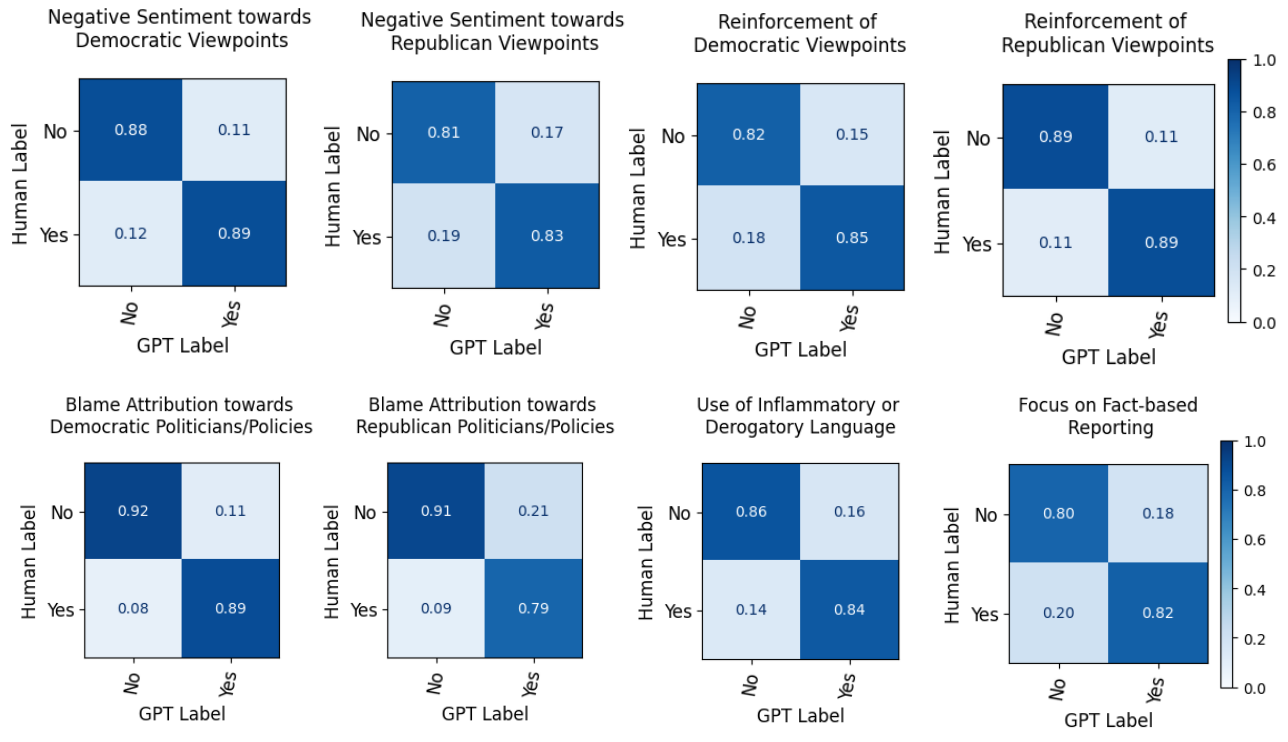}
\caption{\textbf{Confusion matrices showing human-LLM agreement.} Across all eight strategies, agreement between the binary LLM-labels and the majority vote among three trained human coders on 250 news segments was high -- F1 scores ranged between 0.705 to 0.853. Additionally, the degree of LLM-human agreement was indistinguishable from inter-human agreement for each strategy (bootstrap test $p$-values: 0.487-0.736).}
\label{fig:biasmarkers_gpt_human_agreement}
\end{figure*}

On the set of 250 human-coded news segments, GPT-4o shows strong agreement with human labels on each of the eight strategies (Table~\ref{tab:human_gpt_feature_agreement}, F1 scores
0.705-0.853, precision 0.794-0.889, recall 0.611-0.821). A majority vote among humans was taken for each segment, and ties were resolved uniformly at random. See Table~\ref{tab:human_gpt_feature_agreement} and Figure~\ref{fig:biasmarkers_gpt_human_agreement} for more details on the LLM-human agreement on each of the eight strategies.

Similar to the teacher–student framework for measuring partisanship, we train neural networks, one for each rhetorical strategy, to predict the presence or absence of the strategy given the input text embedding of a news segment. The same train and test sets are used as before (5000 episodes in each set). The trained neural networks show strong predictive performance on the test set (Table~\ref{tab:feature_classifier_performance}; F1 scores between 0.571-0.819). The relatively lower performance on the use of inflammatory
or derogatory language (F1 = 0.571) is largely attributable to class imbalance (89.8\% negative vs. 10.2\% positive). For reference, with such an imbalance, a stratified random guesser achieves F1 $\approx 0.10$.

\begin{table*}[h]
\centering
\small
\begin{tabular}{|l|c|c|c|}
\hline
\textbf{Rhetorical Strategy} & \textbf{F1 Score} & \textbf{Precision} & \textbf{Recall} \\
\hline
Negative Sentiment towards Democratic Viewpoints & 0.810 & 0.886 & 0.747 \\
Negative Sentiment towards Republican Viewpoints & 0.705 & 0.833 & 0.611 \\
Reinforcement of Democratic Viewpoints           & 0.714 & 0.846 & 0.618 \\
Reinforcement of Republican Viewpoints           & 0.833 & 0.886 & 0.787 \\
Blame Attribution towards Democratic Politicians/Policies & 0.853 & 0.889 & 0.821 \\
Blame Attribution towards Republican Politicians/Policies & 0.777 & 0.794 & 0.761 \\
Use of Inflammatory or Derogatory Language       & 0.727 & 0.842 & 0.640 \\
Focus on Fact-based Reporting                    & 0.717 & 0.816 & 0.639 \\
\hline
\end{tabular}
\vspace{2pt}
\caption{Agreement between our LLM-based annotation method and the human majority vote label across eight rhetorical strategies. Ties were broken uniformly at random.}
\label{tab:human_gpt_feature_agreement}
\end{table*}

\begin{table*}[h]
\centering
\small
\begin{tabular}{|l|c|c|c|}
\hline
\textbf{Rhetorical Strategy} & \textbf{F1 Score} & \textbf{Precision} & \textbf{Recall} \\
\hline
Negative Sentiment towards Democratic Viewpoints & 0.745 & 0.740 & 0.749 \\
Negative Sentiment towards Republican Viewpoints & 0.752 & 0.751 & 0.752 \\
Reinforcement of Democratic Viewpoints & 0.607 & 0.583 & 0.633 \\
Reinforcement of Republican Viewpoints & 0.755 & 0.744 & 0.766 \\
Blame Attribution towards Democratic Politicians/Policies & 0.687 & 0.726 & 0.653 \\
Blame Attribution towards Republican Politicians/Policies & 0.719 & 0.661 & 0.788 \\
Use of Inflammatory or Derogatory Language & 0.571 & 0.529 & 0.620 \\
Focus on Fact-based Reporting & 0.819 & 0.807 & 0.832 \\
\hline
\end{tabular}
\vspace{2pt}
\caption{Performance of binary neural network classifiers trained on LLM-derived annotations for each of the eight features, evaluated on the held-out test set. The relatively lower performance on the use of inflammatory or derogatory language is largely attributable to class imbalance (89.8\% negative vs.\ 10.2\% positive). For reference, with such an imbalance, a stratified random guesser achieves $F1 \approx 0.10$.}
\label{tab:feature_classifier_performance}
\end{table*}

\FloatBarrier

\subsection{Robustness check for the rhetorical-strategy regression}
\label{si:dsl}

\textcolor{black}{The association between the partisanship score and the eight rhetorical strategies is measured by fitting a linear regression where the eight LLM-labeled binary features are the covariates and the LLM-labeled partisanship score (0-100) is the outcome variable (Table~\ref{tab:regression_results} in main text). To confirm that the observed association is not an artifact of using the same LLM for labeling both the covariates and the outcome variable, we re-estimated the same model against human labels using design-based
supervised learning (DSL)~\cite{egami2023using, egami2024using}. DSL uses a small set of human-coded
segments, together with the LLM labels available for the full corpus, to estimate
the regression that would be obtained if human labels were available corpus-wide.}

\textcolor{black}{We used the 250 human-coded segments, taking the mean of the three coders' ratings as the outcome and their majority label
for each strategy. These segments were drawn by stratified sampling over
partisanship-score bins and topics, so each carries a known selection probability
$\pi_i = n_d/(T_d\,N_{dt})$, where $n_d$ is the number of human-coded segments in
score-bin $d$, $T_d$ the topics represented in that bin, and $N_{dt}$ the corpus size
of cell $(d,t)$.}

\textcolor{black}{The regression results are largely consistent with our findings from the LLM-based model (Table~\ref{tab:dsl}). All six attitudinal
strategies retain their sign and significance under DSL, with reinforcement of
Republican ($\beta = 21.389$) and Democratic ($\beta = -11.204$) viewpoints remaining
the strongest predictors of conservative and liberal slant. Standard errors are
wider, as expected when estimation is anchored on 250 human-labeled segments, and the two stylistic
strategies---already the weakest predictors---are not statistically significant under DSL. The design-weighted leave-one-out $R^2$ of the DSL model\footnote{We report leave-one-out $R^2$ because all 250 human-coded segments are used in the DSL fit; the design weights make this an estimate of corpus-level fit
rather than of fit within the stratified sample of 250 segments.} is 0.847 (bootstrap 95\% CI
$[0.811, 0.900]$), while that for the LLM-labeled model, computed on
the same 250 segments under the same design weights, is 0.756 (bootstrap 95\% CI
$[0.700, 0.829]$). The human-labeled model via DSL therefore fits somewhat better than the LLM-labeled one, so the association reported
in Table~\ref{tab:regression_results} of the main text is, if anything, conservative.}

\begin{table}[h]
\centering
\small
\color{black}   
\begin{tabular}{l S[table-format=-2.3] @{}l c S[table-format=-2.3] @{}l c}
\toprule
Rhetorical Strategy & \multicolumn{2}{c}{$\beta$} & {SE} & \multicolumn{2}{c}{$\beta$ (DSL)} & {SE (DSL)} \\
\midrule
\multicolumn{7}{l}{\textit{Attitudinal Strategy}} \\
\quad Neg.\ sentiment toward D.\ viewpoints  & 8.622   &\textsuperscript{***}& (0.183) & 8.981   &\textsuperscript{*}  & (3.645) \\
\quad Neg.\ sentiment toward R.\ viewpoints  & -8.221  &\textsuperscript{***}& (0.156) & -10.616 &\textsuperscript{***}& (1.752) \\
\quad Reinforcement of D.\ viewpoints        & -10.354 &\textsuperscript{***}& (0.143) & -11.204 &\textsuperscript{***}& (1.938) \\
\quad Reinforcement of R.\ viewpoints        & 17.131  &\textsuperscript{***}& (0.158) & 21.389  &\textsuperscript{***}& (2.825) \\
\quad Blame attribution toward D.            & 6.765   &\textsuperscript{***}& (0.175) & 7.183   &\textsuperscript{*}  & (3.537) \\
\quad Blame attribution toward R.            & -9.098  &\textsuperscript{***}& (0.159) & -10.554 &\textsuperscript{***}& (2.542) \\
\multicolumn{7}{l}{\textit{Stylistic Strategy}} \\
\quad Use of inflammatory/derogatory language & 1.183  &\textsuperscript{***}& (0.156) & -3.516  &                     & (2.513) \\
\quad Focus on fact-based reporting          & -0.013  &                     & (0.117) & 2.070   &                     & (2.121) \\
\midrule
Intercept                                    & 50.434  &\textsuperscript{***}& (0.109) & 46.230  &\textsuperscript{***}& (2.253) \\
\bottomrule
\end{tabular} 

\vspace{2pt}
{\footnotesize $^{*}p<0.05$, $^{**}p<0.01$, $^{***}p<0.001$}

\vspace{5pt}
\caption{\textcolor{black}{Rhetorical-strategy regression estimated under LLM labels (second and third columns) and under human labels via DSL (fourth and fifth columns).}}

\label{tab:dsl}
\end{table}

\subsection{Rhetorical strategies across presidencies}
\label{sec:framing_bias_analysis_presidencies}
\begin{figure*}[h!]
\centering
\includegraphics[width=\textwidth]{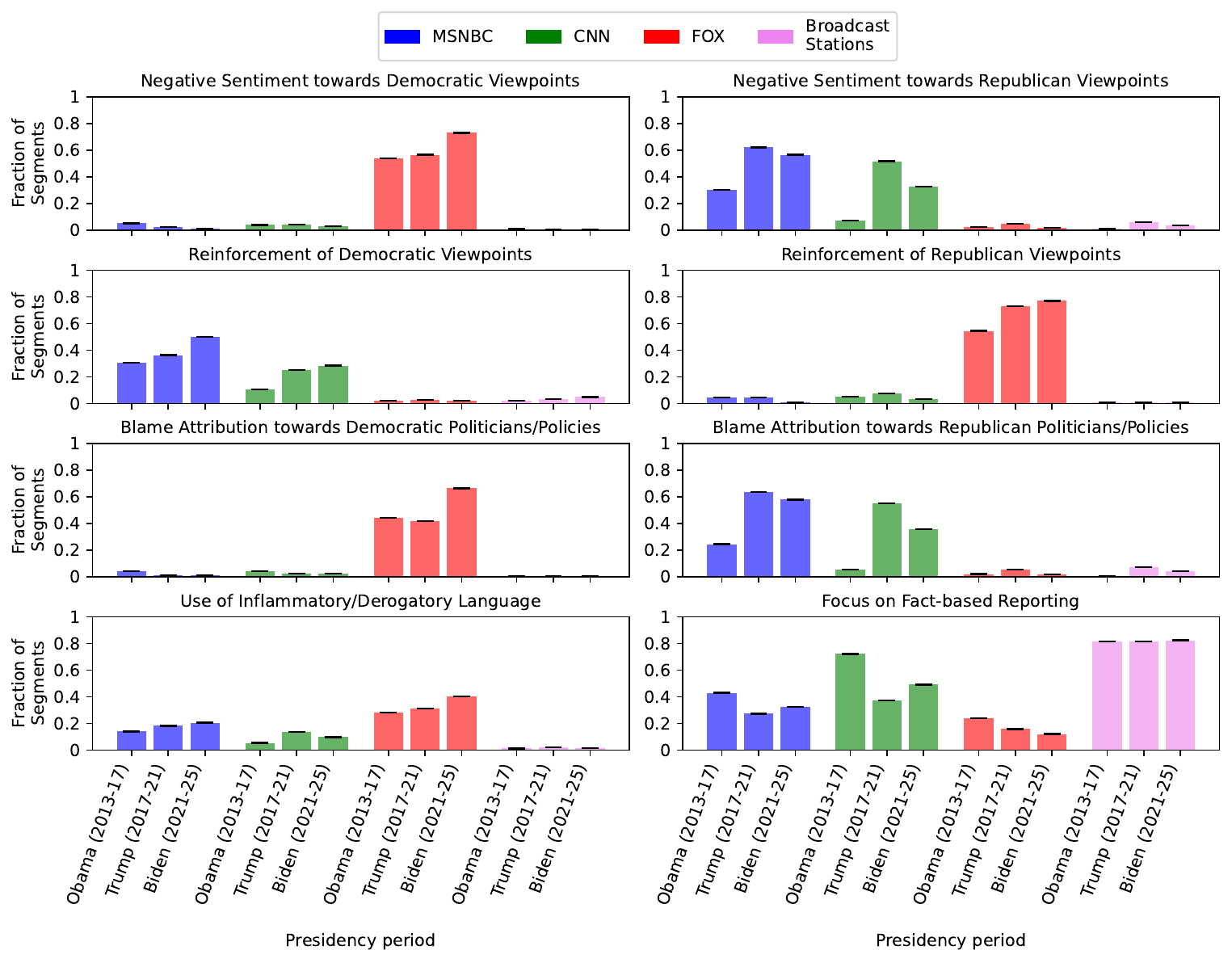}
\caption{\small \textbf{Prevalence of rhetoric in the news stations across different presidencies.} Fraction of segments (y-axis) with the presence of each of the eight rhetorical strategies in MSNBC, CNN, Fox News, and the Broadcast stations (ABC, CBS, NBC combined) across three different presidencies (x-axis). Error bars represent 95\% confidence intervals.} 
\label{fig:framing_marker_histograms}
\end{figure*}

\begin{figure*}[h!]
\centering
\includegraphics[width=0.75\textwidth]{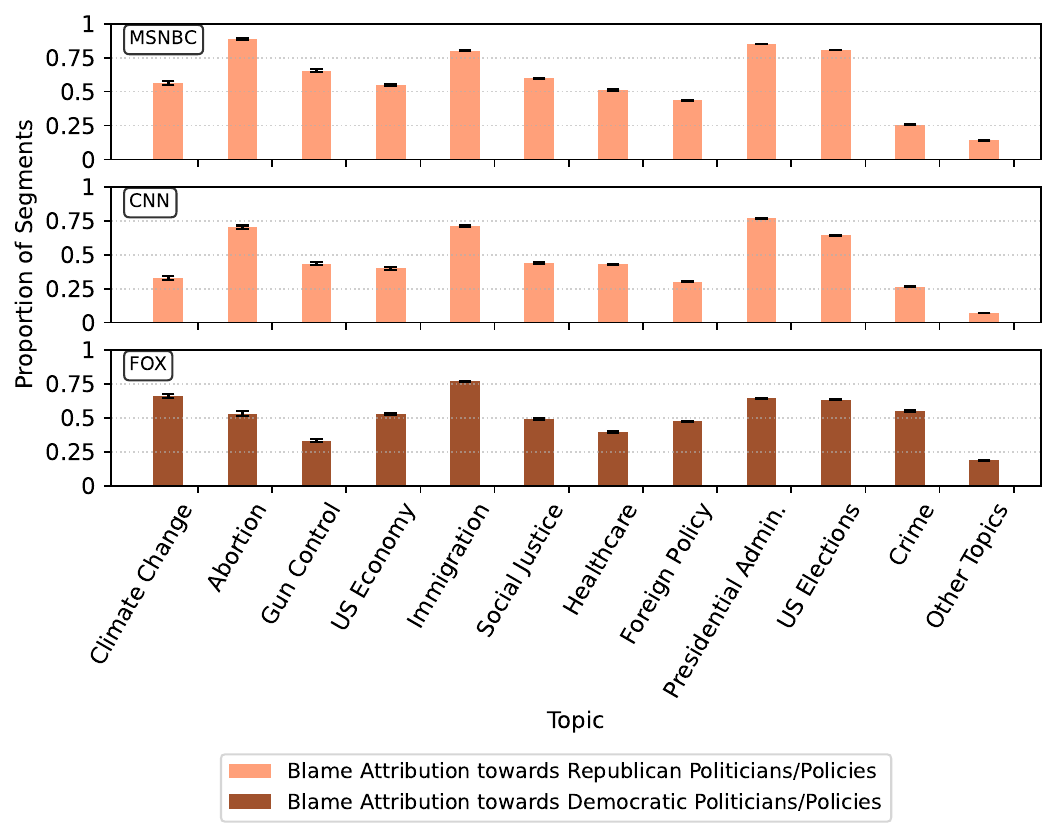}
\caption{\textbf{Topic-specific variation in blame attribution strategies across cable news stations.} }
\label{fig:blame_interactions}
\end{figure*}

To complement the time series depicted in Figure~\ref{fig:feature_timetrends} in main text, here we show histograms (Figure~\ref{fig:framing_marker_histograms}) with the presence of each of the eight strategies in MSNBC, CNN, Fox News, and the broadcast stations, aggregated across three presidencies that span our dataset: Obama (2013-01--2017-01), Trump (2017-02--2021-01), and Biden (2021-02--2025-01). For Fox News, we observe that negative sentiment towards democratic viewpoints and reinforcement of republican viewpoints increased, in aggregate, across the three presidencies. For CNN and MSNBC, the reinforcement of democratic viewpoints increased, in aggregate, across the three presidencies, although the time series shows more nuanced trends about how these strategies shifted right after the presidential electoral outcomes. However, there are significantly fewer news segments that contain this strategy in MSNBC and CNN than the corresponding strategy in Fox News. Furthermore, for MSNBC and CNN, negative sentiment towards Republican viewpoints peaked during the Trump presidency (2017-2020), and then again decreased during Biden's presidency, thus showing variation with respect to the party in power. The blame attribution strategies show qualitatively similar trends to the negative sentiments (Figure~\ref{fig:blame_interactions}). In terms of the stylistic strategies, we observe that Fox News exhibited the highest use of inflammatory/derogatory language among all the stations, and the prevalence of such language, in aggregate, has increased over time. On the contrary, Fox News's focus on fact-based reporting has been declining over the years, whereas for CNN and MSNBC, it was least present during the Trump presidency (2017-2020) and higher during Obama and Biden presidencies.

\subsubsection{In-party reinforcement versus out-party criticism across topics and presidencies}
\label{sec:balance_presidency}

\textcolor{black}{We contrast the relative balance between in-party reinforcement and out-party criticism for each station, topic, and presidency by defining the index $\Delta$ = $P$(in-party reinforcement) -- $P$(out-party criticism), where in-party reinforcement is the share of segments reinforcing the station's aligned party (Republican for Fox News; Democratic for MSNBC and CNN) and out-party criticism is the share expressing negative sentiment toward the opposing party. Positive values ($\Delta > 0$) indicate a tilt toward reinforcing one's own side; negative values ($\Delta < 0$) indicate a tilt toward criticizing the other side.}

\begin{figure*}[h!]
\centering
\includegraphics[width=0.94\textwidth]{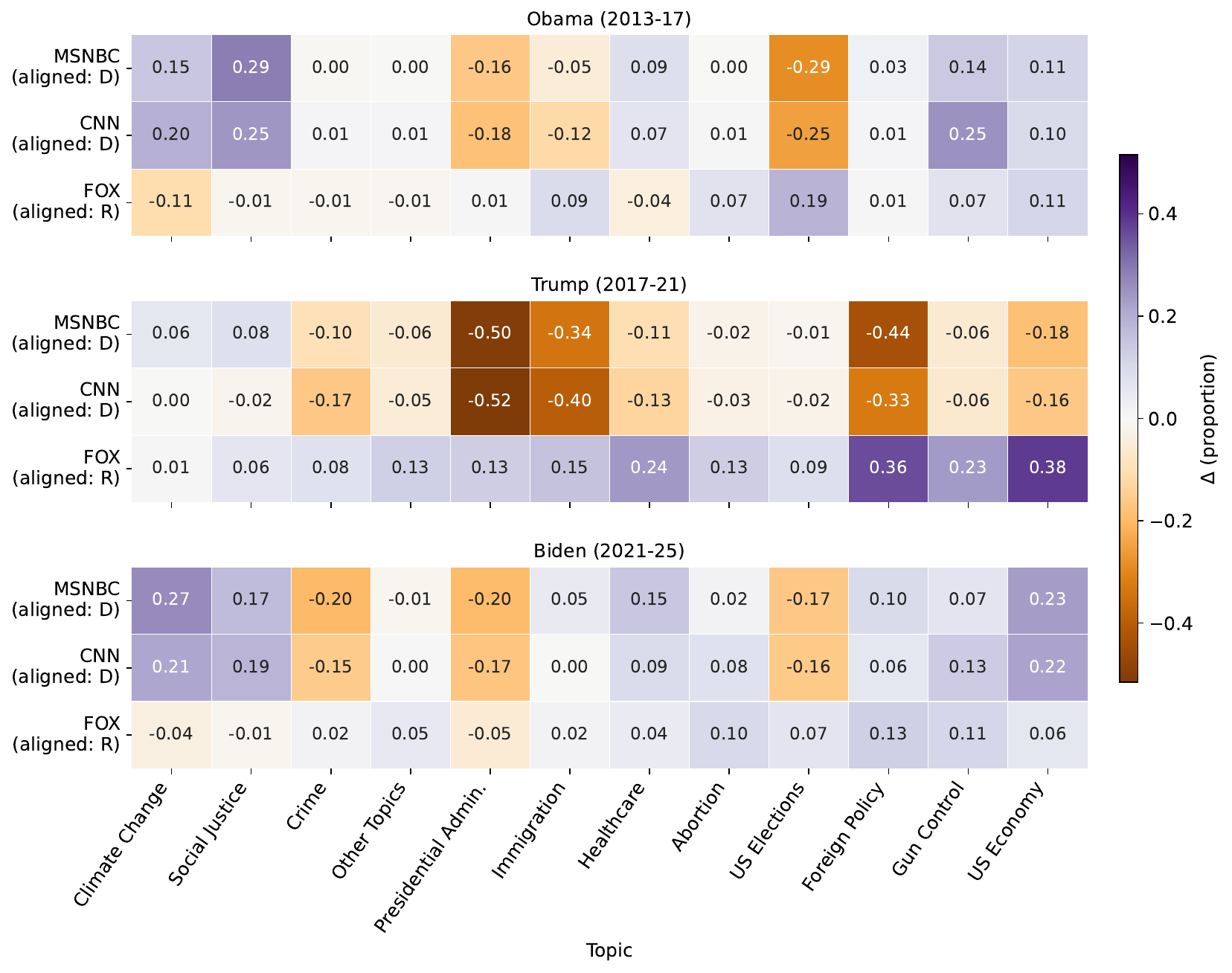}
\caption{\textcolor{black}{\textbf{In-party reinforcement versus out-party criticism, by station, topic, and presidency.} For each station, topic, and presidency, $\Delta$ is the difference between the share of segments that reinforce the station's aligned party and the share that express negative sentiment toward the opposing party (aligned party: Democratic for MSNBC and CNN, Republican for Fox News). Positive values (purple) indicate a tilt toward reinforcing the aligned party; negative values (orange) indicate a tilt toward criticizing the opposing party. Fox News is reinforcement-first across nearly all topics and all three presidencies (except for climate change), whereas MSNBC's and CNN's focus on outparty criticism is most pronounced during the Trump presidency, attenuating or reversing under Obama and Biden.}}
\label{fig:balance_topic}
\end{figure*}

\textcolor{black}{Figure~\ref{fig:balance_topic} shows this index by topic and presidency, and reveals the temporal origin of the topic asymmetry reported in the main text (Figure~\ref{fig:topicfeature_interactions}a in main text). Fox News is reinforcement-tilted across nearly all topics and all three presidencies, with only climate change consistently departing from this pattern. MSNBC and CNN, by contrast, are criticism-tilted overall, and this tilt is strongly concentrated in the Trump presidency: negative sentiment most exceeds reinforcement on Presidential Administration ($\Delta = -0.50$ and $-0.52$), Foreign Policy (-0.44 and -0.33), and Immigration (-0.34 and -0.40) during 2017--2020. Under the Obama and Biden presidencies, these attenuate substantially or reverse, with several topics---climate change, social justice, and the U.S.\ economy---tilting toward Democratic reinforcement. Fox News exhibits no comparable swing from reinforcement-first to criticism-first: its reinforcement tilt persists across presidencies and in fact strengthens on several topics during the Trump years (e.g., U.S.\ economy, $\Delta$=0.38; foreign policy, $\Delta$=0.36).}

\textcolor{black}{Together, these patterns indicate that the out-party criticism vs. in-party reinforcement dynamics of MSNBC and CNN are largely contingent on which party holds power, whereas Fox News maintains its emphasis on in-party reinforcement over out-party across topics and presidencies.}

\FloatBarrier

\FloatBarrier
\FloatBarrier

\begin{figure*}[h!]
\centering
\includegraphics[width=\textwidth]{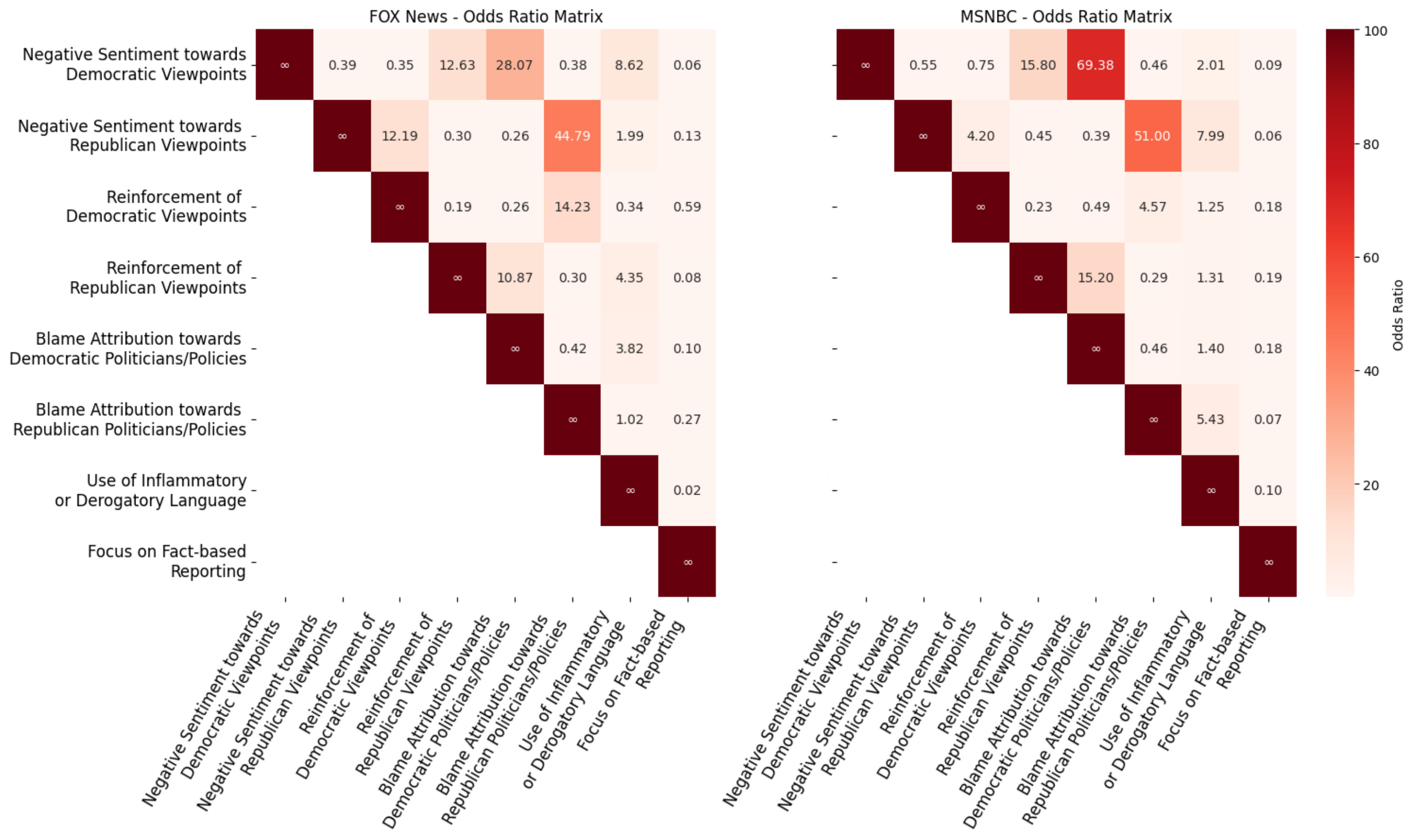}
\caption{\small \textbf{Matrices showing the odds-ratio of every pair of rhetorical features in Fox News and in MSNBC.} The odds ratio is calculated as the ratio of the odds of one strategy occurring when another strategy is present to the odds of it occurring when that other strategy is absent.} 
\label{fig:cooccurance_oddsratio}
\end{figure*}
\FloatBarrier

\subsection{Association between strategies}
\label{sec:co_occurence_analysis}
We analyze the strength of association between all the pairs of rhetorical strategies, particularly in Fox News and in MSNBC, by computing the odds ratio between the strategies (Figure~\ref{fig:cooccurance_oddsratio}). The odds ratio (OR) is calculated as the ratio of the odds of one strategy occurring when another strategy is present to the odds of it occurring when that other strategy is absent. We find that in case of Fox News, the association between negative sentiment towards Democratic views and reinforcement of Republican views (OR = 12.63, $p < 0.001$) was significantly higher than the association in MSNBC between negative sentiment towards Republicans and reinforcement of Democratic views (OR = 4.20, $p < 0.001$; difference in log~OR: $Z = 2.98, p < 0.001$), implying that the ``attack-and-praise" pattern was about three times more pronounced in Fox News than in MSNBC.

\section{Topic Analysis}
\label{sec:topic_analysis}

\begin{figure*}[t!]
\centering
\includegraphics[width=\textwidth]{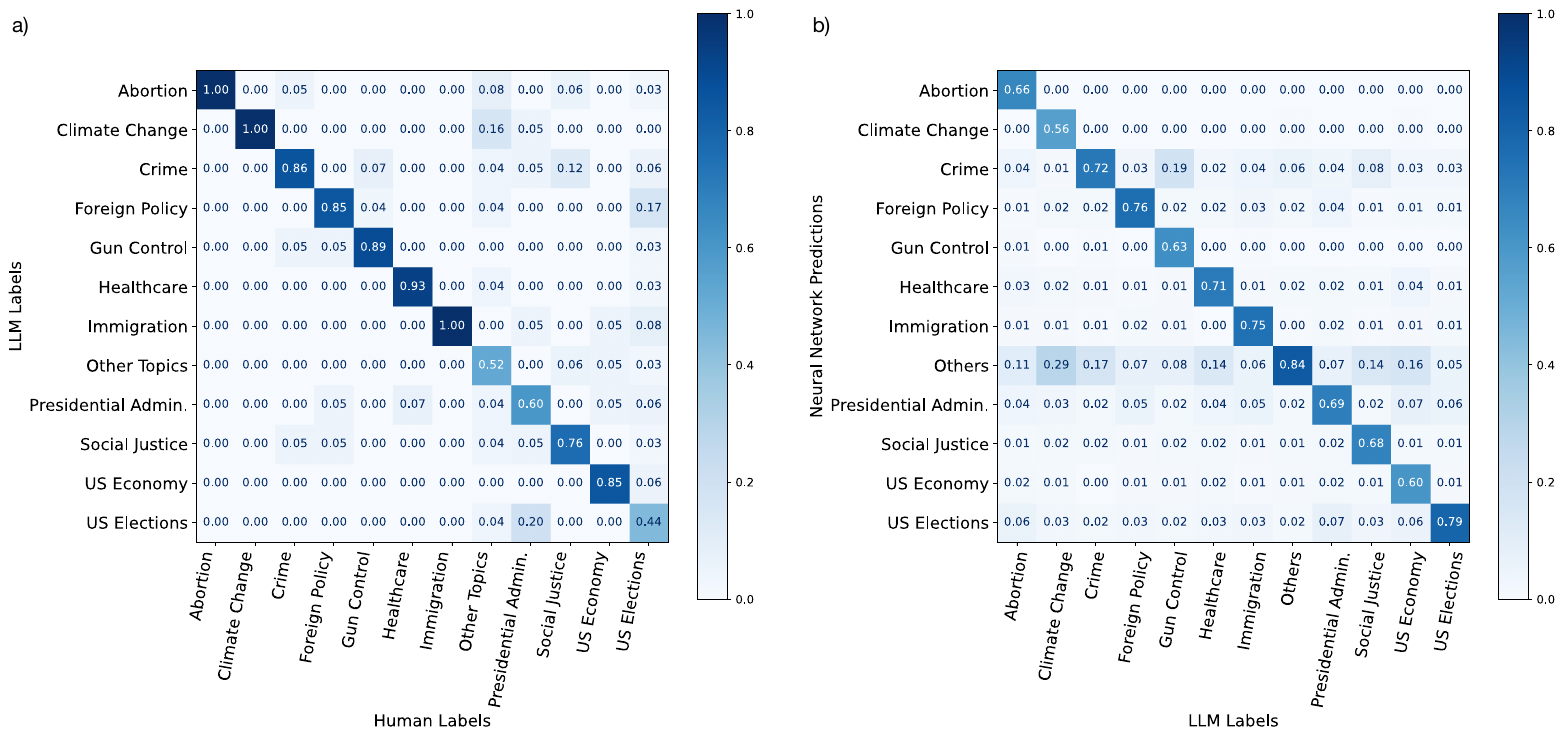}
\caption{\textbf{Confusion matrices showing model agreements on topic label.} (a) Agreement between the majority voted topic label among human annotators and GPT-4o topic labels on 250 news segments. Macro F1 = 0.78, macro precision = 0.77, macro recall = 0.81. For reference, a random guesser in a 12-way classification task would achieve F1 $\approx 0.08$. (b) Agreement between GPT-4o topic labels and topic predictions from the neural network model on 43,452 news segments. Macro F1 = 0.714, macro precision = 0.732, macro recall = 0.699.}
\label{fig:topics_humanlabelcomparisons}
\end{figure*}

Each news segment in the train and test sets was annotated by GPT-4o with one of twelve topics, spanning hallmark culture war issues (e.g., climate change, abortion and reproductive rights, gun control, immigration), core policy domains (e.g., the U.S. economy, healthcare, foreign policy), institutional themes (e.g., presidential administration, US elections, crime) and broader sociopolitical concerns (e.g., social justice and racial inequality) (see Figure~\ref{fig:prompt_topics} for prompt details). A twelfth category (`Other Topics') captured segments that did not match any of the predefined topics, which could include non-political news (e.g., weather, sports, natural disasters) and other relatively less politicized topics (e.g., education debates, transportation). 
We selected topics that consistently feature in partisan discourse and are frequently framed through identity-based or ideological language in cable news. While issues like education and transportation do enter political debate, they tend to be less intensely and less consistently politicized in televised news coverage and were therefore excluded from our predefined set. Our stratified sample of 250 news segments was also labeled for one of the 12 topic categories by the three trained research assistants. The LLM-generated topic labels showed strong agreement with human raters (see Figure~\ref{fig:topics_humanlabelcomparisons}a; macro F1 = 0.78, macro precision = 0.77, macro recall = 0.81). Furthermore, a bootstrap test with 10,000 resamples shows that agreement between the LLM-generated topic labels and the individual raters ($\kappa = 0.7045$) is statistically indistinguishable ($p$-value = 0.4958) from inter-rater agreement among human raters ($\kappa =0.6591$).

\begin{figure}[t!]
\centering
\begin{promptbox}

\footnotesize	
    \vspace*{0.1cm}
    Consider the following TV news transcript for Topic classification:
    \\
    Transcript:
    \\
    \{Transcript\}
    \\
    
    \vspace*{0.3cm}
    
    Based on its content, determine the single most relevant topic for the following TV news transcript and classify it into only one of the predefined categories:
    \\
    (1) \textcolor{blue}{Climate Change} (for e.g., climate action plans, environmental regulations, renewable energy initiatives, debates over natural resource conservation)
    \\
    (2) \textcolor{blue}{Gun Control} (for e.g., gun laws, mass shootings, background checks, NRA influence, debates on gun safety)
    \\
    (3) \textcolor{blue}{Immigration} (for e.g., border security, immigration reform, asylum policies, DACA, debates over immigration status)
    \\
    (4) \textcolor{blue}{Abortion and Reproductive Rights} (for e.g., Abortion laws and access, Roe v.\ Wade and its aftermath, state-level restrictions, reproductive healthcare)
    \\
    (5) \textcolor{blue}{Social Justice and Racial Inequality} (for e.g., racial justice, policing, LGBTQ+ rights, affirmative action, movements like Black Lives Matter)
    \\
    (6) \textcolor{blue}{US Economy} (for e.g., economic growth, unemployment, inflation, minimum wage debates, labor rights, taxation policies)
    \\
    (7) \textcolor{blue}{Foreign Policy} (for e.g., US relations with global powers, military interventions, terrorism, cybersecurity threats, defense spending)
    \\
    (8) \textcolor{blue}{US Elections} (for e.g., presidential and congressional elections, voter suppression, election laws, redistricting)
    \\
    (9) \textcolor{blue}{Healthcare} (for e.g., healthcare policy, Medicare, Medicaid, the Affordable Care Act, COVID-19 pandemic responses, mental health access)
    \\
    (10) \textcolor{blue}{Crime} (for e.g., violent crime, property crime, law enforcement actions, criminal court cases, debates over crime rates)
    \\
    (11) \textcolor{blue}{Presidential Administration} (e.g., actions, appointments, policies, or controversies related to the sitting or incoming U.S. president or their cabinet and staff)
    \\
    
    \vspace*{0.3cm}
    
    Select only one topic - the one that is most substantially discussed in the transcript. A topic should only be assigned if it is substantially discussed, not simply mentioned once. If multiple topics from the above list are present, pick the one most substantially covered in the transcript. If none of the above-listed topics are discussed, label it as "Others". Also, provide a short explanation for your label in fewer than 30 words. Do not make up new topics.\\~\\
    
    Answer only in the following format: \\
    Topic: <single topic name>\\
    Explanation: <explanation>
    
    \vspace*{0.1cm}
    
\end{promptbox}
\caption{\textbf{Prompt for topic classification.} Prompt used with GPT-4o model to label topics in TV news transcripts. The ordering of the 11 topics in the prompts was randomized in every prompt so that the output is not biased by the topic order in the prompt.}
\label{fig:prompt_topics}
\end{figure}
\FloatBarrier

\FloatBarrier

We then trained a supervised multi-class neural network model for topic prediction, which achieved a high predictive performance on our held-out test set of 43,452 news segments (see Figure~\ref{fig:topics_humanlabelcomparisons}b; macro F1 = 0.714, macro precision = 0.732, macro recall = 0.699), as well as on our human-labeled set of 250 news segments (macro F1 = 0.620, macro precision = 0.651, macro recall = 0.619), demonstrating the reliability of our framework in labeling topics in TV news segments at scale.

\begin{figure*}[h]
\centering
\includegraphics[width=0.9\textwidth]{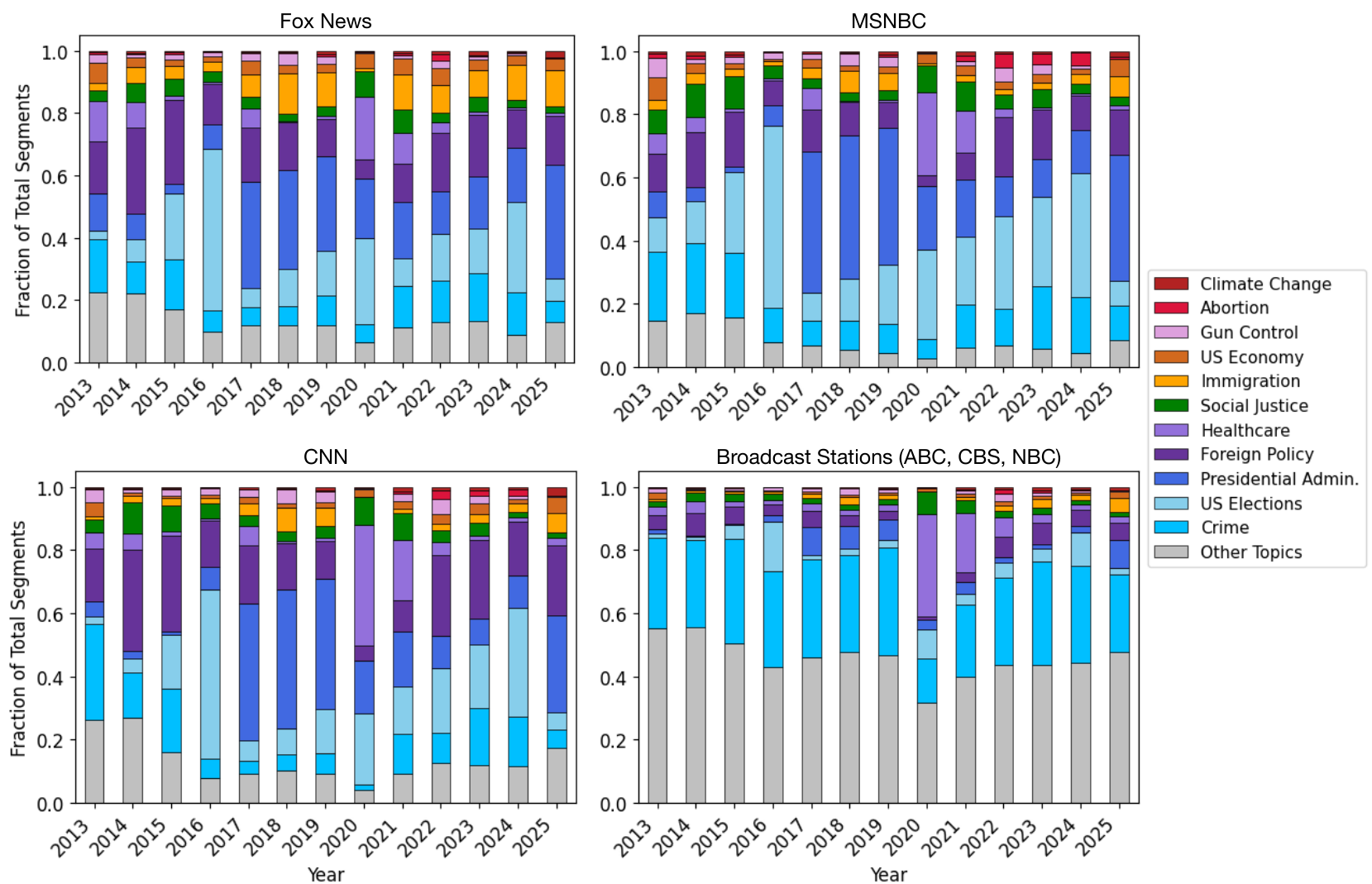}
\caption{\textbf{Coverage of 12 topic categories across years in Fox News, MSNBC, CNN, and all Broadcast stations combined.} }
\label{fig:topic_coverage_across_years}
\end{figure*}

Figure~\ref{fig:topic_coverage_across_years} presents the distribution of topic categories assigned to news segments from Fox News, MSNBC, CNN, and the three broadcast networks (ABC, CBS, NBC combined) across the full dataset (2013–2025). The observed coverage patterns largely align with known political and social events. For example, all stations devoted a substantial share of their segments to U.S. elections during the election years of 2016 and 2024, followed by increased coverage of presidential administrations in the years immediately following (2017 and 2025, respectively). Coverage of healthcare peaked sharply in 2020, coinciding with the onset of the COVID-19 pandemic. Similarly, abortion received relatively limited coverage in most years, but saw a notable increase in 2022, aligning with the Supreme Court decision in Dobbs v.\ Jackson Women’s Health Organization (597 U.S.\ 215, 2022), which overturned Roe v.\ Wade.

To assess the robustness of our topic classification method, we conducted a secondary analysis that allowed for multi-label topic classification. The original topic assignment scheme restricted each news segment to a single dominant topic. However, because transcripts were segmented into blocks of approximately 800 words uniformly, it is possible that individual segments spanned more than one story. To account for this, we drew a stratified sample of 2,000 news episodes from 2013 to 2025 and re-annotated the segments using a modified prompt that permitted the assignment of multiple topics—each corresponding to a distinct section of the transcript.

\begin{figure*}[h]
\centering
\includegraphics[width=0.9\textwidth]{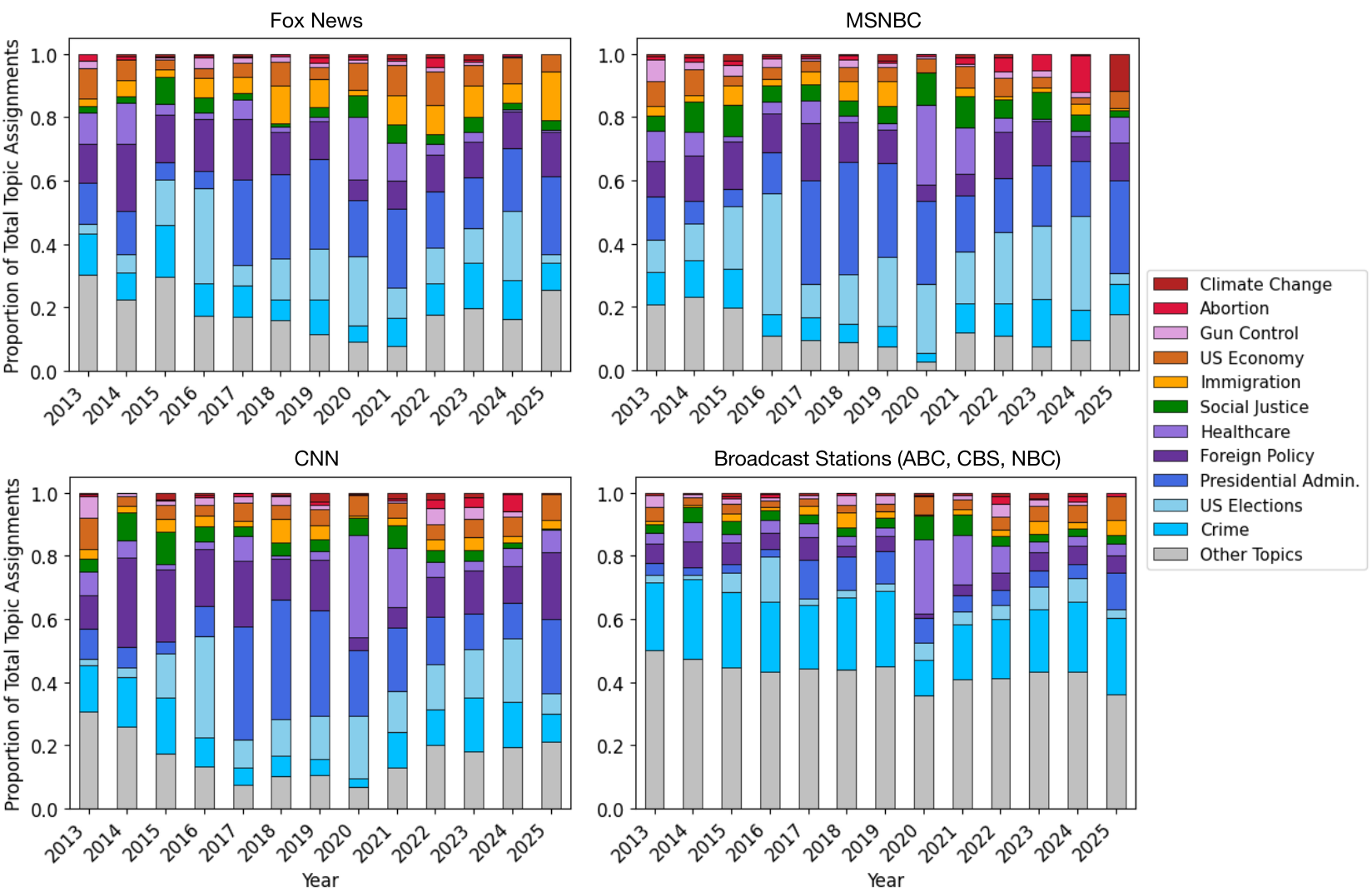}
\caption{\textbf{Coverage of 12 topic categories across years in Fox News, MSNBC, CNN, and all Broadcast stations combined, when multiple topics can be assigned to the same news segment.} }
\label{fig:topic_coverage_multipletopics}
\end{figure*}

Figure~\ref{fig:topic_coverage_multipletopics} shows that the resulting topic prevalence distributions under the multi-label scheme are qualitatively consistent with those from the single-label classification, thereby validating the stability of the observed topic trends. Because the overall patterns remain stable, we retain the single-topic classification for our main analyses to preserve clarity and comparability across topics. 

\begin{figure*}[t!]
\centering
\includegraphics[width=0.9\textwidth]{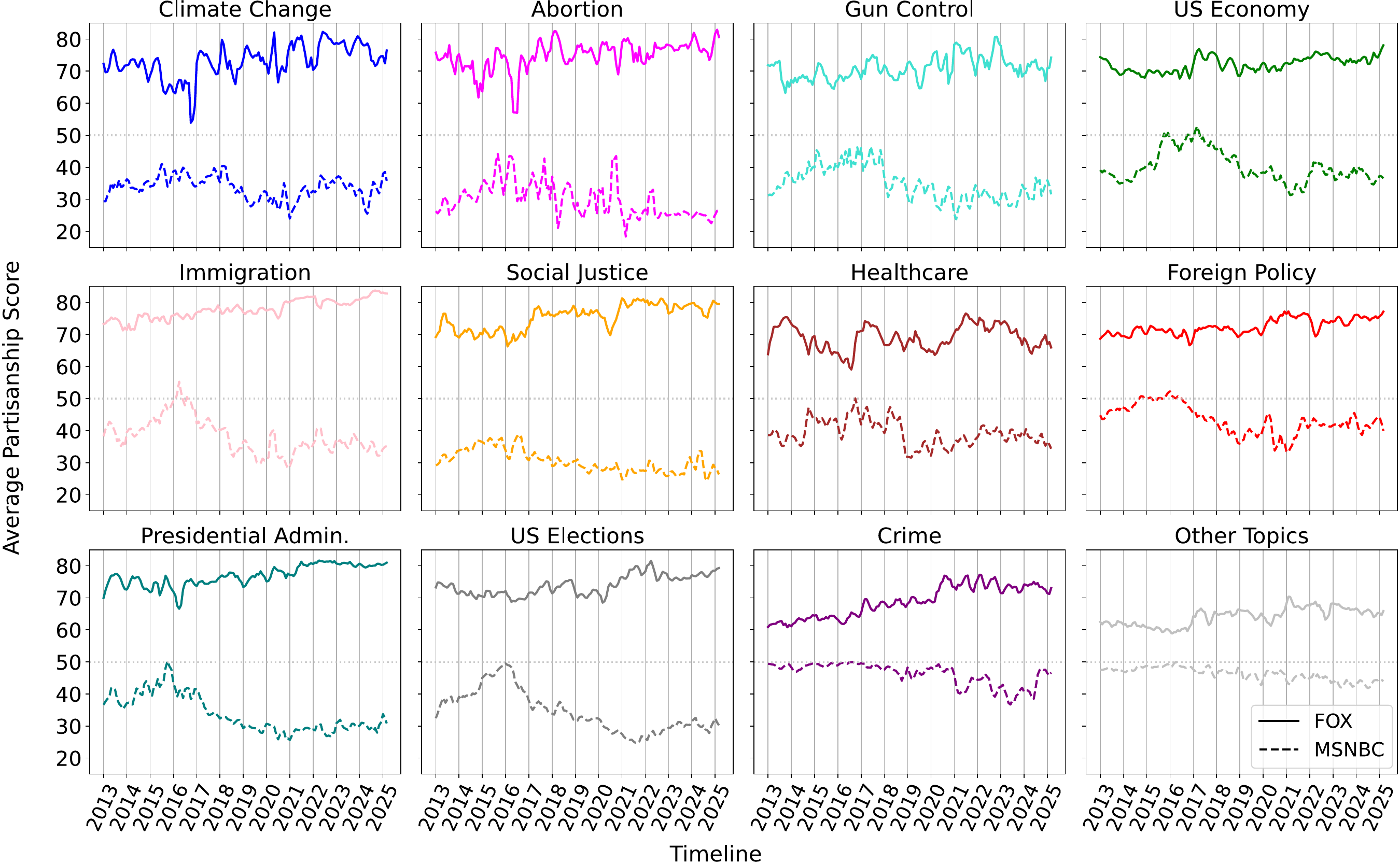}
\caption{\textbf{Average monthly partisanship of Fox News and MSNBC across 12 topic categories.} }
\label{fig:topic_level_partisanship_foxmsnbc}
\end{figure*}

Over the years, across nearly all topics, Fox News exhibited consistently more ideologically slanted coverage on average than MSNBC (Figure~\ref{fig:topic_level_partisanship_foxmsnbc}). This partisan asymmetry is especially pronounced in topics like immigration, the US economy, climate change, and foreign policy. Over time, while Fox News has grown steadily more conservative, MSNBC showed a marked shift toward more liberal coverage post-2016, leading to an increasingly pronounced partisan divide on topics such as U.S. elections, presidential administration, abortion, immigration, and social justice.

\FloatBarrier

\FloatBarrier

\FloatBarrier

\section{Robustness tests with alternative large language models}
\label{sec:robustness_analysis}

Because our measurement strategy relies on large language models (LLMs) to assign continuous partisanship scores, it is important to establish that our findings are not artifacts of a particular model’s latent political orientation. Prior work suggests that many state-of-the-art LLMs may exhibit detectable ideological tendencies on some tasks, with meaningful variation across architectures and training regimes~\cite{westwood2025measuring, santurkar2023whose, feng2023pretraining, haller2025leveraging}. In the main text, we report results based on OpenAI’s GPT-4o. In this section, we examine whether our conclusions are robust to alternative language models that differ in both ownership structure (proprietary versus open-weight) and ideological tendencies. These models include two closed, proprietary systems---Grok 4.1 Fast (xAI) and Gemini 2.5 Flash-Lite (Google)---and two open-weight models---Llama 4 Scout (Meta) and DeepSeek-v3.1 (DeepSeek). Existing benchmarking studies and audit reports suggest that these models differ in their aggregate political leanings. Grok models have been described as exhibiting a conservative bias relative to other models~\cite{rozado2024political, yasseri2025similar, valeriepoliticization2025}, whereas Gemini has been found to be more centrist~\cite{choudhary2024political, chandra2026evaluation}. Among the open-weight models, Llama-4 has been found to show political lean comparable to Grok~\cite{llama4releaseMeta}, while  DeepSeek-v3.1 has been found to subtly favor authoritarianism and a China-centric perspective~\cite{gupta2025comparative, deepseekcensorshipguardian}, without fitting on a standard U.S. liberal-conservative spectrum. By selecting models that span proprietary and open-weight architectures and that differ in their reported political orientations, we provide a stringent test of whether our empirical findings depend on the ideological bias of any single system.

\subsection{Agreement with human-coded partisanship scores}
\label{sec:llm_human_agreement}

\begin{figure*}[t!]
\centering
\includegraphics[width=0.99\textwidth]{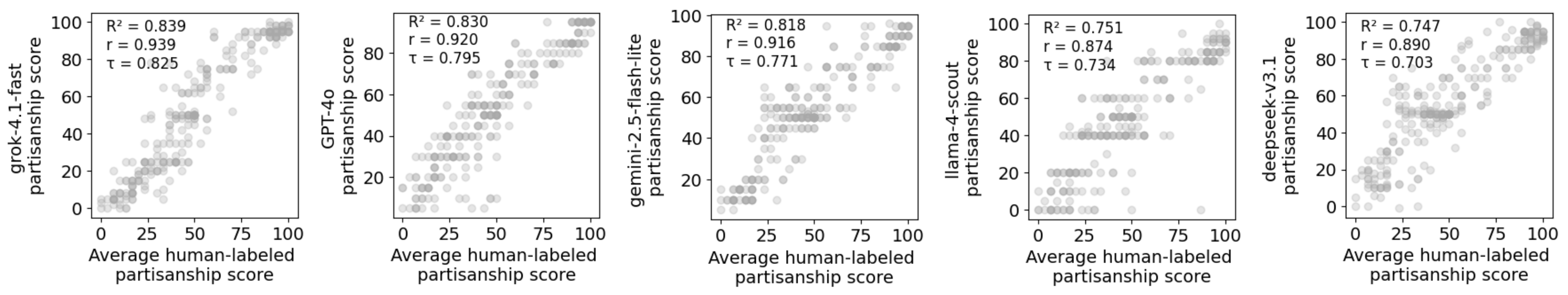}
\caption{\textbf{Agreement between LLM-labeled and human-labeled scores across five language models}. Coefficient of Determination $R^2$, Pearson's correlation $r$, and Kendall's $\tau$ rank correlation between LLM labels and average human-coded labels. Each point represents the model-assigned and the average human-labeled partisanship score for a television news segment. The proprietary models --- Grok-4.1-fast, GPT-4o, and Gemini-2.5-flash-lite --- show the strongest alignment with human annotations, while the open-weights models --- Llama-4-scout and Deepseek-v3.1 --- exhibit a comparatively weaker, although still high, alignment.}
\label{fig:llm_comparisons}
\end{figure*}

First, we evaluated the four alternative large language models on the set of 250 news segments for which we had collected human-coded partisanship labels. All models were evaluated using the same prompt to ensure consistency in how partisanship was assessed across LLMs. As shown in Figure~\ref{fig:llm_comparisons}, all the closed, proprietary models (Grok-4.1-fast, GPT-4o, and Gemini-2.5-flash-lite) show similar alignment with human annotations --- with $R^2$ values ranging from 0.818--0.839 --- despite differences in their reported political bias in past work. The open-weight models --- LLaMA-4-scout and Deepseek-v3.1 --- show comparatively lesser, although still high, alignment with human labels, with $R^2$ values of 0.747--0.751. Overall, these results demonstrate that high-quality partisanship measurement closely aligning with human annotations is not exclusive to the GPT-4o model, and that other state-of-the-art LLMs can also serve as reliable alternatives.

\FloatBarrier

\subsection{Replication of main results using alternative LLMs}
\label{sec:llm_comparison_appendix}

As a test of robustness, we re-estimated the partisanship scores, the topic labels, and the labels for the eight rhetorical strategies, using each of the four alternative LLMs, for the same sample of 10,000 episodes as described in \hyperref[sec:methods]{Materials and Methods}. Each model was prompted using the identical prompt used for GPT-4o (see Figure~\ref{fig:prompt_partisanship}, ~\ref{fig:prompt_topics}, ~\ref{fig:prompt_features} for prompt details), ensuring that any differences across models arise from differences in model behavior rather than differences in prompting or sampling. 

\begin{figure*}[t!]
\centering
\includegraphics[width=0.9\textwidth]{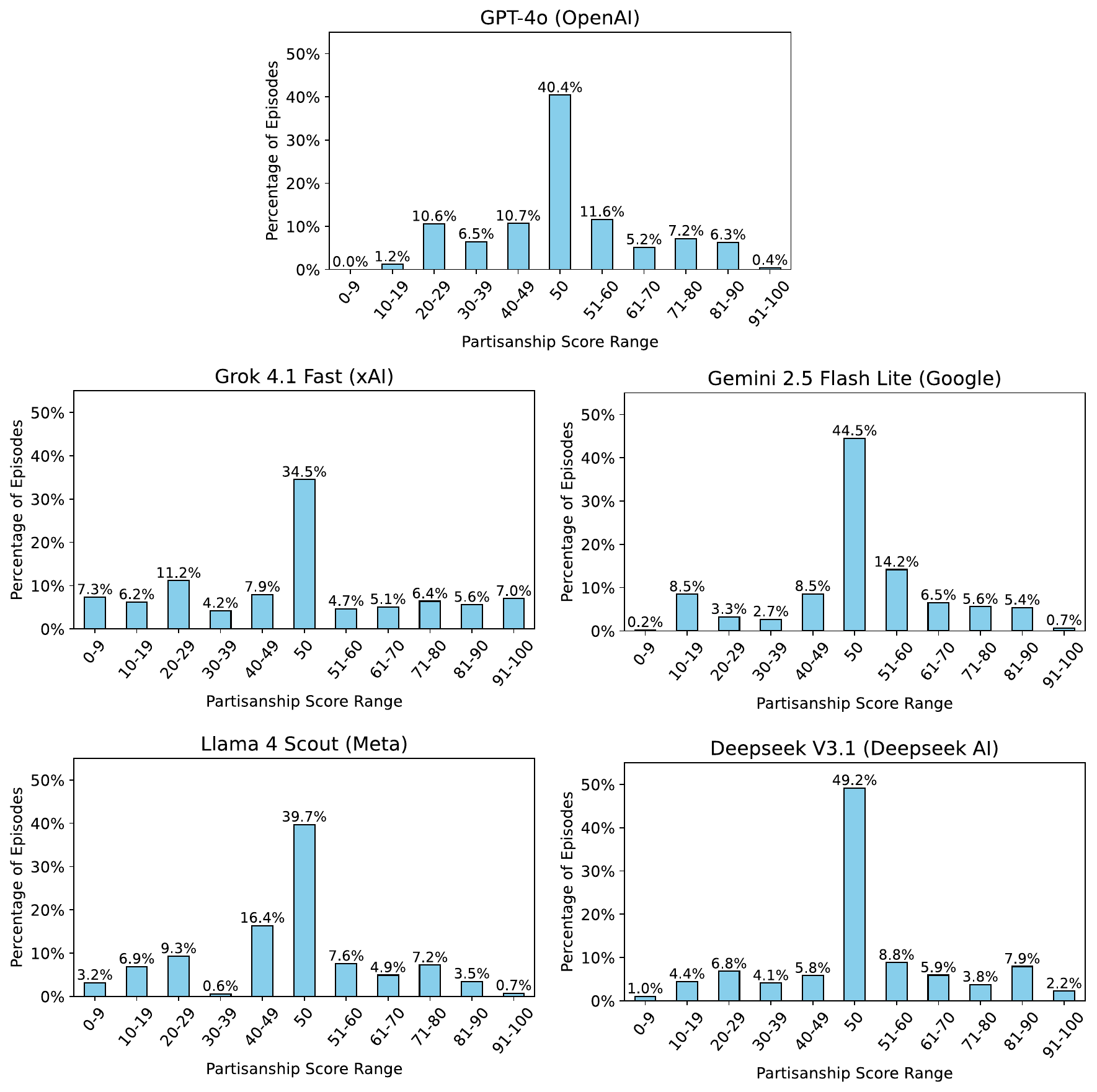}
\caption{\textbf{Distribution of partisanship scores of 10,000 uniformly sampled news episodes as labeled by five different LLMs.} }
\label{fig:appendix_llm_comparison_partisanship_histograms}
\end{figure*}

Across models, the overall distribution of episode-level partisanship scores remains highly similar, with only modest differences in dispersion and central tendency (Fig. \ref{fig:appendix_llm_comparison_partisanship_histograms}). Relative to GPT-4o, Grok-4.1-fast and Llama-4-scout produce somewhat flatter distributions, allocating slightly more mass to the tails of the ideological spectrum. In contrast, Gemini-2.5-flash-lite and DeepSeek-v3.1 generate more centered distributions. Despite these differences, the qualitative shape of the distributions remains substantively unchanged.

\begin{figure*}[t!]
\centering
\includegraphics[width=0.99\textwidth]{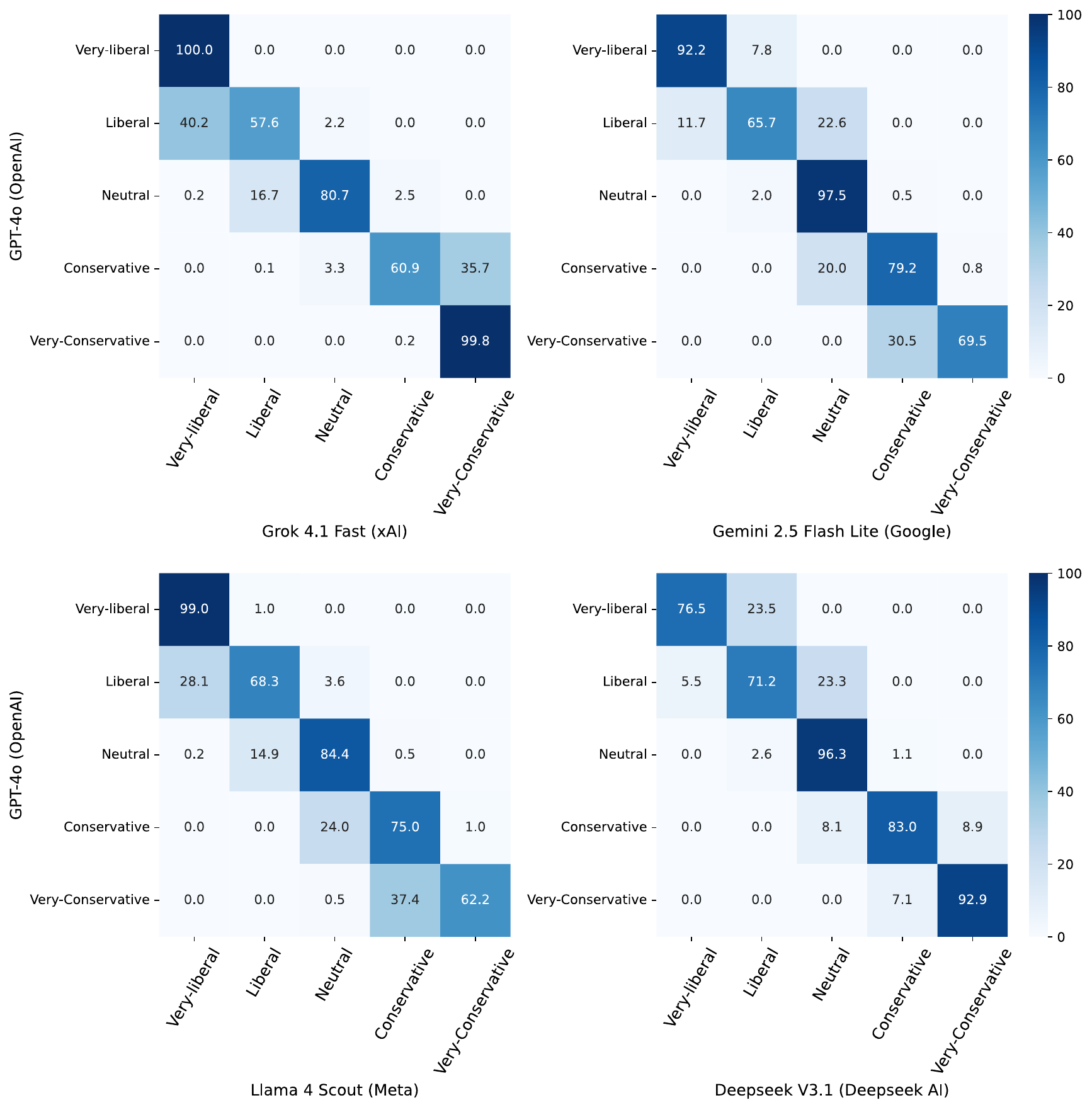}
\caption{\textbf{Confusion matrices showing the agreement between GPT-4o and 4 alternative LLMs on partisanship labels of 10,000 episodes.} The rows are normalized to sum up to 1. Relative to GPT-4o, Grok-4.1-fast tends to shift scores toward more extremes at both ends of the ideological spectrum, Gemini-2.5-flash-lite and DeepSeek-v3.1 pull the scores toward the center, while Llama-4-scout produces a modest leftward shift.}
\label{fig:appendix_llm_comparison_partisanship_confusionmatrix}
\end{figure*}

Pairwise comparisons further indicate high agreement across models, with some systematic deviations (Fig. \ref{fig:appendix_llm_comparison_partisanship_confusionmatrix}). Grok-4.1-fast tends to shift GPT-4o’s scores toward more extreme values at both ends of the ideological spectrum. Gemini-2.5-flash-lite systematically pulls scores toward the center, yielding fewer extreme ratings. Llama-4-scout produces a modest leftward shift relative to GPT-4o across the distribution. DeepSeek-v3.1 also centers scores relative to GPT-4o, though to a lesser extent than Gemini-2.5-flash-lite. Primarily, these differences reflect changes in scale or dispersion rather than reversals in relative ordering.

\begin{figure*}[t!]
\centering
\includegraphics[width=\textwidth]{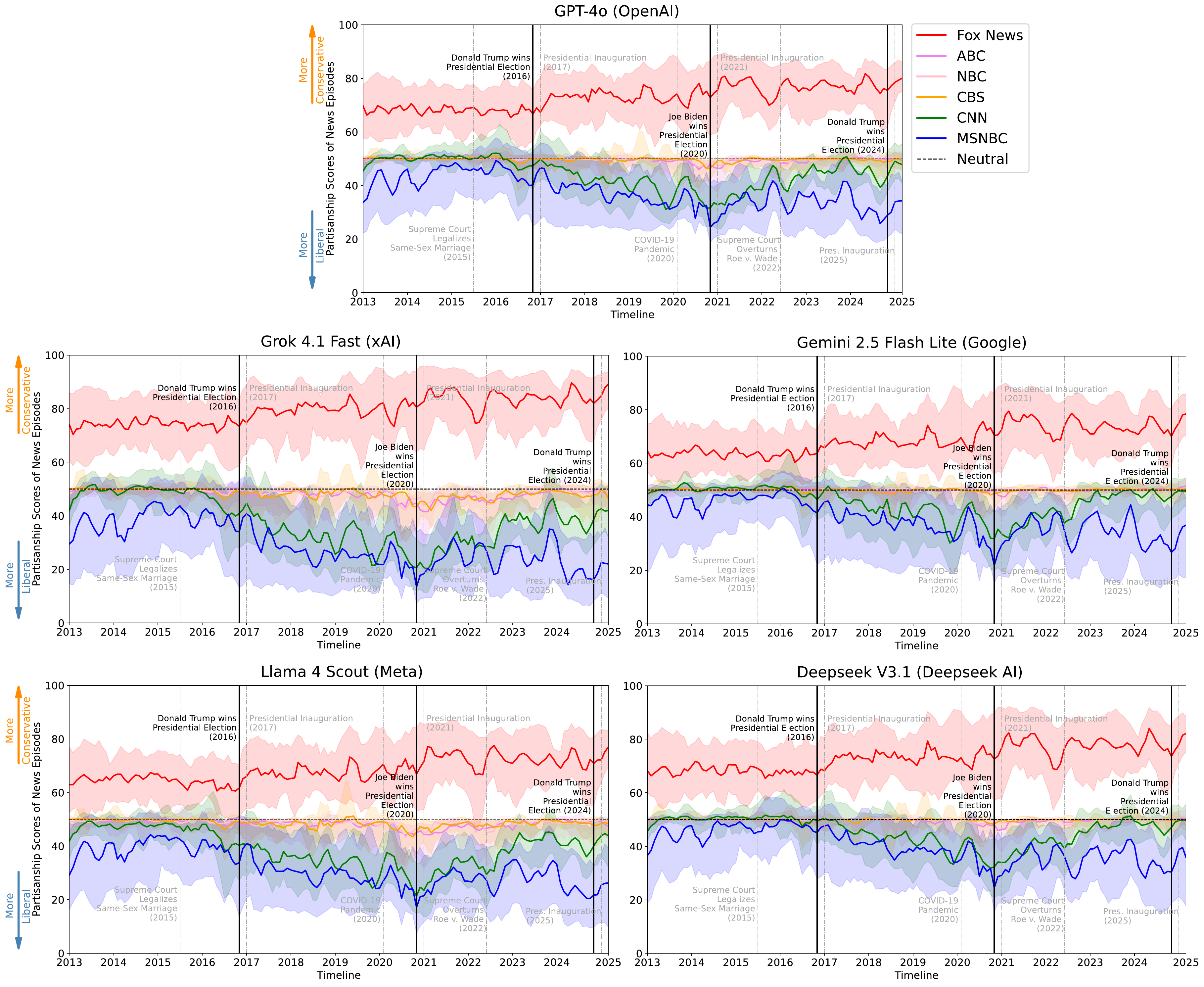}
\caption{\textbf{Partisanship patterns in U.S. television news across time, stations, and programs, using 10,000 episodes labeled by five different LLMs.} }
\label{fig:appendix_llm_comparison_partisanship_timeseries}
\end{figure*}

\begin{figure*}[t!]
\centering
\includegraphics[width=0.7\textwidth]{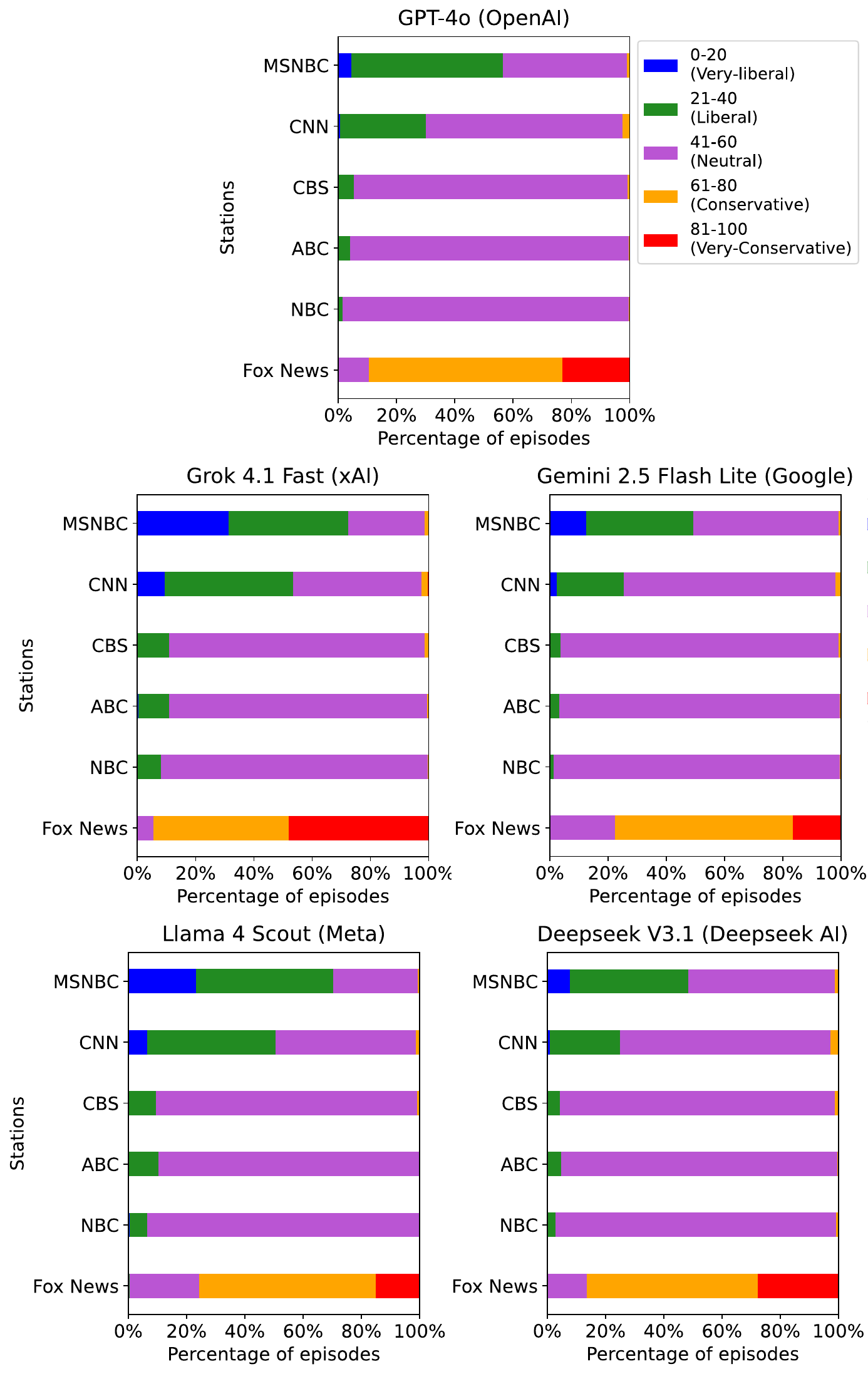}
\caption{\textbf{Distribution of episode-level partisanship scores of 10,000 episodes labeled by five different LLMs, categorized into five ideological bins.} }
\label{fig:appendix_llm_comparison_partisanship_stackedbars}
\end{figure*}

\begin{figure*}[t!]
\centering
\includegraphics[width=\textwidth]{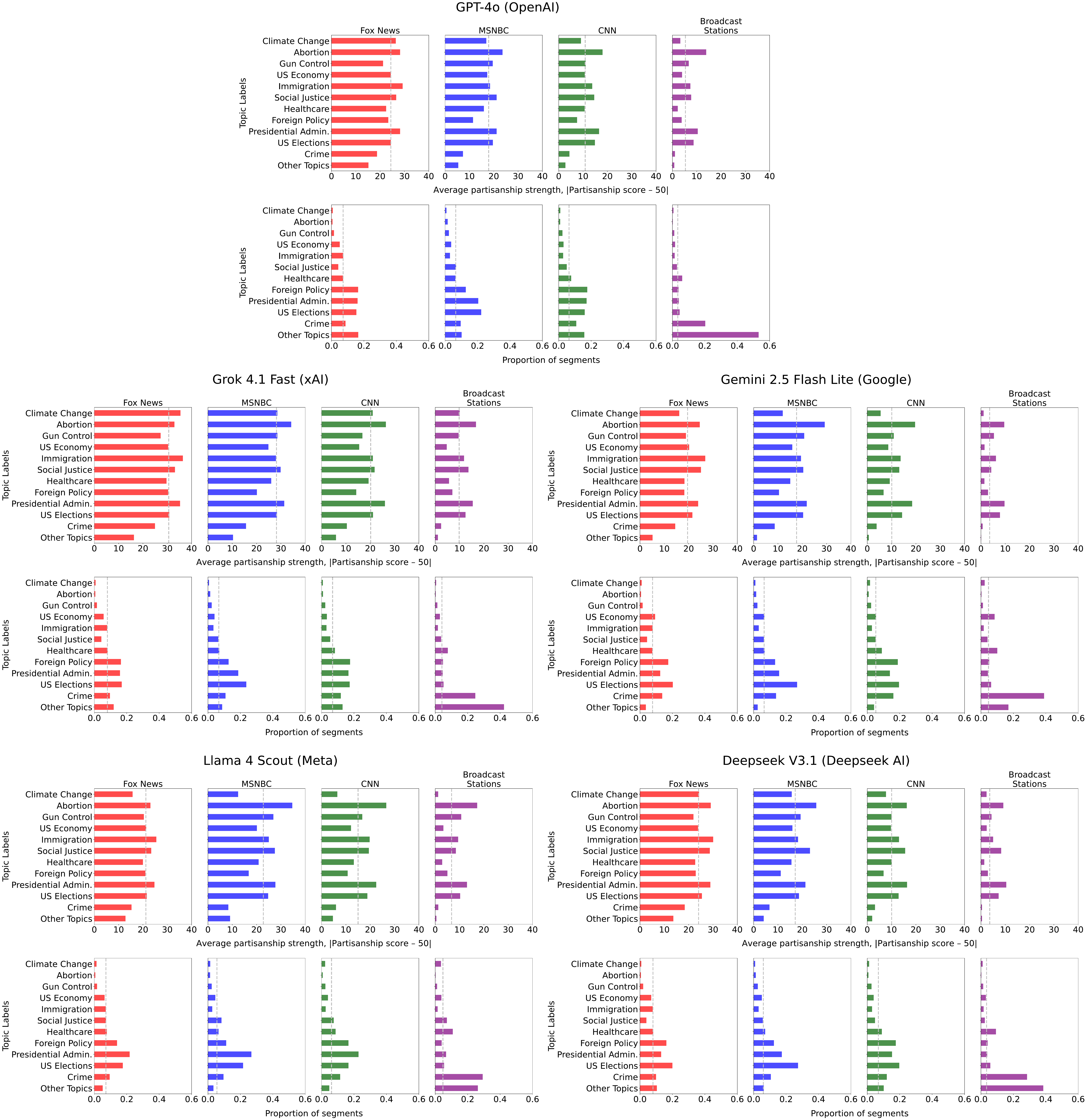}
\caption{\textbf{Topic distribution and topic-specific partisanship of 10,000 episodes across news stations (2013–2025), as labeled by five different LLMs.} }
\label{fig:appendix_llm_comparison_topics}
\end{figure*}

\begin{figure*}[t!]
\centering
\includegraphics[width=\textwidth]{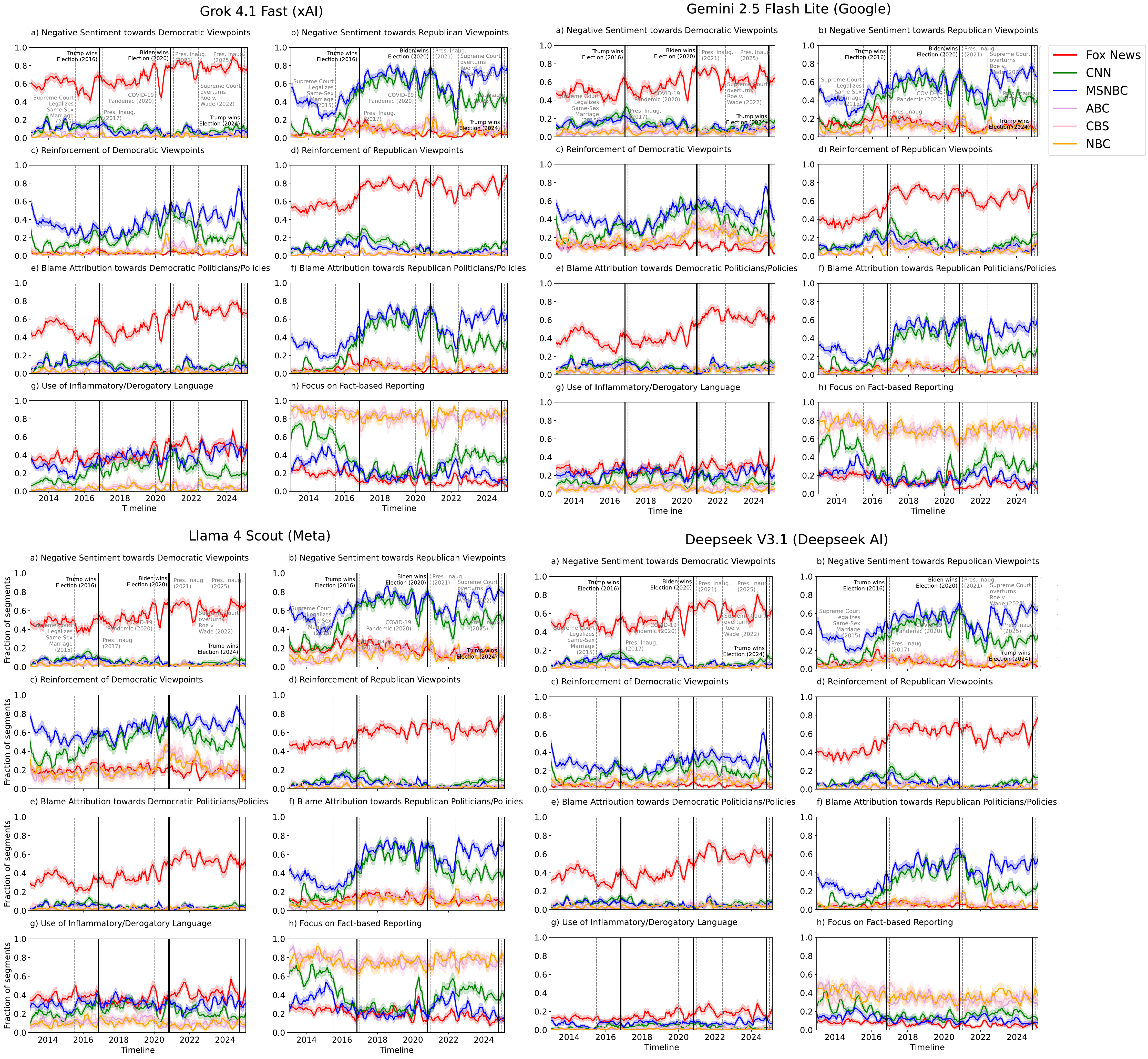}
\caption{\textbf{Temporal trends in rhetorical strategies of 10,000 episodes across broadcast and cable TV news, as labeled by four different LLMs.} The trends with GPT-4o are not shown here due to space constraints. Refer to main text Figure~\ref{fig:feature_timetrends} for trends with GPT-4o.}
\label{fig:appendix_llm_comparison_biasmarkers_timeseries_V2}
\end{figure*}

\begin{figure*}[t!]
\centering
\includegraphics[width=0.9\textwidth]{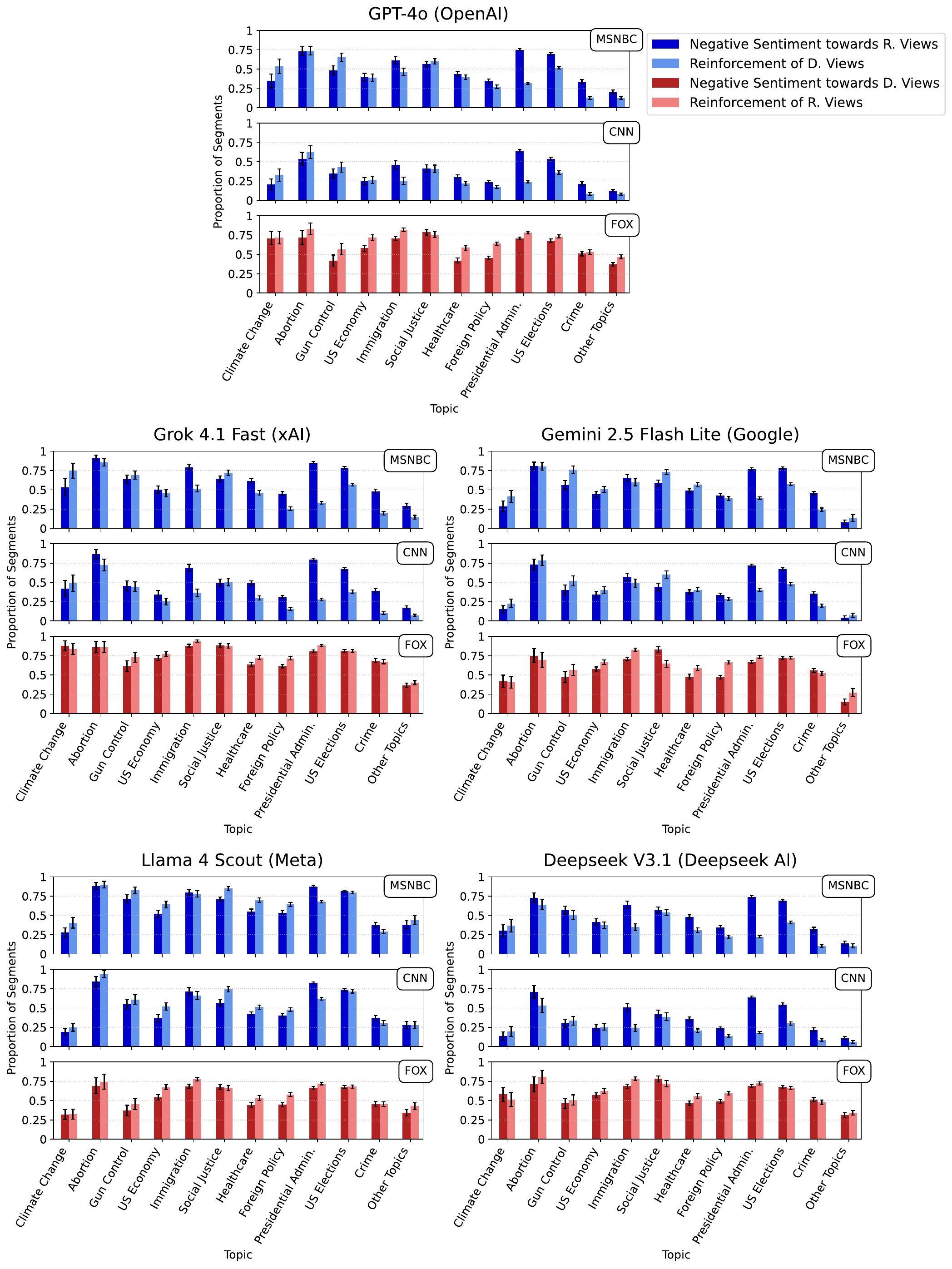}
\caption{\textbf{Topic-specific variation in attitudinal and stylistic strategies across cable news stations, as labeled by five different LLMs.}}
\label{fig:appendix_llm_comparison_biasmarkers_topicinteraction}
\end{figure*}

Most importantly, all substantive conclusions reported in the main text on trends in partisanship, topic, and rhetorical
strategies (Figs.\ \ref{fig:partisanship_trend}--\ref{fig:topicfeature_interactions}) replicate across the four alternative LLMs (see Figures~\ref{fig:appendix_llm_comparison_partisanship_timeseries}, \ref{fig:appendix_llm_comparison_partisanship_stackedbars}, \ref{fig:appendix_llm_comparison_topics}, \ref{fig:appendix_llm_comparison_biasmarkers_timeseries_V2}, \ref{fig:appendix_llm_comparison_biasmarkers_topicinteraction}). For example, across all models, Fox News is estimated to be more conservative than MSNBC is liberal. Following the 2016 presidential election, MSNBC and CNN exhibited a marked increase in liberal alignment, followed by moderation after the 2020 election; by contrast, Fox News increased its conservative alignment after both elections, irrespective of electoral outcomes. Similarly, across models, we find that MSNBC and CNN devoted greater emphasis to criticizing Republican positions than to affirming Democratic positions, whereas Fox News reinforced Republican viewpoints substantially more than it attacked Democratic positions.

Taken together, these analyses demonstrate that our empirical findings do not depend on the specific LLM. The consistency of results across proprietary and open-weight models with varying political orientations provides strong evidence that our conclusions are not artifacts of our choice of model.

\FloatBarrier

\FloatBarrier

\FloatBarrier

\FloatBarrier

\FloatBarrier

\FloatBarrier

\FloatBarrier

\end{document}